\documentclass[useAMS,usenatbib,usegraphicx]{mn2e}
\usepackage{times,amssymb,hyperref,subfigure,epsfig,aas_macros}

\newcommand{\fdfssym}{\xi}

\begin{document}

\title[The exo-Zodi Luminosity Function]{The bright end of the exo-Zodi luminosity
  function: Disk evolution and implications for exo-Earth detectability}

\author[G. M. Kennedy, M. C. Wyatt]
{G. M. Kennedy\thanks{Email:
    \href{mailto:gkennedy@ast.cam.ac.uk}{gkennedy@ast.cam.ac.uk}} \& M. C. Wyatt \\
  Institute of Astronomy, University of Cambridge, Madingley Road, Cambridge CB3 0HA,
  UK\\
}
\maketitle

\begin{abstract}
  We present the first characterisation of the 12$\mu$m warm dust (``exo-Zodi'')
  luminosity function around Sun-like stars, focussing on the dustiest systems that can
  be identified by the WISE mission. We use the sample of main-sequence stars observed by
  Hipparcos within 150pc as an unbiased sample, and report the detection of six new warm
  dust candidates. The ages of five of these new sources are unknown, meaning that they
  may be sites of terrestrial planet formation or rare analogues of other old warm dust
  systems. We show that the dustiest old ($>$Gyr) systems such as BD+20 307 are 1 in
  10,000 occurrences. Bright warm dust is much more common around young ($<$120Myr)
  systems, with a $\sim$1\% occurrence rate. We show that a two component \emph{in situ}
  model where all stars have initially massive warm disks and in which warm debris is
  also generated at some random time along the stars' main-sequence lifetime, perhaps due
  to a collision, can explain the observations. However, if all stars only have initially
  massive warm disks these would not be visible at Gyr ages, and random collisions on the
  main-sequence are too infrequent to explain the high disk occurrence rate for young
  stars. That is, neither component can explain the observations on their own. Despite
  these conclusions, we cannot rule out an alternative dynamical model in which comets
  are scattered in from outer regions because the distribution of systems with the
  appropriate dynamics is unknown. Our \emph{in situ} model predicts that the fraction of
  stars with exo-Zodi bright enough to cause problems for future exo-Earth imaging
  attempts is at least roughly 10\%, and is higher for populations of stars younger than
  a few Gyr. This prediction of roughly 10\% also applies to old stars because bright
  systems like BD+20 307 imply a population of fainter systems that were once bright, but
  are now decaying through fainter levels. Our prediction should be strongly tested by
  the Large Binocular Telescope Interferometer, which will provide valuable constraints
  and input for more detailed evolution models. A detection fraction lower than our
  prediction could indicate that the hot dust in systems like BD+20 307 has a cometary
  origin due to the quirks of the planetary dynamics. Population models of comet delivery
  need to be developed to help distinguish between different possible origins of warm
  dust.
\end{abstract}


\section{Introduction}\label{s:intro}

Perhaps the most important long term goal of humankind is to find and communicate with
aliens, the inhabitants of extra-Solar Systems. This goal is a difficult one, with many
challenges to overcome before it is technically possible to even image an Earth-like
planet around another star. When it is eventually attempted, imaging detection of these
planets could still be thwarted by the presence of exo-Zodiacal dust located in or near
the terrestrial zone, which with sufficient dust levels can mask the presence of
exo-Earths
\citep[e.g.][]{2006ApJ...652.1674B,2010ASPC..430..293A,2012PASP..124..799R}. Therefore,
in preparation for future missions that will target specific stars, a critical piece of
information is whether those stars harbour dust at a level that precludes the detection
of an Earth-like planet.

While such knowledge will be crucial in specific cases, stars searched for warm dust may
not be targeted in an unbiased way that yields information on the underlying distribution
and its origins. For example, the most promising targets may be deemed to be those
without cool Edgeworth-Kuiper belt analogues, or with other specific properties that may
unbeknownst to us correlate with the presence of faint warm dust. The results cannot
therefore be used to construct the brightness distribution (a.k.a. luminosity function,
defined formally at the start of section \ref{s:xs}) of warm dust belts, or to make
predictions for stars that were not observed or about which no dust was
detected. Therefore, an equally important goal is to characterise the luminosity
function, with the overall aim of understanding warm dust origin and evolution.

Evolution is of key importance here, because the brightest warm dust sources seen around
nearby stars cannot necessarily maintain the same brightness indefinitely. The observed
emission comes from small particles that are eventually lost from the system, and the
very existence of emission for longer than terrestrial zone orbital timescales---a few
years---means that mass in small grains is being lost and replenished, and the small
particles necessarily originate in a finite reservoir of larger objects. The question is
whether those parent bodies are located at the same distance from the star as the dust
(what we call an \emph{``in situ''} scenario) or at a much more distant location (what we
call a ``comet delivery'' scenario).

The possibility of such different origins means that the rate at which warm dust decays
is unknown. For example, if the observed dust has its origins in an outer
Edgeworth-Kuiper belt analogue and is scattered inward by planets \citep[i.e. a comet
delivery scenario,][]{2010ApJ...713..816N,2012MNRAS.420.2990B,2012A&A...548A.104B}, the
warm dust could be extremely long lived. Given the lack of evolution seen for
main-sequence exo-Kuiper belts around Sun-like stars \citep[e.g.][]{2008ApJ...674.1086T},
dust levels may not appear to decay at all during the stellar main-sequence lifetime. If,
however, the dust originates near where it is observed (i.e. \emph{in situ}), it would
decay due to a decreasing number of parent bodies via collisions in a way that is
reasonably well understood
\citep[e.g.][]{2003ApJ...598..626D,2007ApJ...658..569W,2008ApJ...673.1123L}. In this
scenario it also follows that the brightest warm disks are necessarily the tip of an
iceberg of fainter disks; the decay timescale depends on the amount of mass that is
present so systems decay as $1/$time \citep[e.g.][]{2003ApJ...598..626D}, and for every
detection of bright warm dust that was created relatively recently, many more exist that
are older and fainter, and are slowly grinding themselves to oblivion. Another conclusion
from this kind of evolution is that the bright warm dust systems seen around relatively
old stars cannot arise from \emph{in situ} decay that started when the star was born;
projecting the $1/$time decay back to early times predicts implausibly large initial
brightnesses, and delaying the onset of this decay for $\sim$Gyr timescales within a
steady-state scenario requires implausibly large or strong planetesimals
\citep{2007ApJ...658..569W}.

These two different origins present very different structural and dynamical pictures of
planetary systems, which may manifest differently in the warm dust luminosity
function. For example, an observed luminosity function similar to that expected from
$1/$time decay would be a smoking gun for \emph{in situ} origin and evolution. However,
because the luminosity function in the comet delivery scenario is at least in part set by
the (unknown) distribution of systems with the appropriate structural and dynamical
properties, and the source regions may themselves be decaying as $1/$time, the
observation of such a function would not be absolute proof. As a further complication, it
may be that the processes that generated the dust in BD+20 307-like systems (i.e.  dusty
warm systems whose host stars are $\sim$Gyr old) are completely different to those that
result in the faintest systems. It could for example be that they are the result of late
(i.e. $\sim$Gyr) dynamical instabilities \citep{2005Natur.436..363S}, while faint
Asteroid belt and Zodiacal cloud-like systems are the decay products of young massive
belts that emerged from the protoplanetary disk phase or arise due to comet
delivery. Therefore, our work here must be complemented by studies that for example aim
to discover whether warm dust is correlated with the presence of cooler dust
\citep[e.g.][]{Absil13}.

Here we use the stars observed by Hipparcos \citep{2007A&A...474..653V} as an unbiased
sample to characterise the bright end of the warm dust luminosity function using mid-IR
photometry from the WISE mission \citep{2010AJ....140.1868W}. A secondary goal is to
discover new bright warm disks, though few discoveries are expected among these stars due
to previous IRAS and AKARI studies with a similar goal
\citep[e.g.][]{1991ApJ...368..264A,1992A&AS...96..625O,2005Natur.436..363S,2010ApJ...714L.152F,2012ApJ...749L..29F,2012Natur.487...74M}. The
key difference between those studies and ours is that we count the non-detections, and
therefore quantify the rarity of bright warm dust systems. We first outline our sample
(\S \ref{s:sample}) and excess (\S \ref{s:xs}) selection, and then cull disk candidates
from these excesses (\S \ref{s:cand}). We go on to compare the luminosity function with a
simple \emph{in situ} evolutionary model and make some basic predictions within the
context of this model for future surveys for faint warm dust (\S
\ref{ss:model}). Finally, we discuss some of the issues with our model and comet delivery
as an alternative (but equally plausible) exo-Zodi origin (\S \ref{s:disc}).

\section{Sample definition}\label{s:sample}

Stars are selected from the Hipparcos catalogue \citep{2007A&A...474..653V} using an
observational HR diagram. Because warm dust and planets have been discovered over a wide
range of spectral types that might be taken as ``Sun-like'', we use a fairly broad
definition of $0.3 < B_T - V_T < 1.5$, in addition only keeping stars with $B_T<11$ to
ensure the Tycho-2 photometric uncertainties are not too large
\citep{2000A&A...355L..27H}, particularly for faint cool M types and white dwarves that
otherwise creep into the sample. Using the Tycho-2 Spectral Type Catalogue
\citep{2003AJ....125..359W}, the main-sequence spectral types whose most common $B_T-V_T$
colours are 0.3 and 1.5 are A8 and K5 respectively. While stars have a range of colours
at a given spectral type, the bulk of the spectral types in our sample will lie between
these bounds. To ensure that we do not select heavily reddened or evolved stars that are
wrongly positioned in the HR diagram, we only retain stars within 150pc, and restrict the
parallax S/N of stars to be better than 5. Finally, we exclude stars within 2$^\circ$ of
the Galactic plane, as we found that many WISE source extractions, even for relatively
bright stars, are unreliable there. With these cuts 27,333 stars remain.

\begin{figure}
  \begin{center}
    \hspace{-0.5cm} \includegraphics[width=0.5\textwidth]{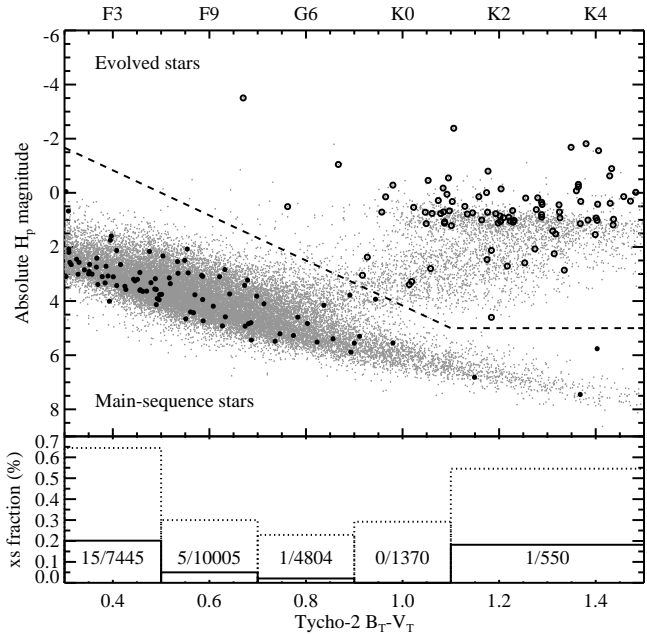}
    \caption{\emph{Top:} HR diagram showing stars with $0.3 < B_T-V_T < 1.5$, split into
      Sun-like main-sequence stars and evolved stars by the dashed line. The top axis
      shows spectral types calculated using synthetic photometry of PHOENIX stellar
      atmosphere models \citep{2005ESASP.576..565B}. Stars with $W1-W3$ greater than
      0.1mag are shown as circles; open for evolved stars and filled for main-sequence
      stars. \emph{Bottom:} Fraction of main-sequence stars with $W1-W3>0.1$. The dotted
      bars show all stars with $W1-W3>0.1$ (i.e. filled dots in the top panel), while the
      solid bars and numbers show those with plausible warm dust emission after
      individual checking (see \S \ref{s:cand}).}\label{fig:hr}
  \end{center}
\end{figure}

We exclude evolved stars, which may show excesses due to circumstellar material
associated with mass loss, by excluding stars with $M_{H_p} > 5$ when $B_T-V_T > 1.1$,
and with $M_{H_p} > 25/3 (B_T-V_T - 0.5)$ when $B_T-V_T \le 1.1$ (3,159
stars).\footnote{$H_p$ is a Hipparcos-specific bandpass, whose response peaks at about
  5000\AA~with a full width at half-maximum sensitivity of about 2000\AA. See
  \citet{2005ARA&A..43..293B} for a comparison with other passbands.} This cut is based
purely on by-eye inspection of Figure \ref{fig:hr}, but as long as it lies in the
Hertzsprung gap the sample size is only weakly sensitive to actual position because there
are few stars there. The line is broken at 1.1 because a single line does not
simultaneously exclude most stars crossing the Hertzsprung gap and include
pre-main-sequence stars such as HD 98800 (the rightmost filled circle at $B_T-V_T=1.4$,
$H_p=6$). Because they are selected solely using optical photometry and parallax, the
24,174 stars selected by these criteria are unbiased with respect to the presence of warm
dust. The Hipparcos HR diagram for the sample is shown in Figure \ref{fig:hr}, which
shows all 27,333 stars within the $B_T-V_T$ range, the cut used to exclude evolved stars
(dashed line), and those found to have 12$\mu$m excesses by the criteria described below
(large dots and circles).


\subsection{Age distribution}\label{ss:ages}

\begin{figure}
  \begin{center}
    \hspace{-0.5cm} \includegraphics[width=0.5\textwidth]{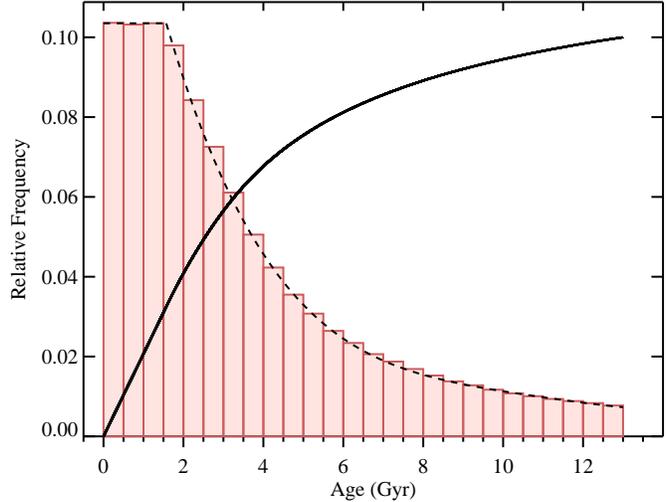}
  \end{center}
  \caption{Spectral type-weighted age distribution adopted for our sample. The line shows
    the cumulative fraction (divided by ten). Thus, about 50\% of our stars are younger
    than 2.5Gyr. The dashed line shows the constant relative frequency of 0.1035 and the
    fourth order polynomial fit used to generate synthetic age populations (see Appendix
    \ref{s:ages}).}\label{fig:agehist}
\end{figure}

It will become clear later that knowing the age distribution of our sample stars is
useful. The simplest assumption would be that the ages are uniformly
distributed. However, given the tendency for stars to have earlier types than the Sun
(Fig \ref{fig:hr}), this assumption yields an unrealistically high fraction of
$\sim$10Gyr old main-sequence stars (i.e. spectral types later than the Sun). Our sample
spans a sufficiently wide range of spectral types that the main-sequence lifetimes vary
significantly. For example, the earliest spectral types are late A-types, with $\sim$Gyr
main-sequence lifetimes, while the latest spectral types are late K-types, with
main-sequence lifetimes exceeding the age of the Universe. Stars are therefore assumed to
have ages distributed uniformly between zero and their spectral type-dependent
main-sequence lifetime, with the overall age distribution derived using the distribution
of $B_{\rm T} - V_{\rm T}$ colours (i.e. the colours are used as a proxy for the
distribution of spectral types, a full description is given in Appendix
\ref{s:ages}). The final age distribution is shown in Figure \ref{fig:agehist}, which we
assume throughout. Using a more simplistic assumption of uniformly distributed ages
between 0-10Gyr for all stars does not alter any of our conclusions.

\section{Excess selection}\label{s:xs}

If warm dust is sufficiently bright relative to the host star, it will manifest as a
detectable mid-infrared (i.e. 10-20$\mu$m) excess relative to the stellar photosphere,
which for $\sim$300K dust peaks near 12$\mu$m. The WISE bands are well suited for this
task, and because it has about an order of magnitude better sensitivity to 300K dust
compared to W4 \citep[see Fig 2 of][]{2012MNRAS.426...91K} we use the 12$\mu$m $W3$
band.\footnote{The WISE bands are known as $W1$, $W2$, $W3$, and $W4$, with wavelengths
  of 3.4, 4.6, 12, and 22$\mu$m respectively
  \citep{2010AJ....140.1868W,2011ApJ...735..112J}.}

Our luminosity function is therefore formally the distribution of 12$\mu$m disk-to-star
flux density ratios, which we will use in cumulative form. We use the term ``luminosity
function'' largely for conciseness. The disk luminosity of course not only depends on the
12$\mu$m flux ratio, but also the disk temperature and stellar properties. If the ratio
of the observed to stellar photospheric flux density is $R_{12} = F_{12}/F_\star$, and
the disk to star flux ratio $\fdfssym_{12} = R_{12} - 1$, then the luminosity function is
the fraction of stars with $\fdfssym_{12}$ above some level.

To derive the 12$\mu$m luminosity function, we would ideally have a sample for which the
sensitivity to these excesses is the same for all stars. Then the distribution is simply
the cumulative excess counts divided by the total sample size. Our use of Hipparcos stars
with the above parallax requirement ensures that this ideal is met; all stars in our
overall sample have better than 6\% photometry at 1$\sigma$ in $W3$, and 98\% have better
than 2.5\% photometry at 1$\sigma$. Given that the calibration uncertainty in this band
is 4.5\% \citep{2011ApJ...735..112J}, the fractional photometric sensitivity for all
sample stars is essentially the same at about 5\% 1$\sigma$. Therefore, the disk
sensitivity is ``calibration limited'' for our purposes
\citep[see][]{2008ARA&A..46..339W}, with a 5\% contribution from the WISE photometry, and
a $\sim$2\% contribution from the stellar photosphere. That is, we are limited to finding
3$\sigma$ 12$\mu$m excesses brighter than about 15\% of the photospheric level (flux
ratios of $R_{12} \ge 1.15$). This property means that we can select excesses by choosing
stars with a red $3.4-12$$\mu$m ($W1-W3$) colour.

Though our excess limit is a uniform 15\% due to the sample stars being bright, this
brightness brings other potential problems. Stars brighter than 8.1, 6.7, 3.8, and
-0.4mag are saturated in $W1-4$ respectively, so nearly all stars are saturated in the W1
band to some degree, while only about a hundred are saturated in W3. However, comparison
with 2MASS photometry shows that the WISE W1 source extraction is reliable well beyond
saturation and shows no biases up to about 4.5mag, and only about 400 stars are brighter
than this
level.\footnote{\href{http://wise2.ipac.caltech.edu/docs/release/allsky/expsup/sec6\_3c.html}{http://wise2.ipac.caltech.edu/docs/release/allsky/expsup/sec6\_3c.html}}
Therefore, saturation actually affects a very small fraction of our stars. However,
bright stars are the nearest to Earth and therefore potentially important for future
studies if warm dust can be discovered, so we need to be sure that we are not biased
against true excesses, i.e. that excesses selected based on $W1-W3$ are not
systematically missed due to some problem associated with $W3$. We did not find any signs
of $W3$-specific issues, but noticed that the S/N of $W1$ photometry depends fairly
strongly on stellar brightness, with the brightest (i.e. more saturated) stars having
lower S/N (i.e. stars brighter than about 7.25mag in $V_{\rm T}$ have S/N$<20$). The
increased $W1-W3$ scatter for these stars means some excesses may be missed for those
with $W1$ that happens to be scattered brighter than the actual brightness. This issue
affects $\sim$1000 stars, and we show in Section \ref{ss:missing} that we can rule out
missing excesses for the worst of these cases. Given that there is no evidence for a
problem with using $W1-W3$ to identify excesses, we now proceed and leave possible issues
with the WISE photometry to be identified in the individual source checking stage below.

\begin{figure}
  \begin{center}
    \hspace{-0.5cm} \includegraphics[width=0.5\textwidth]{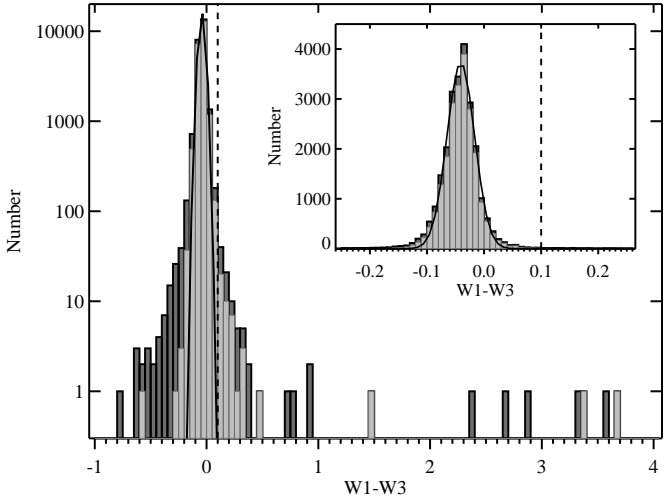}
    \caption{Histograms of $W1-W3$ colours for 24,174 main-sequence stars (i.e. those
      below the dashed line in Figure \ref{fig:hr}). The inset shows a smaller range of
      $W1-W3$ with a linear count scale. Dark grey bars show all sources, and light grey
      bars show only sources where the WISE extension flag is not set
      (i.e. \texttt{ext\_flg}=0). Curves show Gaussian fits to the overall sample. The
      dashed line shows the threshold of $W1-W3 = 0.1$mag for stars to become excess
      candidates (filled dots in Figure \ref{fig:hr}).}\label{fig:w1w3}
  \end{center}
\end{figure}

The $W1-W3$ distribution of the 24,174 stars is shown in Figure \ref{fig:w1w3}.  A
Gaussian fit centered on $W1-W3=-0.041$ with dispersion $\sigma=0.024$ is shown, the fit
being consistent with the photometric accuracy described above. The slightly negative
average colour arises because our stars are later spectral types than the reference A0V
spectrum used for Vega magnitudes. The logarithmic scale (main plot) shows that
systematic effects spread the distribution further for a relatively small number of
sources. We therefore looked for properties that appear to predict $W1-W3$ colours
significantly different to -0.041 as indicators of systematic effects. Stars with
companion sources at separations of 5-16\arcsec in the Hipparcos catalogue are more
likely to have $W1-W3<0.3$, as are WISE-saturated sources. We found that the best
indicator of negative $W1-W3$ was the WISE extension flag (\texttt{ext\_flg}), which
indicates sources not well described by the WISE point spread function, or those near
extended sources in the 2MASS Extended Source Catalogue. Aside from a few very bright
sources, there is little brightness dependence on $W1-W3$ colour, so it seems that the
bulk of sources with blue WISE colours are due to poor source extractions for confused
sources. Five of the warm dust candidates found below have $\texttt{ext\_flg}=1$, but
four are already known to host warm dust. For this reason, and because we subsequently
check them individually we cull candidates from the full list of 24,174 stars.

The WISE $W3$ calibration-limited sensitivity of about 5\%, combined with the 2 average
$W1-W3$ colour of -0.04 means that a sensible threshold to use for true 12$\mu$m excesses
that can be found to be significant is about 0.1mag. While we could choose a slightly
lower threshold to include more sources that may turn out to be robust after more
detailed analysis, this threshold is also practical in that it limits the number of
sources to check by hand to a reasonable number; as we show below the false-positive rate
already is fairly high and we expect this rate to increase strongly as the threshold
moves closer to -0.04 (e.g. decreasing the threshold from 0.15 to 0.1 roughly doubles the
number of sources to check, but only results in 25\% more warm dust candidates, and
decreasing the threshold to 0.05 would yield 181 more sources to check). Stars later than
about K0 have slightly redder $W1-W3$ than average (tending to about zero colour for
$B_T-V_T=1.5$). This trend has the potential to add more false positives, but in practise
does not because it only affects a few hundred stars. We therefore identify as promising
12$\mu$m excess candidates the 96 main-sequence stars with $W1-W3$ above 0.1, which are
marked as large filled dots in Figure \ref{fig:hr}.
Stars this red may show a 3$\sigma$ $W3$ excess based on the photometric uncertainties of
most stars. To study these 96 sources further we use $\chi^2$ minimisation to fit PHOENIX
stellar photosphere (SED) models \citep{2005ESASP.576..565B} to 2MASS, Hipparcos
\citep{1997ESASP1200.....P}, Tycho-2 \citep{2000A&A...355L..27H}, AKARI
\citep{2010A&A...514A...1I}, and WISE $W1-2$ photometry \citep{2010AJ....140.1868W} to
provide a more accurate prediction of the $W3$ flux density to compare with the WISE
measurements. The SED fitting method has previously been validated through work done for
the \emph{Herschel} DEBRIS survey \citep[e.g.][]{2012MNRAS.421.2264K,2012MNRAS.426.2115K}

\section{Warm dust candidates}\label{s:cand}

We now check the 96 stars with candidate excesses for plausibility by inspecting the SEDs
and WISE images for each source. The issues that arise are all due to WISE source
extraction, but for several different reasons. Some are affected by bright nearby sources
and image artefacts, while others are moderately separated binary systems where the WISE
source extraction has only measured a single source. In some cases the $W1$ photometry is
shown to be underestimated, generally due to saturation. The results of the SED fitting
alone are not sufficient for recognising spurious excesses and image inspection is
generally required. While we inspected all of the images for the few candidates
identified here, there are commonly indications in the WISE catalogue---primarily
apparent variability---that might be used in a larger study. Notes on individual sources
are given in Appendix \ref{s:notes}. We find 25 systems with plausible and significant
12$\mu$m excesses, of which all but seven were previously known to host significant
levels of warm dust. These seven should be considered as promising candidates and require
further characterisation, but are considered here to be real. The 25 are listed in Table
\ref{tab:notes} and detailed SEDs are shown in Appendix \ref{s:seds}.

\begin{table*}
  \begin{tabular}{rlrlllll}
    \hline
    HIP & Name & $R_{12}$ & $T_{\rm BB}$ & Spty & Age & Group & Comments \\
    \hline
    8920 & BD+20 307 & 27.4 & 440 & G0 & $\sim$1Gyr & & Close binary, no cold dust
    (1,2)\\ 
    11696 & HD 15407A & 5.4 & 560 & F5 & 80Myr & AB Dor & Wide binary, no cold dust (3,4,5)\\ 
    $^\star$14479 & HD 19257 & 2.1 & 250 & A5 & ? & \\
    17091 & HD 22680 & 1.2 & 200 & G & 115Myr & Pleiades & (6) \\ 
    17401 & HD 23157 & 1.2 & 180 & A5 & 115Myr & Pleiades & (6) \\ 
    $^\star$17657 & HD 23586 & 1.4 & 290 & F0 & ? & & \\
    $^\star$53484 & HD 94893 & 1.2 & 150 & F0 & ? & LCC? & Membership uncertain (7)
    \\ 
    56354 & HD 100453 & 91.6 & -- & A9 & 17Myr &  LCC & Herbig Ae (8,9,10) \\ 
    55505 & HD 98800B & 7.6 & 170 & K4 & 8Myr &  TWA & Close binary
    in quadruple system. Possible transition disk. (11,12) \\ 
    58220 & HD 103703 & 1.4 & 260 & F3 & 17Myr & LCC & (8,13) \\ 
    59693 & HD 106389 & 1.3 & 290 & F6 & 17Myr & LCC & (8,13) \\ 
    61049 & HD 108857 & 1.4 & 230 & F7 & 17Myr & LCC & (8,13) \\ 
    63975 & HD 113766A & 1.4 & 300 & F4 & 17Myr & LCC & Excess around primary in 160AU
    binary (8,14) \\ 
    64837 & HD 115371 & 1.2 & 238 & F3 & 17Myr & LCC &  Membership probability 50\%
    (15) \\ 
    73990 & HD 133803 & 1.2 & 180 & A9 & 15Myr & UCL & (8,9) \\ 
    78996 & HD 144587 & 1.4 & 220 & A9 & 11Myr & US & (8,16) \\ 
    79288 & HD 145263 & 13.6 & 240 & F0 & 11Myr & US & (8,9,17) \\ 
    79383 & HD 145504 & 1.2 & 206 & F0 & 17Myr & US & (15) \\ 
    79476 & HD 145718 & 110.0 & -- & A8 & 17Myr & US & Herbig Ae (8,18) \\ 
    81870 & HD 150697 & 1.2 & 234 & F3 & 3.2Gyr? & ? & Near edge of
    $\rho$ Oph star-forming region so may be younger (19,20) \\ 
    $^\star$83877 & HD 154593 & 1.2 & 285 & G6 & ? & & \\
    86853 & HD 160959 & 1.6 & 210 & F0 & 15Myr? & UCL? & Membership probability 50\%
    due to excess (20), may be older (20) \\ 
    $^\star$88692 & HD 165439 & 1.2 & 200 & A2 & 11Myr? & & Age uncertain? (22) \\ 
    89046 & HD 166191 & 45.9 & ? & F4 & ? & & Protoplanetary disk? (suggests
    young age)\\
    $^\star$100464 & HD 194931 & 1.2 & 188 & F0 & ? & & \\
    \hline
  \end{tabular}  
  \caption{Significant 12$\mu$m excess candidates, new excesses are marked with a
    $^\star$. $T_{\rm BB}$ is the blackbody temperature fitted to the warm dust emission
    (in K). See Appendix \ref{s:notes} for notes on all 96 sources considered. Spectral types
    are from SIMBAD. References:
    1: \citet{2008ApJ...688.1345Z}, 2: \citet{2011ApJ...726...72W}, 3: \citet{2012ApJ...759L..18F}, 4: \citet{2012ApJ...749L..29F}, 5: \citet{2010ApJ...717L..57M}, 6: \citet{2010ApJ...712.1421S}, 7: \citet{2005ApJ...634.1385M}, 8: \citet{1999AJ....117..354D}, 9: \citet{2012ApJ...756..133C}, 10: \citet{1992A&AS...96..625O}, 11: \citet{1993ApJ...406L..25Z},  12: \citet{2012ApJ...744..121Y}, 13: \citet{2011ApJ...738..122C}, 14: \citet{2012MNRAS.422.2560S}, 15: \citet{2011MNRAS.416.3108R}, 16: \citet{2009ApJ...705.1646C}, 17: \citet{2004ApJ...610L..49H}, 18: \citet{1992AJ....103..549G}, 19: \citet{2000ApJ...538L.155F}, 20: \citet{2009A&A...501..941H}, 21: \citet{2012MNRAS.421L..97R}, 22: \citet{2011MNRAS.410..190T}.}\label{tab:notes}
\end{table*}

The inclusion or exclusion of targets from the list of 25 in the context of our goal here
merits some discussion. Because we wish to find the warm dust luminosity function for
main-sequence stars, our sample should include any object that plausibly looks like a
debris disk, even if it appears to be an extreme specimen. Though the most extreme disks,
such as BD+20 307 are very rare, this rarity is unsurprising, at least in an \emph{in
  situ} scenario; brighter (i.e. more massive) disks decay more rapidly due to more
frequent collisions, resulting in a decreased detection probability at the peak of their
activity.

\subsection{Protoplanetary disks}

We do not include protoplanetary disks in our luminosity function because they represent
a qualitatively different phase of evolution to the debris disk phase \citep[for example,
models of their decay are not simply power-laws,][]{2001MNRAS.328..485C}. Including
protoplanetary disks would require that models of the luminosity function include a
prescription for the poorly understood transition from this phase to the debris disk
phase, which would be very uncertain and limit our interpretation.

In terms of photometry the main feature that distinguishes protoplanetary disks from
debris disks is the breadth of the disk emission spectrum. Protoplanetary disks cover a
wide range of radii and extend right down to near the stellar surface, and therefore have
near, mid, and far-IR excesses. On the other hand, debris disks are usually well
described by blackbodies, sometimes with the addition of spectral features. In some cases
\citep[e.g. $\eta$ Corvi, HD
113766A,][]{2004MNRAS.348.1282S,2005ApJ...620..492W,2011ApJ...730L..29M,2013A&A...551A.134O},
debris disks are poorly fit by a single temperature component, suggesting that
populations of both warm and cool dust exist, and may argue for a comet delivery scenario
for the warm dust with the cool component acting as the comet reservoir
\citep{2012MNRAS.420.2990B}. These systems are however easily distinguished from
protoplanetary disks because the cool debris disk components always have much lower
fractional luminosities than protoplanetary disks \citep[$L_{\rm disk}/L_\star \lesssim
10^{-4}$ compared to $\sim$10$^{-1}$, e.g.][]{2008ApJ...674.1086T}.

A more useful discriminant for our purposes is that debris disks never have large near-IR
excesses. This lack of near-IR emission is first seen as the protoplanetary disk is
dispersed in so called ``transition disks'' \citep{1990AJ.....99.1187S}. We therefore
formalise the protoplanetary-debris disk distinction by considering the presence of a
near-IR (3.4$\mu$m) excess, using the $K_{\rm s}-W1$ colour. While all sources were
selected to have a red $W1-W3$ colour, the addition of a sufficiently large $W1$ excess
indicates that the source is probably a protoplanetary disk. The $K_{\rm s}-W1$ colours
for the sources in Table \ref{tab:notes} are all less than 0.2, with three notable
exceptions. The first two are the Herbig Ae stars HD 100453 and HD 145718 ($K_{\rm s}-W1
\approx 0.9$), and the third is a new potential warm dust source, HD 166191 ($K_{\rm
  s}-W1 = 0.5$), which we discuss further below. The SEDs of all three are shown in
Appendix \ref{s:seds}, and with both near and far-IR excesses are clearly different from
the other 22. The detection of significant far-IR excesses by IRAS indicates emission
from cold material but we did not use this as a discriminant because it would not have
been detectable around all sources, even if a protoplanetary disk was present.

HD 166191 has only recently been noted as a potential warm dust source
\citep{2013A&A...550A..45F}, though was reported as having an IR excess from both IRAS
and MSX \citep{1992A&AS...96..625O,2005MNRAS.363.1111C}.\footnote{None of the other new
  disk candidates were detected by IRAS, but see Appendix \ref{s:notes} notes for HIP
  14479.}  In addition, an excess was detected with AKARI-IRC, but not AKARI-FIS (though
the upper limit from AKARI-FIS is about 10Jy, so not strongly constraining). An excess
was found in the IRAS 60$\mu$m band, suggesting that the excess emission extends into the
far-infrared. A large excess over a wide range of infrared wavelengths suggests that the
emission is at a wide range of temperatures, and therefore that HD 166191 in fact
harbours a protoplanetary disk. However, a nearby (1\farcm7) red source was detected by
WISE, MSX, and AKARI-IRC, opening the possibility that because IRAS had poor resolution
the far-IR emission in fact comes from the nearby source and not HD 166191. A preliminary
conclusion from new \emph{Hershel} PACS \citep{2010A&A...518L...1P,2010A&A...518L...2P}
observations is that the disk spectrum is more consistent with a protoplanetary disk;
given the WISE photometry, both the 70$\mu$m flux density ($\approx$1.7Jy) and the
fractional luminosity ($\approx$10\%) are larger than would be expected for a warm debris
disk. A detailed study of this source will be presented elsewhere.

Finally, though it is not excluded by its $K_{\rm s}-W1$ colour, HD 98800 merits some
discussion because whether it is a young debris disk or in the transition between the
protoplanetary and debris disk phases is unclear. The disk emission spectrum is well
modelled by a blackbody, as are most debris disks. The fractional luminosity is very high
at around 10\%, meaning that a significant fraction of the star's emission is being
reprocessed by the disk. However, the far-IR emission may be decreased compared to that
expected from a protoplanetary disk by truncation by the outer binary
\citep{2010ApJ...710..462A}. In addition, the detection of H$_2$ emission suggests that
HD 98800 may still harbour some circumstellar gas \citep{2012ApJ...744..121Y}, though CO
emission has not been detected \citep{1995Natur.373..494Z,2010ApJ...710..462A}. Based on
the age, blackbody-like emission, and lack of CO detection and near-IR excess we include
HD 98800 in our luminosity function, with the expectation that it will decay in a
deterministic manner and that it is therefore both a plausible warm dust source and
direct progenitor of fainter warm dust systems.

\subsection{Warm debris disks}

Having excluded three sources from the original list of 25, we now consider some
properties of the remaining 22. For each source, after fitting a stellar photosphere
model we fit a blackbody disk model to any significant infra-red excesses. In general
blackbody models are a reasonable match to debris disk emission, though the agreement is
poorest for warm dust systems, which commonly have non-continuum silicate features
\citep[e.g.][]{2005Natur.436..363S,2010ApJ...714L.152F}. However, the temperatures
derived are representative, and more complicated models are unwarranted because we
typically only have 2-3 data points to fit.\footnote{We use a ``modified'' blackbody,
  where the emission beyond $\lambda_0=210\mu$m is multiplied by $\lambda_0/\lambda$
  \citep[see][]{2008ARA&A..46..339W} to account for inefficient emission from small
  grains. While this modification is not needed, it ensures that sub-mm fluxes predicted
  by the models are more realistic than a pure blackbody.}

The warm dust sources have a range of temperatures, which correspond to a range of radial
distances. While we are not explicitly looking for habitable-zone dust, it is instructive
to compare the temperatures to the width of the ``habitable zone'', which is naively the
radial distance at which the equilibrium temperature is $\sim$280K (i.e. the temperature
of a blackbody at 1AU from the Sun). The width of the habitable zone is of course very
uncertain, and is probably at least a factor of two wide in radius
\citep{1993Icar..101..108K,2013arXiv1301.6674K}. Therefore, the temperature range allowed
is at least a factor $\sqrt{2}$ wide, giving a range from approximately 230-320K. In
addition, there is considerable uncertainty in the location of dust belts inferred from
SED models for several reasons; such as the likely presence of non-continuum spectral
features \citep[e.g. all warm dust targets considered transient by][have silicate
features]{2007ApJ...658..569W}. In addition, belts may be a factor $\sim$2 more distant
than inferred from blackbody models due to grain emission inefficiencies at long
wavelengths \citep[relative to their sizes, e.g.][]{2013MNRAS.428.1263B}. However, this
factor, which has been derived from cool Kuiper-belt analogues, may not apply to fainter
warm dust if larger grains dominate the emission. Finally, the picture of a ring of warm
dust is probably over-simplified, particularly at faint levels where Poynting-Robertson
drag becomes important. Based on this discussion, we retain all 12$\mu$m excess sources
in what follows. In any case, removal of the coolest few sources (those below 200K say)
would make little difference to our analysis. Temperature is of course worthy of future
study, and for example the luminosity function could (with sufficient detections) be
extended to include it as a third dimension.

It is clear from Table \ref{tab:notes} that the 12$\mu$m excess systems are
preferentially young, with at least 11 (and possibly 13) being members of the
Scorpius-Centaurus association in Lower Centaurus Crux (LCC), Upper Centaurus Lupus
(UCL), and Upper Scorpius (US) (i.e. $<$20Myr old). In addition, HD 98800 belongs to the
8Myr old TW Hydrae association (TWA), HD 165439 was derived to be 11Myr based on
isochrone models \citep{2011MNRAS.410..190T}, HD 15407A is a member of the AB Doradus
moving group \citep{2010ApJ...717L..57M}, and HD 22680 and HD 23157 belong to the 115Myr
old Pleiades. Of the 7 stars that are not known to be young, only BD+20 307 is known to
be older than about 1Gyr. HD 160959 was suggested to be a member of Upper Centaurus Lupus
with 50\% probability \citep[i.e. 15Myr old][]{2012MNRAS.421L..97R}, but also has an
isochrone-derived age of 1.4Gyr \citep{2009A&A...501..941H}. Based on the likelihood of
stars with excesses to be young, the former age seems more likely to be correct, but
merits further study. Similarly, HD 150697 is suggested to be 3.2Gyr old based on
isochrones \citep{2009A&A...501..941H}, but with a similar sky location and distance as
the $\rho$ Ophiuchi star-forming region \citep{2000ApJ...538L.155F} requires further
study to verify the age.

For a typical main-sequence lifetime of 10Gyr and stars spaced randomly throughout, there
is a 1\% chance of a star being randomly younger than 100Myr. However, 15 (possibly 18)
of the 22 systems are younger than 120Myr, implying that the warm disk occurrence rate
for stars this young is at least 5\%. Therefore, as with longer wavelengths
\citep[i.e. 24$\mu$m,][]{2007ApJ...654..580S}, there appears to be a significant decrease
in the incidence of 12$\mu$m emission with time. However, this decay is unlikely to be
universal; BD+20 307 is at least 1Gyr old, so assuming 1/time decay should have been
$\sim$100 times brighter at 10Myr \citep[i.e. verging on physically impossible given that
the fraction of the host star's light that is captured by the dust in this system is
currently 3.2\%,][]{2011ApJ...726...72W}. Therefore, the more reasonable conclusion is
that either the dust level in this system is not evolving, as might be expected in a
comet delivery scenario, or became very bright some time well after the protoplanetary
disk was dispersed, as might be expected for a recent collision
\citep{2005Natur.436..363S}.

The stars in Table \ref{tab:notes} are biased towards earlier spectral types, though the
histogram in Figure \ref{fig:hr} shows that this could be explained by more early-type
stars in the sample distribution. Based on the histogram there is some evidence that
earlier spectral types are more likely to host 12$\mu$m excesses. The likely explanation
is that later-type stars in the sample are biased somewhat against having younger ages,
as it is only the brighter (earlier-type) stars in the nearby young associations that
were selected for the Hipparcos sample.

\subsection{Variability}\label{ss:var}

Given the possibly transient nature of warm dust, particularly in old systems such as
BD+20 307, we made a basic search for variability using the WISE data. This search is
motivated by the likely orbital timescales being of order years, and the ability of
individual collisions to affect the overall dust level
\citep{2005AJ....130..269K,2007ApJ...658..569W,2012ApJ...751L..17M}. Indeed, evidence for
$\sim$year timescale variability has been seen in several relatively young warm dust
systems \citep{2012Natur.487...74M,2012ApJ...751L..17M}.

\begin{figure}
  \begin{center}
    \hspace{-0.5cm} \includegraphics[width=0.5\textwidth]{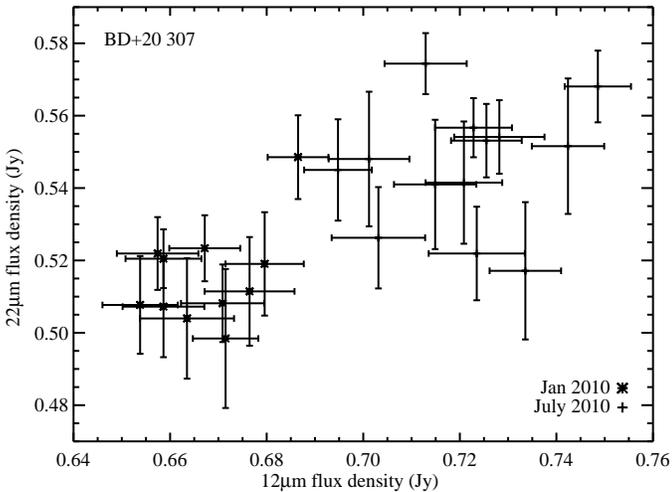}
    \caption{Evidence for a positive increase in the BD+20 307 mid-IR emission from 19
      Jan 2010 to 26 July 2010. Data are from the WISE single exposure catalogue and
      exclude the last image from each period, which have clear artefacts.}\label{fig:bd}
  \end{center}
\end{figure}

Though stars were only observed over a day or so in a single WISE pass, another was made
6 months later for sources that were in the appropriate RA range.\footnote{WISE did not
  survive an entire year, so the whole sky was not observed twice
  \citep{2010AJ....140.1868W}. Here we have not used the so called ``3-Band Cryo'' data
  release as any variability found would remain in question due to the change in cooling
  after the end of the full cryogenic mission.} Of the 22 warm dust sources, only BD+20
307 and HD 103703 show evidence for WISE variability. HD 103703 is marked as variable in
$W3$ in but not W4. While it may be possible for $W3$ to vary while W4 stays constant due
to changes in the strength of silicate features or the emission from each being dominated
by emission at different radial dust locations, the change of $\sim$0.1mag ($\pm$0.02mag)
in only one band is suggestive but requires verification. In contrast, BD+20 307 shows a
brightening of 6-8\% in both $W3$ and $W4$ over a six month period \citep[see
also][]{2012ApJ...751L..17M}, and as shown in Figure \ref{fig:bd}, these are correlated
in the sense that an increase in $W3$ brightness is accompanied by an increase at
W4. This increase suggests that while the total mass (i.e. including parent bodies) must
be decreasing (or at least be constant), effects such as individual collisions, size
distribution variations and mineralogical changes can increase the amount of dust as
measured by the infrared excess on $\sim$year long timescales.

\subsection{Missing warm dust sources?}\label{ss:missing}

An apparent omission from our warm dust list is HD 69830. As one of the nearest few dozen
Sun-like stars and host to both planets and warm dust, it is of key importance
\citep{2005ApJ...626.1061B}. The reason it does not appear here is in fact simple; the
excess at 12$\mu$m is not large enough to be formally detected with WISE. A detailed SED
model that includes all available photometry, including that from the \emph{Spitzer}
Infra-Red Array Camera \citep{2004ApJS..154....1W,2004ApJS..154...10F}, shows that the
12$\mu$m excess is only at the 5\% level, which also agrees with IRAS \citep[see also
Figure 2 in][]{2011ApJ...743...85B}. Another key warm dust source is the F2V star $\eta$
Corvi, which has $W1-W3$ of 0.14 and a $W3$ excess significance of only 2$\sigma$. The
``disappearing disk'' TYC 8241-2652-1 \citep{2012Natur.487...74M} is not included here
because it is not in the Hipparcos catalogue. Similarly, well known A-type warm dust
sources such as $\beta$ Pictoris and HD 172555 have significant WISE 12$\mu$m excesses,
but with $B_T-V_T$ colours of around 0.2 were not part of the initial sample.

We made a further check for potential warm dust sources within the Unbiased Nearby Stars
(UNS) sample, which includes five samples, each having the nearest $\sim$125
main-sequence stars with spectral types of A, F, G, K, and M \citep[i.e. 125 A-stars, 125
F-stars, etc.,][]{2010MNRAS.403.1089P}. UNS stars meeting the criteria outlined above are
a subset of the larger sample in question here, and SEDs for this sample have been
studied in great detail as part of the \emph{Herschel} DEBRIS Key Programme
\citep[e.g.][]{2012MNRAS.424.1206W,2012MNRAS.421.2264K}. With the inclusion of IRAS
\citep{1993yCat.2156....0M} and AKARI \citep{2010A&A...514A...1I} mid-IR photometry, and
in many cases \emph{Spitzer} Infra-Red Spectrograph
\citep[IRS,][]{2004ApJS..154...18H,2011ApJS..196....8L} spectra, this subset therefore
provides a good check for the sources that are subject to the worst saturation
effects. We found no 12$\mu$m excesses in this subset that should have resulted in a
significant detection with WISE.


\subsection{The 12$\mu$\lowercase{m} luminosity function}\label{s:lf}

\begin{figure}
  \begin{center}
    \hspace{-0.5cm} \includegraphics[width=0.5\textwidth]{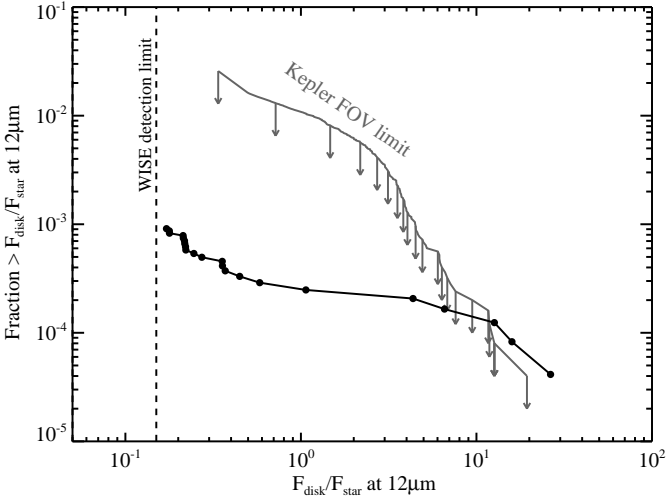}
    \caption{Luminosity function at 12$\mu$m for Sun-like Hipparcos stars (black line
      with dots). The black dashed line shows the WISE detection limit of 0.15. Also
      shown are the limits from the Kepler field \citep[grey
      line,][]{2012MNRAS.426...91K}. The Hipparcos line most likely lies above the Kepler
      field line due to small number variation.}\label{fig:cum}
  \end{center}
\end{figure}

Figure \ref{fig:cum} shows the 12$\mu$m luminosity function that results from the 22
unbiased warm dust detections described above. The distribution is generated simply by
assuming that these excesses could have been detected around all 24,174 stars, and
dividing the cumulative disk to star flux ratio distribution by this number. Thus, while
the brightest warm dust systems such as BD+20 307 are known to be rare, we have
quantified this rarity to be of order 1 in 10,000.

Figure \ref{fig:cum} also compares the luminosity function with the limit previously
derived for the \emph{Kepler} field \citep{2012MNRAS.426...91K}. Stars in the
\emph{Kepler} field are sufficiently faint that the number and distribution of detected
excesses is consistent with that expected from chance alignments with background
galaxies. The distribution derived here lies slightly above the limit from the
\emph{Kepler} field at the bright end, which most likely arises due to small number
variation, though could also be suggesting that the brightest excesses in the
\emph{Kepler} field are in fact real but not very robust against background galaxy
confusion.

\begin{figure*}
  \begin{center}
    \hspace{-0.5cm} \includegraphics[width=0.5\textwidth]{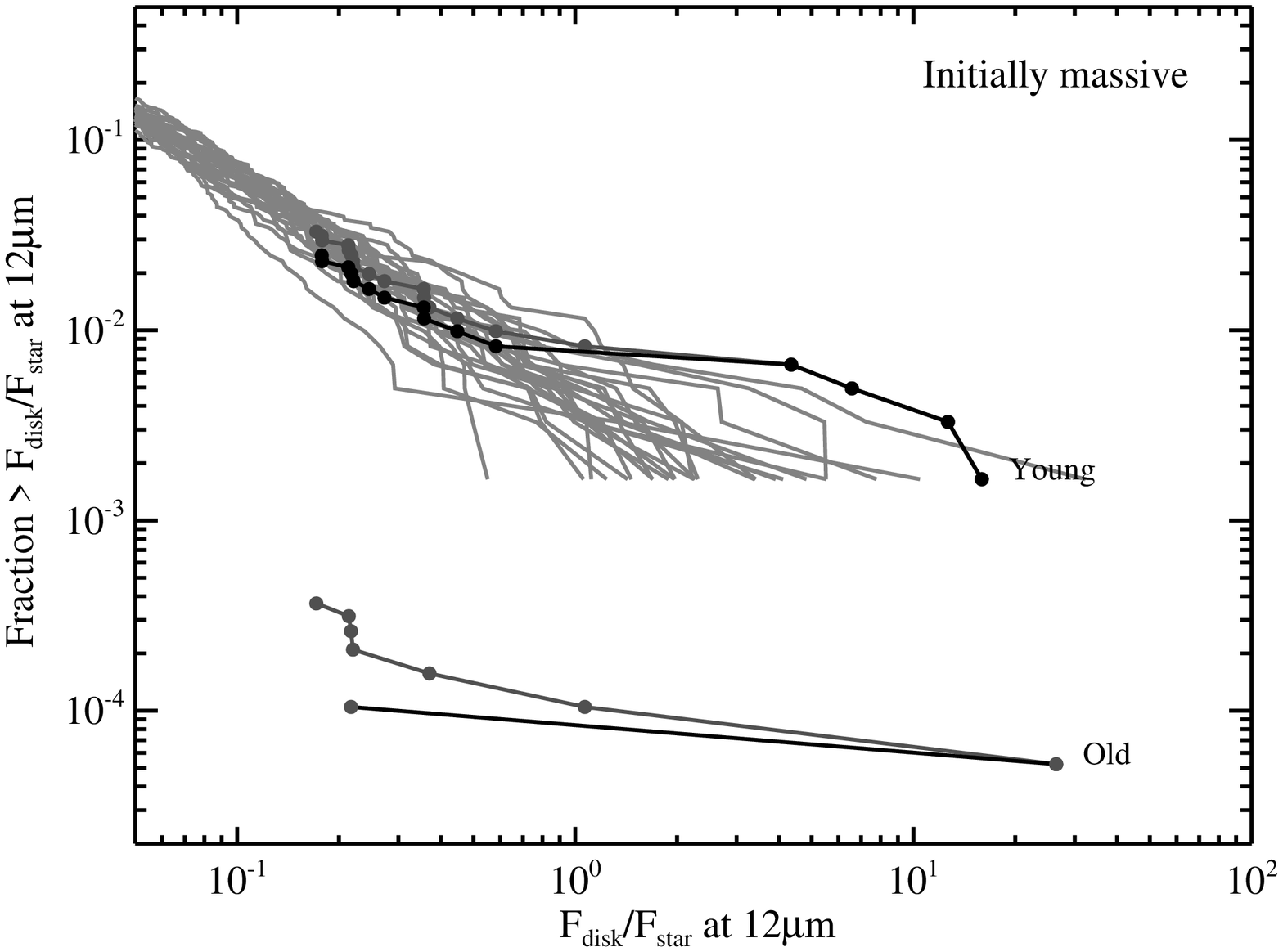}
    \includegraphics[width=0.5\textwidth]{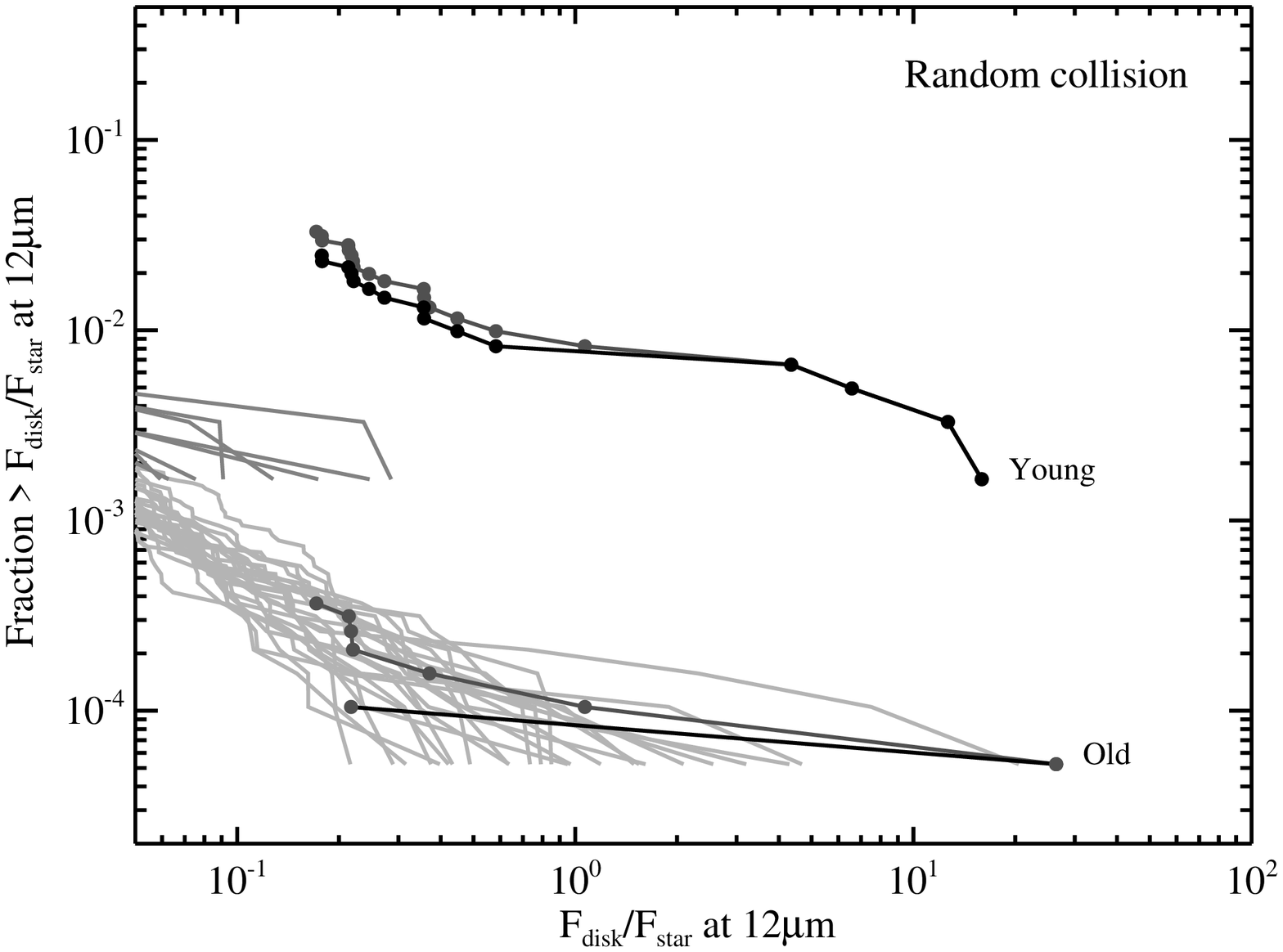}
    \caption{Monte-Carlo evolution simulations. Lines with dots show the observed
      luminosity function, and grey lines show 25 realisations of the model in the young
      (darker grey, 0-120Myr) and old (lighter grey, 1-13Gyr) age bins. \emph{Left
        panel:} Initially massive scenario, showing that young disks are well explained
      but that old disks are not expected at a level visible in this plot. \emph{Right
        panel:} Random collision scenario, showing that if old disks are explained, young
      stars do not have a sufficient number of bright disks to match those observed
      (i.e. the handful of dark grey lines near $F_{\rm disk}/F_\star \sim 10^{-2}$ are
      well below the observed line for young stars).}\label{fig:cumsim}
  \end{center}
\end{figure*}

Given the clear bias towards young stars among our excess candidates, the dark lines with
added dots in both panels of Figure \ref{fig:cumsim} show our luminosity function split
by age (the grey lines are models, described below). ``Young'' stars are those younger
than 120Myr, and ``old'' stars are those older than 1Gyr (our warm dust sample has no
stars between 120Myr and 1Gyr, see also \S \ref{ss:evol}). This split requires us to use
the age distribution shown in Figure \ref{fig:agehist}, for which the number of young
stars is 607, and the number of old stars 19,114. The distributions in Figure
\ref{fig:cumsim} therefore divide the detections by these total number estimates. Stars
with unknown ages are added to each group to illustrate the possible uncertainty in each
distribution. The main point here is that old warm dust systems are extremely rare, while
young ones are not.

\section{\emph{In situ} evolution}\label{ss:model}

One powerful advantage of knowing the distribution of warm excesses is that it contains
information about how disks evolve. For example, a population of disks whose brightness
decays as $1/{\rm time}$ that is observed at random times will be distributed with a
slope of -1 in Figure \ref{fig:cum}. A disk spends ten times longer in each successive
decade of luminosity than it did in the previous one, so every bright disk that is
discovered is the tip of an iceberg of fainter disks that are decaying ever more slowly.

Our basic assumption in now making a model to compare to our derived luminosity function
is that warm dust arises from some \emph{in situ} process, for example a collision or
collisions between parent bodies that normally reside where the dust is observed. In the
alternative comet delivery scenario, where objects are injected from elsewhere
(i.e. larger radial distances), no well informed population evolution model can currently
be made because the behaviour depends on the detailed dynamics of each individual system,
how those dynamics change over time, and how systems with the appropriate dynamics are
distributed. Of course, this difficulty does not mean that comet delivery is a less
viable scenario; we discuss it further in Section \ref{ss:comet}.

\subsection{Model description}\label{ss:mod}

Our goal is therefore to use an \emph{in situ} evolution model and apply it to the
observed 12$\mu$m luminosity function, testing different scenarios and making predictions
for surveys that probe fainter dust. The model described below has been outlined in
previous works \citep{2007ApJ...658..569W,2007ApJ...663..365W,2008ARA&A..46..339W}. The
basic premise is that a size distribution of objects is created in catastrophic
collisions between a reservoir of the largest objects. The resulting ``collisional
cascade'' size distribution remains roughly constant in shape
\citep{1969JGR....74.2531D}, so the dust level (i.e. the number of smallest objects) is
proportional to the number of the largest objects. To connect the mass with the observed
dust level requires a relation between the total mass $M_{\rm tot}$ and the total surface
area $\sigma_{\rm tot}$, which depends on the maximum and minimum object sizes $D_{\rm
  min}$ and $D_{\rm c}$, and the slope of the size distribution $q$, where the number of
objects between $D$ and $D+dD$ is $n(D) = K D^{2-3q}$ \citep{2007ApJ...663..365W}
\begin{equation}\label{eq:mtot}
  M_{\rm tot} = 2.5 \times 10^{-9} \frac{3q-5}{6-3q} \rho \sigma_{\rm tot} D_{\rm min}
  \left( 10^9 D_{\rm c}/D_{\rm min} \right)^{6-3q}
\end{equation}
where $M_{\rm tot}$ is in units of $M_\oplus$, $\rho$ is the planetesimal density in kg
m$^{-3}$, $\sigma_{\rm tot}$ is in AU$^2$, $D_{\rm min}$ is in $\mu$m, and $D_{\rm c}$ in
km. With $q=11/6$, the total mass is dominated by the largest objects and the total
surface area by the smallest grains. To now connect the surface area with the observed
dust level, we assume that the dust emits as a pure blackbody, so
\begin{equation}
  F_{\rm \nu,disk} = 2.4 \times 10^{-11} B_\nu(\lambda,T_{\rm disk}) \sigma_{\rm tot} d^{-2}
\end{equation}
where $B_\nu$ is the Planck function in Jy sr$^{-1}$ and $d$ is the distance to the star
in pc. We approximate the stellar emission as
\begin{equation}
  F_{\nu,\star} = 1.8 B_\nu(\lambda,T_\star) L_\star T_\star^{-4} d^{-2}
\end{equation}
where the stellar luminosity $L_\star$ is in units of $L_\odot$ and the effective
temperature $T_\star$ is in K. We define $\fdfssym_\nu$ as the disk to star flux ratio at
some wavelength
\begin{equation}\label{eq:eps}
  \fdfssym_\nu \equiv \frac{ F_{\rm \nu,disk} }{ F_{\nu,\star} } = 1.3 \times 10^{-11} 
  \frac{  B_\nu(\lambda,T_{\rm disk}) }{ B_\nu(\lambda,T_\star) }
  \sigma_{\rm tot} T_\star^4 L_\star^{-1}
\end{equation}
With equations (\ref{eq:mtot}) and (\ref{eq:eps}) and some assumed planetesimal
properties we are therefore able to link an excess observed at some wavelength and
temperature with the total mass in the disk.

Because the size distribution is such that most of the total mass $M_{\rm tot}$ is
contained in the largest objects, the dust level decays at a rate proportional to the
large-object collision rate, which has a characteristic timescale $t_{\rm coll}$ (in Myr)
\begin{equation}\label{eq:tcoll}
  t_{\rm coll} = 1.4 \times 10^{-9} r^{13/3} (dr/r) D_{\rm c} \left. Q_{\rm D}^\star
  \right.^{5/6} e^{-5/3} M_\star^{-4/3} M_{\rm tot}^{-1}
\end{equation}
where $dr$ is the width of the belt at $r$ (both in AU), $Q_{\rm D}^\star$ is the object
strength (J kg$^{-1}$), and $e$ is the mean planetesimal eccentricity, and $M_\star$ is
the stellar mass. The disk radius can be expressed in terms of temperature with
\begin{equation}
  T_{\rm disk} = 278.3 L_\star^{1/4} r^{-1/2}
\end{equation}
The mass loss rate is $dM_{\rm tot}/dt \propto M_{\rm tot}^2$ and the mass decays as
\citep{2003ApJ...598..626D}
\begin{equation}
  M_{\rm tot} = M_{\rm tot,0} /(1+t/t_{\rm coll,0}) \, ,
\end{equation}
where $t$ is the time since the evolution started and $M_{\rm tot,0}$ was the mass
available for collisions at that time (with a consequent collision timescale $t_{\rm
  coll,0}$). This time may be the age of the host star $t_{\rm age}$, but may be smaller,
for example if the belt was generated from a collision event that was the result of a
dynamical instability well after the star and planets were formed. An equivalent equation
also applies to the disk to star flux ratio
\begin{equation}
  \fdfssym_\nu =  \fdfssym_{\nu,0} / ( 1+t/t_{\rm coll,0} )
\end{equation}

Though numerical models find that the evolution can be slower than $1/{\rm time}$
\citep[e.g.][]{2008ApJ...673.1123L,2013ApJ...768...25G}, these results are based on disks
at large semi-major axes where the largest objects are not in collisional
equilibrium. For example, though \citet{2008ApJ...673.1123L} state that their disks decay
as $t^{-0.3 {\rm~to~} -0.4}$, they also note that both the disk and dust masses tend to
$1/{\rm time}$ decay at times that are sufficiently late that the largest objects have
reached collisional equilibrium. \citet{2013ApJ...768...25G} find a much slower
$t^{-0.08}$ mass decay for their reference model. However, the slow decay appears to be
because the largest (1000km) objects in their size distribution have not started to
collide, even at the latest times, and hence the total mass (which is dominated by these
objects) is only decaying due to the destruction of smaller objects \citep[the lack of
1000km-object evolution for their reference model can be seen in Fig 1
of][]{2012ApJ...754...74G}. Because the warm disks we consider here are very close to
their central stars (i.e. a few AU) and the collision rate scales very strongly with
radius (eq \ref{eq:tcoll}), collisional equilibrium is reached in only a few Myr (which
we also confirm \emph{a posteriori}). Therefore, we consider that while our analytic
model is necessarily simplified, a $1/{\rm time}$ evolution is justifiably realistic.

\subsection{Model implementation and interpretation}\label{ss:interp}

To implement the model we simply assume some initial disk brightness $\fdfssym_\nu$ and
that the collision timescale has the same form as equation (\ref{eq:tcoll})
\begin{equation}
  t_{\rm coll,0} = C / \fdfssym_{\nu,0}
\end{equation}
so $C$ has units of time in Myr and is shorter for disks that are initially more massive,
being the collision time for a disk of unity $\fdfssym_{\nu,0}$ (and some equivalent
$M_{\rm tot,0}$ that can be calculated using equations \ref{eq:mtot} and
\ref{eq:eps}). For our model $C$ is the only variable parameter.

To interpret $C$ in terms of physical parameters we combine equations (\ref{eq:mtot}),
(\ref{eq:eps}), and (\ref{eq:tcoll}), and rewrite the answer in terms of $C$, giving
\begin{eqnarray}\label{eq:c1}
  C & = &  7.4 \times 10^{-12} \frac{r^{13/3}}{\rho} \frac{dr}{r}
  \left( \frac{1}{10^9} \frac{D_{\rm c}}{D_{\rm min}} \right)^{3q-5} \times \nonumber \\
  && \frac{ T_\star^4 }{ M_\star^{4/3} L_\star }
  \frac{ B_\nu(\lambda,T_{\rm disk}) }{ B_\nu(\lambda,T_\star) }
  \frac{6-3q}{3q-5} {Q_{\rm D}^\star}^{5/6} e^{-5/3}
\end{eqnarray}
Because the collision timescale depends on the disk mass (or equivalently brightness),
$C$ does not depend on the initial disk mass \citep[see][]{2007ApJ...658..569W}, but on a
number of other parameters, most of which we can estimate values for. We assume a
Sun-like host star with $L_\star=1L_\odot$, $T_\star = 5800$K, and $M_\star=1M_\odot$. We
assume that the disk lies at 1AU and hence $T_{\rm disk}=278.3$K, and that $dr=1$. We
assume a size distribution slope of $q=11/6$. Finally, we include the wavelength of
observation, $\lambda=12$. With these assumptions equation (\ref{eq:c1}) reduces to
\begin{equation}\label{eq:c2}
  C = 3.1 \times 10^{-7} \sqrt{\frac{D_{\rm c}}{D_{\rm min}}} {Q_{\rm D}^\star}^{5/6} e^{-5/3}
\end{equation}
The remaining parameters are therefore the maximum and minimum planetesimal sizes, their
strengths, and their eccentricities. We may further assume that the minimum grain size is
$D_{\rm min} = 1\mu$m, the approximate size at which grains are blown out by radiation
pressure from Solar-type stars. The coefficient for this equation does not change
strongly with spectral type, and for example is $5 \times 10^{-7}$ for an early F-type
star, assuming that the dust remains at the same temperature (i.e. has a larger radius of
about 2.5AU) and that the minimum grain size has increased to 4$\mu$m due to the
increased luminosity.

\subsection{Excess evolution}\label{ss:evol}

In order to test the expectations for disk evolution we use the evolution model described
above to construct two simple Monte-Carlo models of 12$\mu$m evolution. In the first
``initially massive'' scenario, all stars have relatively massive and bright warm disks
at early times (see \S \ref{ss:insitu} for further discussion of how this scenario could
be interpreted). In this picture the time since the onset of decay is simply the stellar
age. All stars have disks regardless of age, but these become progressively fainter for
populations of older stars. In the second ``random collision'' scenario, stars have a
single dust creation event at some random time. Though we have a range of spectral types,
we set the epoch of this event to be a random time between 0-13Gyr because it is unlikely
to be related to stellar evolution (the time distribution of these events is however
revisited in \S \ref{ss:insitu}). Our age distribution means that many stars will be
observed before such a collision happens (see \S \ref{s:lf}), though this detail does not
affect the model because it is degenerate in the sense that multiple (instead of single)
collisions could be invoked and offset by a shorter collision time to yield the same
results.  Only those stars that happen to have a collision at a young age and are
observed at a young age will have a detectable excess.

We assume that stars have the age distribution shown in Figure \ref{fig:agehist}. The
initial distribution of flux ratios for both models is assumed to be log-normal with zero
mean and unity dispersion in $\log(F_{\rm disk}/F_\star)$, so covers the bright end of
the luminosity function. We found that the choice of initial distribution does not
strongly influence the results, and for example a power-law distribution weighted towards
fainter levels but with some bright disks yields very similar results.

We generate disks according to each scenario using two age bins; those that are ``young''
($<$120Myr) and those that are ``old'' ($>$1Gyr), using the same numbers of stars noted
in \S \ref{s:lf} (607 young stars and 19,114 old stars). We restrict the old stars to be
older than 1Gyr because BD+20 307 is $\sim$1Gyr old, whereas placing the cut at $>$120Myr
results in disk detections in the old group that are not much older than 120Myr, so not
as old as the stars we wish to compare the model with. While there are other possible
choices of bin locations, the point is that warm excesses around young stars are common,
and those around old stars are not, which allows us to distinguish between the models
outlined above. In each bin we generate 25 model realisations to illustrate the scatter
due to small numbers. The only model parameter is $C$, which sets how rapidly disks
decay, and is varied by hand so that the synthetic distributions have approximately the
same level as those observed.

The results from the two models are shown in Figure \ref{fig:cumsim}. The left panel
shows the initially massive warm belt scenario, while the right panel shows the random
collision scenario. Looking first at the initially massive panel, while reasonable
agreement is obtained for young stars, no old stars are seen to have large excesses as
the dust has decayed significantly by Gyr ages (these disks have $F_{\rm disk}/F_\star
\sim 10^{-3}$, see 1-2Gyr dotted line for initially massive warm disks in Figure
\ref{fig:cumpredage}). This result echoes the conclusions of \citet{2007ApJ...658..569W},
who found that bright debris disks around old stars cannot arise from the \emph{in situ}
evolution considered in this scenario. The observed young-star distribution is usually
flatter than most of the models, which might be explained by extra physical processes not
included in our model. We note a few of these in section \ref{ss:insitu}.

Turning now to the right panel of Figure \ref{fig:cumsim}, the random collision scenario
reasonably reproduces the luminosity function of warm dust around old stars, but fails to
match those seen around young stars. The chance that at least some of these are young,
and the possibility that HD 150697 is in fact young, means that a somewhat larger value
of $C$ (slower evolution) could be used to find better agreement with BD+20 307
(i.e. shift the model lines upward, thereby making dust from random collisions more
common). This slower evolution would however predict about ten times more systems with
$F_{\rm disk}/F_\star \gtrsim 0.15$ that are not observed. While a few young systems are
seen to have excesses due to random collisions (about 1\% at $F_{\rm disk}/F_\star \sim
0.1$), these are too infrequent to match the relatively high occurrence rate that is
observed. The conclusion is perhaps obvious; if collisions occur randomly over time it is
unlikely that a young star will have a collision that is observed soon afterwards. The
rapid collisional evolution means that even if they are, the dust levels are unlikely to
be extreme.

The only significant model parameter is $C$, which is set to 1Myr for both scenarios and
provides reasonable agreement with the observed luminosity function in each case. This
parameter need not be the same, as the scenarios could for example have very different
sized parent bodies or random velocities. As outlined above in \S \ref{ss:interp}, for
various assumptions $C$ can be interpreted in terms of physical system
parameters. Assuming $D_{\rm min}=1$, equation (\ref{eq:c2}) implies that $D_{\rm
  c}^{1/2} \left( Q_{\rm D}^\star \right)^{5/6} e^{-5/3} = 3.2 \times 10^6$. This value
can be compared to previous results for this combination, for example
\citet{2007ApJ...663..365W} found $7.4 \times 10^4$ for the evolution of Kuiper belt
analogues around A-stars, and \citet{2011MNRAS.414.2486K} found $1.4 \times 10^6$ for
Kuiper belt analogues around Sun-like stars (i.e. both were for disks at much larger
radii than those considered here). The latter authors discuss possible reasons for the
large difference, which could arise due to real differences between planetesimal
populations around A-type and Sun-like stars, or due to differences in the observables
such as the 24-70$\mu$m colour temperature, which could change due to different grain
blowout sizes for example. The point here is that the resulting $C$ is sensible compared
to previous results. If it were significantly different the plausibility of the \emph{in
  situ} scenario would be questionable.

\begin{figure}
  \begin{center}
    \hspace{-0.5cm} \includegraphics[width=0.5\textwidth]{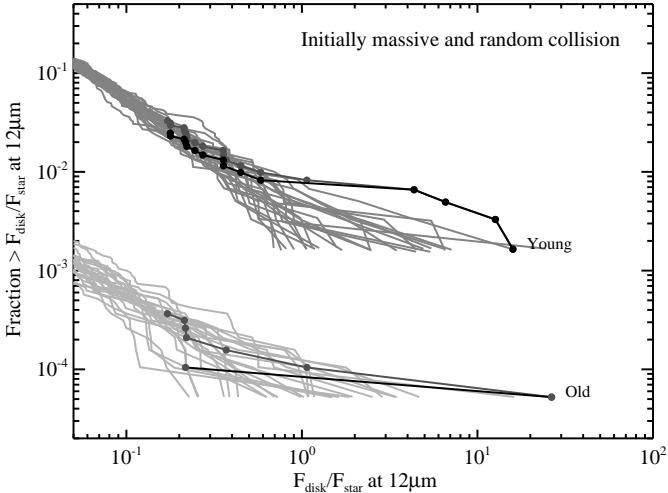}
    \caption{Same as Figure \ref{fig:cumsim}, but for a model that combines both
      initially massive warm belts and a single collision at a random time during the
      main sequence lifetime.}\label{fig:cumsim2}
  \end{center}
\end{figure}

To create an \emph{in situ} picture that explains significant levels of warm dust around
both young and old stars clearly requires some combination of our two scenarios. We
therefore make a simple combined model where systems start with the same initial
distribution of warm excesses and also have a single random collision at some point
during their lifetimes. This model has the same value of $C=1$Myr as before, and is shown
in Figure \ref{fig:cumsim2}. Given that each respective scenario dominates the young and
old populations with little effect on the other, that the model is in reasonable
agreement with the observed luminosity function is unsurprising. One or two random
collisions may be present in the young population of disks, and all old stars have
initially massive belts that are now very faint. This model could be tuned a little more
by specifying the fraction of systems that are initially massive and/or those that
undergo collisions. However, the agreement shown in Figure \ref{fig:cumsim2} is good
enough to illustrate that such a picture matches the observed distribution
well. Therefore, we conclude that the Monte-Carlo evolution model could be a realistic
(though not uniquely so) description of the origins of warm dust. This model is of course
highly simplified and largely empirical, and we discuss some issues further in Section
\ref{ss:insitu}.

Though we have shown that it is sensible, the assumption of \emph{in situ} decay made by
the above model may not be correct, with material being delivered to the terrestrial
regions from elsewhere, in which case the apparently sensible value of the 
parameter $C$ is coincidental. In a comet delivery scenario it is likely that the
luminosity function is not the same as expected for $1/t$ evolution (though it could
be). One test of our model is therefore to extrapolate to fainter levels, where
detections are possible with much more sensitive observations of relatively few stars.

\subsection{Extrapolation to faint exo-Zodi levels}

The level of dust around a ``typical'' Sun-like star is a key unknown, and crucial
information needed for the future goal of imaging Earth-like planets around other stars
\citep[e.g.][]{2006ApJ...652.1674B,2010ASPC..430..293A,2012PASP..124..799R}. There are
several avenues for finding the frequency of warm dust at relatively low levels; either
to directly detect it around an unbiased sample of stars
\citep[e.g.][]{2011ApJ...734...67M}, or to make predictions based on the expected
evolution of brighter dust. The former approach is clearly preferred as it yields a
direct measure, but finding such faint dust is technically difficult and the focus of
dedicated mid-IR instruments such as the Bracewell Infrared Nulling Camera (BLINC) at the
MMT \citep{2009ApJ...693.1500L}, the Keck Interferometer Nuller
\citep[KIN,][]{2009PASP..121.1120C}, and the Large Binocular Telescope Interferometer
\citep[LBTI,][]{2009AIPC.1158..313H}.\footnote{Several near-IR interferometers have also
  been used to search for exo-Zodiacal dust, though these are more sensitive to dust
  temperatures of $\sim$1000K
  \citep[e.g.][]{2006A&A...452..237A,2011ApJ...736...14M,2011A&A...534A...5D,Absil13}}.
These instruments are more sensitive to warm dust than photometric methods because their
sensitivity is set by their ability to null the starlight and detect the astrophysical
flux that ``leaks'' through the fringes \citep[e.g.][]{2011ApJ...734...67M}. The typical
scales on which these mid-infrared instruments are sensitive is 10-100mas so are well
suited to the terrestrial regions around the nearest stars. The latter approach of
extrapolating results from less sensitive photometric surveys based on evolution models
(i.e. the approach here) is easier in the sense that data for bright warm dust systems is
readily available, but is of course more uncertain due to the unknown origin and
evolution of these systems. A future goal is to fold the results from interferometry back
into the modelling process.

\begin{figure}
  \begin{center}
    \hspace{-0.5cm} \includegraphics[width=0.5\textwidth]{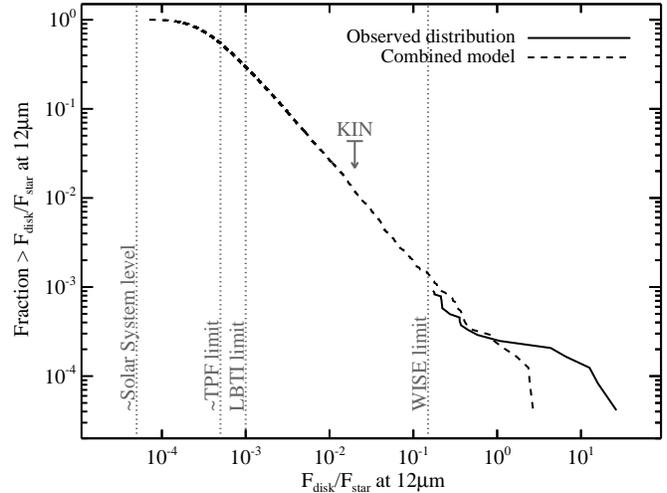}
    \caption{Observed luminosity function (solid line) and a simple prediction for faint
      exo-Zodi for stars between 0 and 13Gyr with a single realisation of the combined
      initially massive and random collision model shown in Figure \ref{fig:cumsim2}
      (dashed line). Grey dotted lines show the limits for LBTI and WISE sensitivity, as
      well as approximate levels for the Solar System Zodiacal cloud and for which
      TPF-like missions will be compromised by exo-zodiacal dust. The upper limit from
      the 23-star KIN survey is also shown (see text).}\label{fig:cumpred}
  \end{center}
\end{figure}

Figure \ref{fig:cumpred} shows the 12$\mu$m luminosity function and one realisation of
the combined initially massive and random collision model for 0-13Gyr old stars over a
wider disk-star flux ratio range than Figure \ref{fig:cumsim}, using our adopted age
distribution. The model under-predicts the number of brightest excesses, as this is the
typical outcome (as shown in Fig \ref{fig:cumsim2}), but varies considerably due to the
small number of such disks. Below $F_{\rm disk}/F_\star \sim 0.1$ all realisations, and
hence the predictions for fainter disks, are very similar.

The right side of the figure is covered by photometric surveys like the one presented
here, whereas fainter disks (left side) can only be detected with more sophisticated
methods, such as those employed by the KIN and LBTI. We have taken the approximate Solar
System 12$\mu$m dust level as $F_{\rm disk}/F_\star=5 \times 10^{-5}$, and the LBTI
sensitivity as 20 times this level \citep{2009AIPC.1158..313H}.  The predicted limit for
exo-Earth detection by a Terrestrial Planet Finder (TPF)-like mission is also shown (at
ten times the Solar System level).\footnote{In the sense that integration times become
  too long for a worthwhile mission for some reasonable assumptions, not that detection
  is impossible \citep{2012PASP..124..799R}.}

The limit set by the KIN survey \citep{2011ApJ...734...67M}, which resulted in no
significant detections, is shown, but merits further discussion. Their sample was
classified into ``high'' and ``low'' dust systems, meaning those with or without
detections of cool Kuiper belt analogues. Of the 25 systems observed, two were high dust
systems ($\eta$ Corvi and $\gamma$ Ophiuchi) and warm dust was confidently detected
around the former. $\eta$ Crv is already well known to have distinct warm and cool dust
belts \citep{2005ApJ...620..492W,2008A&A...485..897S}, while only cool dust has been
detected around $\gamma$ Oph \citep{1986PASP...98..685S,2008ApJ...679L.125S}. We use the
other 23 low dust stars in calculating the KIN limit, with the caveat that if warm dust
levels are positively (or negatively) correlated with cool dust levels, the limit will be
slightly too low (or too high). Given that only $\sim$20\% of stars are seen to have cool
dust (i.e. many low dust systems could still have relatively high dust levels that could
not be detected) this bias will not be very strong.

For stars drawn randomly from our assumed age distribution, the predicted disk fraction
is about 50\% at the LBTI sensitivity. Future work may however show that our assumption
of $1/t$ evolution is not exactly correct; the fraction decreases to 10\% if the decay is
instead $t^{-1.5}$ so the prediction is not extremely sensitive to the decay rate. A very
slow decay, such as that seen in the reference model of \citet{2013ApJ...768...25G} would
appear to be ruled out by the KIN upper limit as it would predict many bright warm disks
(such slow evolution is not expected for warm disks anyway, see the end of Section
\ref{ss:mod}). It is in any case clear that an unbiased LBTI survey of just a few tens of
stars would result in a strong test of our prediction, and has the potential for
detection of a dozen or so disks if the evolution predicted by our model is correct. Such
a result would be a great success for the instrument, but less promising for the study of
exo-Earths; another prediction of our model is that roughly only a few tens of percent of
stars are amenable to exo-Earth detection if our combined model is correct.

\begin{figure}
  \begin{center}
    \hspace{-0.5cm} \includegraphics[width=0.5\textwidth]{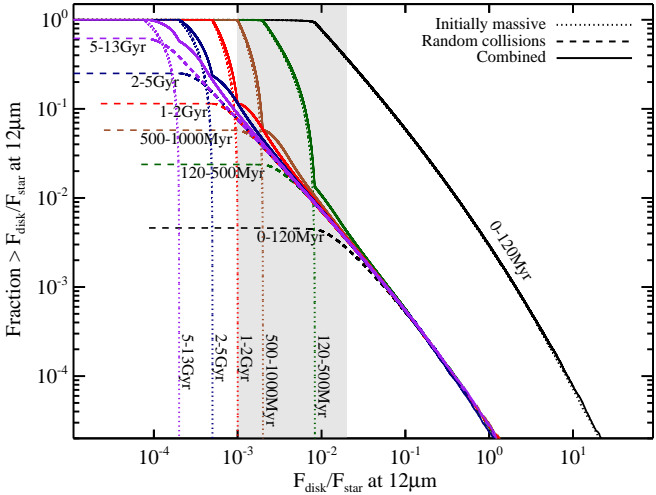}
    \caption{Similar to Figure \ref{fig:cumpred}, but showing the contribution of
      initially massive disks (dotted lines) and random collisions (dashed lines), and
      the combined model (solid lines). Stars are shown in different age bins given by
      boundaries of: 0, 120, 500, 1000, 2000, 5000, and 13,000Myr. The region where the
      predicted LBTI sensitivity improves over the KIN is marked with a grey band. Older
      systems have lower dust levels from the initially massive contribution, and the
      underlying level from random collisions dominates.}\label{fig:cumpredage}
  \end{center}
\end{figure}

We have already shown that the different scenarios depend strongly on the age of the
stellar population that is observed. This effect is shown explicitly in Figure
\ref{fig:cumpredage}, which shows the individual and combined distributions in a series
of age bins. Here we have increased the number of stars in each age bin to $10^7$ so that
the curves are smooth. For both dust creation scenarios, the level of the faintest
systems decreases with time as the stars become older. However, because random collisions
can happen at any time, the luminosity function for these events is otherwise independent
of time. The initially massive population dominates the distribution for stars younger
than about 100Myr, but becomes much less important as time goes on. For stars that are
older than $\sim$500Myr the brightest excesses are already dominated by random collisions
if our model is correct.

The solid lines show the combined model, which generally follow the fraction set by the
model that dominates at a given disk to star flux ratio. At flux ratios where both
components contribute to a similar fraction of systems (i.e. just above the maximum ratio
for initially massive belts), the fraction is somewhat greater than expected just for
random collisions. This increase arises because for these systems the emission from
random collisions is augmented by dust that was initially bright (but is now much
fainter).

This comparison of scenarios shows that for the oldest main sequence stars the predicted
frequency of dust systems is lower than shown in Figure \ref{fig:cumpred}. The faintest
dust may still come from those that were initially massive, but the bulk of the
population are remnants of random collisions. That is, though the random collisions are
very rare and are only bright for a short period of time, their decay through lower
levels means that they may in fact dominate the bulk of the warm dust luminosity function
for main-sequence stars.

\section{Discussion}\label{s:disc}

\subsection{\emph{In situ} evolution}\label{ss:insitu}

We have described a simple model of \emph{in situ} evolution that reproduces the observed
distribution of 12$\mu$m excesses reasonably well. This model has two components, an
initially massive component and a random collision component, both of which are needed to
simultaneously explain relatively frequent warm dust around young stars and extremely
rare warm dust around $\sim$Gyr old stars (Fig \ref{fig:cumsim2}). The predicted
luminosity function changes strongly with age due to the different contribution from each
component (Fig \ref{fig:cumpredage}). However, the $\sim$Gyr old bright warm dust systems
imply an underlying population of fainter warm dust that is present, regardless of
age. Therefore, if our model is correct, the fraction of stars with warm dust levels
sufficient to hinder exo-Earth detection may be a few tens of percent, even if only old
stars are targeted.

By using the term ``initially massive'' for the disk component that is bright at early
times, we have implied that it is a pumped-up version of our own Asteroid belt. While
such an interpretation may be correct, an alternative possibility is that large amounts
of warm dust are generated in the first $\sim$100Myr as a by-product of terrestrial
planet formation
\citep{2004ApJ...602L.133K,2008ApJ...673.1106L,2009ApJ...701.2019L,2012MNRAS.tmp.3462J}. Such
a scenario evolves similarly to an initially massive Asteroid belt analogue, but is
different in a few specific ways; the dust levels may initially increase as large
planetary embryos that stir the planetesimals to destructive collision velocities form,
dust may be visible for longer as mass is released from embryos in planet forming
impacts, and planetesimals and dust may also be depleted by dynamical interaction with
the embryos. These differences mean that the emission from terrestrial planet formation
may not simply decay as $1/{\rm time}$, providing a potential way to distinguish it from
initially massive Asteroid belt analogues \citep[e.g.][]{2008ApJ...672..558C}.

One limitation of our model is that processes beyond the basic collisional evolution will
modify the model luminosity functions. As noted above, dynamical depletion may follow
catastrophic collision events, perhaps with an initially optically thick phase
\citep{2012MNRAS.tmp.3462J}. Significant variability and non-linear decay could also
arise at early stages due to changes in dust mineralogy. For example, the silicate peak
near 12$\mu$m could decrease in strength after a collision event as the smallest grains
are removed by radiation pressure. The overall dust level may also vary following the
initial creation event as the system settles into dynamical and collisional
equilibrium. Such processes might be responsible for the observed luminosity function
being somewhat flatter than the models (Figs \ref{fig:cumsim}-\ref{fig:cumpred}).

A more basic limitation is our implementation of random collisions. With only a few
confirmed warm dust sources around $\sim$Gyr old stars there is little leverage to
constrain models, so this part of our model is necessarily simple, and not for example
linked to the mass remaining from initially massive belts. The largest simplification is
that uniform probability of a collision over 0-13Gyr assumed here is unlikely to be the
case. \citet{2007ApJ...658..569W} showed that if these collisions are between the largest
objects in a continuous size distribution (i.e. just the largest objects in an Asteroid
belt analogue), then the probability of detecting a collision drops as $1/{\rm
  time}^2$. An alternative possibility is that the collisions are between individual
large objects (e.g. planets or planetary embryos) that simply did not collide earlier, or
were perhaps subject to some late instability. These objects may have originally been
part of a continuous size distribution, but were ``stranded'' when only a few remained
and their chances of destruction became much lower
\citep[see][]{2011MNRAS.412.2137K}. However, in these cases collisions are more likely to
occur at earlier times, though perhaps with a weaker dependence compared to the large
asteroid scenario just described.

That the two ``old'' stars hosting warm dust, BD+20 307 and possibly HD 150697 are
between 1 and 4Gyr old can be used to illustrate that there is a rough constraint on the
time dependence of random collisions. This constraint is uncertain because in our sample
there are 5 sources of unknown age, and the ages of some others are poorly
constrained. However, a $1/{\rm time}^2$ dependence is not favoured because the existence
of warm dust around the $\sim$Gyr old stars implies a significant number should exist
around stars with ages from $\sim$100-1000Myr, which are not observed \citep[see
also][]{2010ApJ...717L..57M}. The argument is similar for $1/{\rm time}$, but given the
uncertainties is harder to rule out. Similarly, for the random collision model with
uniform time dependence a few detections are expected in the 4-13Gyr age bin and these
are not observed either. Therefore, if the random collision scenario is correct the time
dependence of collisions must decrease with time fairly weakly, if at all.

\subsection{Comet delivery}\label{ss:comet}

While we so far have discussed our observational results in the context of an \emph{in
  situ} model of evolution, the observed dust may originate elsewhere. Specifically,
material may be delivered from more distant regions after interacting with planets, thus
replenishing the mass that resides in the warm belt
\citep[e.g.][]{2012MNRAS.420.2990B,2012A&A...548A.104B}. With this kind of evolution, the
observed emission need not decay as $1/t$, and the luminosity function need not have the
slope of -1 seen in Figure \ref{fig:cumpred}.

Such a scenario may be acting to replenish the Zodiacal cloud in the Solar System. The
Zodiacal dust was first suggested to arise from particles dragged radially inwards from
the Asteroid belt by Poynting-Robertson drag \citep{1980IAUS...90..211K}. More recently
however, \citet{2010ApJ...713..816N} have shown that most of the Zodiacal cloud as seen
by IRAS can be explained by a sufficient level of spontaneous disruptions of
Jupiter-family comets \citep[though the existence of dust bands means that Asteroid
families still contribute some dust,][]{1984Natur.312..505D}. Because Jupiter-family
comets are themselves thought to originate in the Edgeworth-Kuiper belt
\citep{1997Icar..127...13L}, the location in which the dust is observed (i.e. near Earth)
is only linked to the source region by comet dynamics, which are dominated by the Solar
System's four giant planets.

In this kind of scenario, the amount of warm dust is plausibly linked to the number of
comets in the source region; a change in the number of source objects should be directly
reflected in the number scattered inwards, and therefore be reflected in the observed
dust level. Therefore, if the source region is in collisional equilibrium at all sizes
(i.e. $M_{\rm tot} \propto 1/t$), then the observed dust level should decay as roughly
$1/t$, with some additional contribution from dynamical depletion (that may be strongest
at early times). While the possibility of $1/t$ decay might appear consistent with the
model of section \ref{ss:model}, a very different value for the decay timescale $C$ is
expected because collisional evolution is much slower at larger radial distances
\citep[i.e. at least Gyr instead of Myr,][]{2008ApJ...674.1086T}. Therefore, a naive
picture of warm dust occurring in all systems exclusively due an uniformly initially
bright level of steady delivery throughout the main-sequence does not work because the
brightest systems are not predicted to be rare; their Gyr decay times mean that they
should in fact be ubiquitous. The problem with this picture is that comet delivery will
not always be initially bright; the level is set by the dynamics of individual
systems. It may be that near-constant comet delivery is the dominant warm dust origin and
the luminosity function set entirely by the distribution of systems with the appropriate
dynamics \citep{2012MNRAS.420.2990B}.

Though we have argued for the slow decay of reservoirs from which comets are delivered,
it is also possible that a comet delivery scenario could mimic our initially bright
asteroid belt analogues or terrestrial planet formation due to the planetary system
dynamics in the first $\sim$100Myr. A newly formed planetary system will have small
bodies scattered throughout, and in addition to those in young Asteroid and Kuiper belt
analogues, many will also be on unstable orbits. These unstable objects could contribute
to warm dust levels as they are cleared over the first 100Myr or so, thus appearing to
decay in the same way as initially massive belts.

Clearly, more work is needed to explore the likelihood of \emph{in situ} and cometary
origins for warm dust. For example, it could be that warm dust originates in some cases
from an impact on an object orbiting at $\sim$1AU by an interloping comet
\citep[e.g.][]{2012ApJ...747...93L}, in which case the material is terrestrial in origin
and evolves as such (i.e. \emph{in situ}), but whose probability of occurrence is more
closely related to the comet delivery rate.

\subsection{Solar System comparison}

A basic prediction of our model is that the level of Zodiacal dust seen in the Solar
System is roughly a factor of ten fainter than is typical (Fig \ref{fig:cumpred}). Given
the uncertainties discussed above this predicted rarity is not an issue for our model,
but worth some consideration.

If the bulk of our Zodiacal light comes from the Asteroid belt, the lower than predicted
dust level could arise in several ways. The initial distribution may simply have more
faint belts than we assumed, reflecting the range of outcomes dictated by planetesimal
growth in the protoplanetary disk or other processes such as planet migration. For
example, in the Solar System the Asteroid belt is thought to have been severely
dynamically depleted when Jupiter formed \citep{2001Icar..153..338P}, and a more general
inner Solar System depletion may have occurred if Jupiter was at some point closer to the
Sun that it is now \citep{2011Natur.475..206W}. This difference could be accounted for by
weighting the initial brightness distribution towards lower levels, with a slower disk
decay rate to keep the same number of observed warm disks.

If the bulk of our Zodiacal light comes from comet delivery, it may again be that the
true initial distribution is different to that assumed here, or that the evolution is not
continuous as assumed in our model. For example, it could be that the Solar System's
Zodiacal cloud evolution was interrupted by a late dynamical instability that ejected of
most of the primordial Kuiper belt \citep{2005Natur.435..459T}. Removal of most of the
source region mass would cause the warm dust level to drop in a way not accounted for by
our model. While it is currently unknown if the Kuiper belt has a relatively low mass
\citep[e.g.][]{2010MNRAS.404.1944G}, the relative faintness of the Zodiacal cloud in the
comet delivery picture could be explained if such depletion events are relatively
uncommon around stars in general.

Naturally, many of the possible extensions to our model could involve the effect of
planets on the parent planetesimals, whether they be asteroids or comets. Therefore, in
addition to looking for correlations between warm and cold dust as a signature of
scattering by planets \citep{Absil13}, an important goal is to look for correlations
between warm dust and planets. While a tentative positive correlation has been seen
between low-mass planets and cool Kuiper belt analogues \citep{2012MNRAS.424.1206W}, no
trends have yet been seen for warm dust \citep[e.g.][]{2012ApJ...757....7M}.

\section{Summary and Conclusions}\label{s:sum}

We have presented an analysis of the bright end of the 12$\mu$m warm dust luminosity
function as seen by WISE around nearby Sun-like stars. We report the possible detection
of 6 new warm dust candidates, HD 19257, HD 23586, HD 94893, HD 154593, HD 165439, and HD
194931.

There is a clear preference for excess emission around young stars, with at least 14/22
systems younger than 120Myr, despite the $\sim$10Gyr stellar lifetime. However, one of
the dustiest systems is $\sim$1Gyr old, and another may be $\sim$Gyr old, suggesting
that while youth is an important factor for setting the brightness of warm dust, it is
not the sole factor.

Using a simple \emph{in situ} evolution model to interpret the observed luminosity
function, we show that neither a picture where all systems have initially massive disks,
nor one where all excesses are due to randomly timed collisions, can explain the fact
that warm dust is observed around both young and old stars. A simple combination of these
two \emph{in situ} scenarios reproduces the observed luminosity function. However, we
cannot rule out that warm dust is in fact dominated by comet delivery. In this scenario
comets are scattered from a more distant reservoir, and reliable luminosity function
predictions cannot be made because they depend on the distribution of systems with the
appropriate dynamics.

We have made simple predictions for the number of exo-Zodi that might be detected by an
unbiased LBTI survey. For the expected sensitivity we anticipate a handful of detections
and a strong test of our evolution model. Such tests are a crucial step towards
understanding the origins and evolution of exo-Zodi, and ultimately the imaging of
Earth-like planets.

\section*{Acknowledgements}

We would like to thank the LBTI Science and Instrument teams for fruitful discussions,
Fei Dai for preliminary work, and Olivier Absil for a thorough referee report with
valuable criticisms and suggestions that improved the manuscript. This work was supported
by the European Union through ERC grant number 279973. This research has made use of the
following: The NASA/IPAC Infrared Science Archive, which is operated by the Jet
Propulsion Laboratory, California Institute of Technology, under contract with the
National Aeronautics and Space Administration. Data products from the Wide-field Infrared
Survey Explorer, which is a joint project of the University of California, Los Angeles,
and the Jet Propulsion Laboratory/California Institute of Technology, funded by the
National Aeronautics and Space Administration. Data products from the Two Micron All Sky
Survey, which is a joint project of the University of Massachusetts and the Infrared
Processing and Analysis Center/California Institute of Technology, funded by the National
Aeronautics and Space Administration and the National Science Foundation. The SIMBAD
database and VizieR catalogue access tools, operated at CDS, Strasbourg, France.

\appendix

\section{Notes on individual sources}\label{s:notes}

Notes for the 96 sources identified as having 12$\mu$m excesses based on the criteria
outlined in sections \ref{s:sample} and \ref{s:xs}. The 25 warm dust systems are shown in
bold face. Stars noted as bright and very bright are those where saturation in W1 becomes
an issue, meaning flux densities greater than $\sim$1-10Jy at 3$\mu$m.

\renewcommand{\labelenumi}{\arabic{enumi}. }

\begin{enumerate}

\item HIP 999 (LN Peg): RS CVn. W3 excess not significant.

\item HIP 3333 (NLTT 2305): Spectacularly aligned with a dust lane in Andromeda (M
  31). Excess very likely not real.

\item HIP 8920 (\textbf{BD+20 307}): Well known warm dust source
  \citep{2005Natur.436..363S,2011ApJ...726...72W}. Dust temperature is about 400K.

\item HIP 9269 (HD 12051): No excess. $W1$ underestimated and WISE catalogue shows
  significant variability. May be affected by bright nearby source (BD+32 361).

\item HIP 11696 (\textbf{HD 15407A}): Known warm dust source from AKARI
  \citep{2010ApJ...717L..57M,2012ApJ...749L..29F,2012ApJ...759L..18F}. Age from
  \citet{2009A&A...501..941H} is 2.1Gyr, but \citet{2010ApJ...717L..57M} show that
  Lithium absorption and kinematics argue that HD 15407AB is a member of the AB Doradus
  moving group, with an age of about 80Myr. Dust temperature is about 550K.

\item HIP 14479 (\textbf{HD 19257}): Bright star nearby (IRAS 03038+3018), but at
  1\farcm6 is well separated enough for reliable WISE extraction. WISE images clean so
  excess likely robust. Likely missed by IRAS due to being masked by the bright (0.9 Jy
  at 12$\mu$m, 0.4 Jy at 25$\mu$m) nearby source. Dust temperature is 253K. Spectral type
  is A5 from Simbad, but we find an effective temperature of 7485K, closer to a late-A or
  early F-type.

\item HIP 15244 (HD 20395): W3 excess not significant.

\item HIP 17091 (\textbf{HD 22680}): Known 24$\mu$m excess source in the Pleiades
  \citep{2010ApJ...712.1421S}. Dust temperature is 202K.

\item HIP 17401 (\textbf{HD 23157}): Pleiades member \citep{2010ApJ...712.1421S}. Clean
  detection in WISE images. No nearby sources. Excess appears robust. Dust temperature is
  184K.

\item HIP 17657 (\textbf{HD 23586}): Clean detection in WISE images. No nearby
  sources. Excess appears robust. Dust temperature is 285K.

\item HIP 17827 (HD 23623): W3 excess not significant.

\item HIP 20612 (CPD-77 172B): About 10'' from the primary HIP 20610 (HD 29058). No
  excess; high $W1-W3$ caused by low $W1$ measurement (which is also marked as variable in
  the catalogue). Significant excess at IRAS 12$\mu$m not confirmed by WISE.

\item HIP 21619 (HD 29364): Binary, excess caused by unreliable $W1$ measurement.

\item HIP 22449 (HD 30652): Very bright, W1 unreliable and underestimated. No excess.

\item HIP 22531 (HD 31203): Binary but only single WISE source extracted. Significant $W3$
  variability. Excess probably not real.

\item HIP 22563 (HD 31532): W1 underestimated, no excess.

\item HIP 23835 (HD 32923): Very bright, W1 unreliable and underestimated. No excess.

\item HIP 24065 (HD 33269): W1 underestimated, no excess.

\item HIP 26460 (HD 37495): Very bright, W1 unreliable and underestimated. No excess.

\item HIP 26525 (HD 245710): About 1' from the extremely bright variable M7 giant GP Tau
  (HD 37291). In addition the WISE $W3$ photometry is marked as variable. Very likely
  that the excess is not real and the result of proximity to GP Tau.

\item HIP 26756 (HD 37805): WISE images show very high background because this star lies
  in the L1630 molecular cloud in Orion \citep[called MIR-31
  by][]{2009A&A...507.1485M}. Despite their conclusion that the emission was
  photospheric, their MIPS 24$\mu$m photometry has an excess of $R_{24} \approx
  4-5$. However, inspection of the images suggests that the MIPS excess is for the same
  reason as the excess seen in W4; the bright background.

\item HIP 27284 (HD 38589): Very red bright object nearby in WISE images. Also detection
  in IRAS at 60$\mu$m. Excess unlikely to be related to the star in question.

\item HIP 28415 (HD 233190): Excess caused by halo from nearby bright star V442 Aur (HD
  39864).

\item HIP 30399 (HD 44582): WISE images show high background level, excess probably not
  real.

\item HIP 30953 (HD 46273): $W3$ excess due to saturated $W1$ and not real. Excess
  reported in \citet{2007ApJ...658.1289T} at 70$\mu$m, but this was only 2.7$\sigma$
  significant (i.e. may not be real either). Perhaps worth follow up observations to check.

\item HIP 31160 (HD 47149): No excess. $W1$ underestimated, perhaps due to nearby source
  (BD+12 1221).

\item HIP 32362 (HD 48737): No excess. Poor photospheric prediction due to saturated
  2MASS photometry and unusable $W1-2$ photometry.

\item HIP 35069 (HD 55128): W1 underestimated, no excess.

\item HIP 35564 (WDS J07204-5219AB): Binary. Two WISE sources extracted so $W1$ measurement
  uncertain, the reason for a $W1-W3$ excess, which the SED fitting shows is unlikely to
  be real.

\item HIP 41319 (HD 71030): W1 underestimated, no excess.

\item HIP 41389 (HD 71152): No excess. SED fitting shows that $W1$ is underestimated.

\item HIP 42173 (HD 72946): Two WISE sources extracted so $W1$ measurement uncertain, the
  reason for a $W1-W3$ excess, which the SED fitting shows is unlikely to be real.

\item HIP 47605 (HD 83804): W1 underestimated, no excess.

\item HIP 51459 (HD 90839): Very bright, W1 unreliable and underestimated. No excess.

\item HIP 53484 (\textbf{HD 94893}): Excess appears real, dust temperature is
  151K. \citet{2005ApJ...634.1385M} noted this star in looking for new members of the TW
  Hydrae association, but rejected it due to a large distance (97pc) and $\sim$15$^\circ$
  distance from the association centroid. Possible presence of an excess suggests
  membership of Lower Centaurus Crux should be considered.

\item HIP 54155 (HD 96064): Binary for which only a single source was
  extracted. Significant $W3$ variability indicated by catalogue, so excess unlikely to be
  real.

\item HIP 54745 (HD 97334): W1 underestimated, no excess.

\item HIP 55505 (\textbf{HD 98800}): Well known warm excess source in TW Hydrae
  association \citep[e.g.][]{1988PASP..100.1509W,1993ApJ...406L..25Z}. Age is about
  8Myr. Sometimes classed as a disk in ``transition'' between the protoplanetary and
  debris disk phases \citep{2007ApJ...664.1176F}, with the argument perhaps strengthened
  due to the recent detection of H$_2$ \citep{2012ApJ...744..121Y}. Dust temperature is
  168K.

\item HIP 56354 (\textbf{HD 100453}): Well known isolated Herbig Ae star
  \citep[e.g.][]{2002A&A...392.1039M}.

\item HIP 56809 (HD 101177): Binary but only single WISE source extracted. Significant $W3$
  variability. Excess probably not real.

\item HIP 58220 (\textbf{HD 103703}): Member of Scorpius-Centaurus in Lower Centaurus
  Crux \citep{1999AJ....117..354D}.  The age is therefore about 17Myr.  With no evidence
  for significant far-IR emission \citep{2011ApJ...738..122C}, HIP 58220 can be
  considered a 12$\mu$m excess system with a dust temperature of 264K.

\item HIP 58351 (HD 103941): Simbad spectral type is F2III and listed as a $\delta$ Del
  giant. SED fitting finds $T_{\rm eff}=5900$K but $L_\star = 6 L_\odot$, suggesting that
  this star is not a true dwarf. We therefore do not consider it a warm dust source for
  our purposes here.

\item HIP 59431 (HD 105963B): W1 underestimated, no excess.

\item HIP 59693 (\textbf{HD 106389}): Member of Scorpius-Centaurus in Lower Centaurus
  Crux \citep{1999AJ....117..354D}.  The age is therefore about 17Myr.  With no evidence
  for significant far-IR emission \citep{2011ApJ...738..122C}, HIP 59693 can be
  considered a 12$\mu$m excess system with a dust temperature of 291K.

\item HIP 60197 (BD+27 2115): Binary but only single WISE source extracted. Significant
  $W3$ variability. Excess probably not real.

\item HIP 61049 (\textbf{HD 108857}): Member of Scorpius-Centaurus in Lower Centaurus
  Crux \citep{1999AJ....117..354D}. The age is therefore about 17Myr. With no evidence
  for significant far-IR emission \citep{2011ApJ...738..122C}, HIP 61049 can be
  considered a 12$\mu$m excess system with a dust temperature of 220K.

\item HIP 61174 ($\eta$ Crv, HD 109085): Well known warm dust source. However, the WISE
  excess is not significant and arises here because of W1 underestimation due to the star
  being very bright. Not included in our luminosity function.

\item HIP 62472 (HD 11261): Binary but only single WISE source extracted. Significant $W3$
  variability. Excess probably not real.

\item HIP 63439 (HD 112810): Member of Scorpius-Centaurus in Lower Centaurus Crux
  \citep{1999AJ....117..354D}. The age is therefore about
  17Myr. \citet{2011ApJ...738..122C} find a large 70$\mu$m flux, suggesting that the
  12$\mu$m excess is probably the Wien side of cooler (67K) emission. The detailed SED
  model finds that the 12$\mu$m excess is only 2.4$\sigma$ significant, so this source is
  not considered as a true warm dust source.

\item HIP 63975 (\textbf{HD 113766A}): Well known warm dust source in Lower Centaurus
  Crux \citep{1999AJ....117..354D,1992A&AS...96..625O,2008ApJ...673.1106L}. The age is
  therefore about 17Myr. The excess appears around the primary star in this binary
  system, the companion is about 160AU distant so too distant to affect the warm dust
  unless the eccentricity is very high \citep[see][]{2012MNRAS.422.2560S}.

\item HIP 64184 (HD 114082): Member of Scorpius-Centaurus in Lower Centaurus Crux
  \citep{1999AJ....117..354D}. The age is therefore about
  17Myr. \citet{2011ApJ...738..122C} find a large 70$\mu$m flux, suggesting that the
  12$\mu$m excess is probably the Wien side of cooler (110K) emission. The detailed SED
  model finds that the 12$\mu$m excess is only 2$\sigma$ significant, so this source is
  not considered as a true warm dust source.

\item HIP 64241 (HD 114378J): Very bright, W1 unreliable and underestimated. No excess.

\item HIP 64532 (HD 115043): No excess. SED fitting shows that $W1$ is underestimated. No
  excess found from \emph{Spitzer} IRS \citep{2009ApJ...705...89L}.

\item HIP 64837 (\textbf{HD 115371}): Proposed as a member of Scorpius-Centaurus in Lower
  Centaurus Crux by
  \citet{2011MNRAS.416.3108R,2012MNRAS.421L..97R} with 69\% probability. Hence age is
  likely 17Myr. Disk temperature is 238K.

\item HIP 64995 (HD 115600): Member of Scorpius-Centaurus in Lower Centaurus Crux
  \citep{1999AJ....117..354D}. The age is therefore about
  17Myr. \citet{2005ApJ...623..493C} find a large 70$\mu$m flux, suggesting that the
  12$\mu$m excess is probably the Wien side of cooler (120K) emission. The detailed SED
  model finds that the 12$\mu$m excess is only 1.4$\sigma$ significant, so this source is
  not considered a true warm dust source.

\item HIP 65875 (HD 117214): Member of Scorpius-Centaurus in Lower Centaurus Crux
  \citep{1999AJ....117..354D}. The age is therefore about
  17Myr. \citet{2011ApJ...738..122C} find a large 70$\mu$m flux, suggesting that the
  12$\mu$m excess is probably the Wien side of cooler (110K) emission. The detailed SED
  model finds that the 12$\mu$m excess is only 2$\sigma$ significant, so this source is
  not considered a true warm dust source.

\item HIP 67497 (HD 120326): Member of Scorpius-Centaurus in Lower Centaurus Crux
  \citep{1999AJ....117..354D}. The age is therefore about
  17Myr. \citet{2011ApJ...738..122C} find a large 70$\mu$m flux, suggesting that the
  12$\mu$m excess is probably the Wien side of cooler (105K) emission. The detailed SED
  model finds that the 12$\mu$m excess is only 1$\sigma$ significant, so this source is
  not considered a true warm dust source.

\item HIP 67970 (HD 121189): Member of Scorpius-Centaurus in Upper Centaurus Lupus
  \citep{1999AJ....117..354D}. The age is therefore about 15Myr. Excess found at 24$\mu$m
  by \citet{2011ApJ...738..122C}, but no longer wavelength detection. WISE 12$\mu$m
  excess is only 2.5$\sigma$ significant, so this source is not considered a true warm
  dust source.

\item HIP 69953 (HD 125038): WISE images show filamentary structure, excess unlikely to
  be real.

\item HIP 71957 (HD 129502): Very bright nearby star, SED fit shows excess unlikely to be
  real due to $W1$ saturation.

\item HIP 73241 (HD 131923): No excess. SED fitting shows $W1$ is underestimated and WISE
  catalogue shows significant variability.

\item HIP 73990 (\textbf{HD 133803}): Member of Scorpius-Centaurus in Upper Centaurus
  Lupus \citep{1999AJ....117..354D}. The age is therefore about 15Myr. Excess found at
  24$\mu$m by \citet{2012ApJ...756..133C}, but no longer wavelength detection. Dust
  temperature is 177K.

\item HIP 77190 (HD 327427): WISE images suggest moderate nearby background level, and
  that W4 excess most likely not real. W3 excess questionable (and only 3.3$\sigma$) so
  not considered a true warm dust source.

\item HIP 77545 (HD 141441): WISE images show high background levels so excess most
  likely not real.

\item HIP 78996 (\textbf{HD 144587}): Member of Scorpius-Centaurus in Upper Scorpius
  \citep{1999AJ....117..354D}. The age is therefore about 11Myr. Excess found at 24$\mu$m
  by \citet{2009ApJ...705.1646C}, but no longer wavelength detection. Dust temperature is
  223K.

\item HIP 79054 (HD 144729): W3 excess not significant.

\item HIP 79288 (\textbf{HD 145263}): Member of Scorpius-Centaurus in Upper Scorpius
  \citep{1999AJ....117..354D}. The age is therefore about 11Myr. Excess found at 24$\mu$m
  by \citet{2012ApJ...756..133C}, but no longer wavelength detection. Dust temperature is
  241K.

\item HIP 79383 (\textbf{HD 145504}): Proposed as a member of Scorpius-Centaurus in Upper
  Scorpius by \citet{2011MNRAS.416.3108R} with 74\% probability. Hence age is likely
  17Myr. Disk temperature is 206K.

\item HIP 79476 (\textbf{HD 145718}): Member of Scorpius-Centaurus in Upper Scorpius
  \citep{1999AJ....117..354D}. The age is therefore about 11Myr. Known Herbig Ae star
  \citep{2003AJ....126.2971V,2010A&A...517A..67C}.

\item HIP 79607 (HD 146361): Multiple system. Very bright, $W1$ underestimated and excess
  unlikely.

\item HIP 79977 (HD 146897): Member of Scorpius-Centaurus in Upper Scorpius
  \citep{1999AJ....117..354D}. The age is therefore about
  11Myr. \citet{2011ApJ...738..122C} find a large 70$\mu$m flux, suggesting that the
  12$\mu$m excess is probably the Wien side of cooler (89K) emission. The detailed SED
  model finds that the 12$\mu$m excess is just below 3$\sigma$ significant, so this
  source is not considered a true warm dust source.

\item HIP 80170 (HD 147547): Very bright, SED fitting shows that $W1$ is underestimated
  and excess unlikely.

\item HIP 80759 (HD 148371): Excess in SED appears reliable. WISE images show possible
  contamination due to filamentary background at both W3 and W4, though whether the
  excess is caused by this is hard to tell. The W3 excess is 3.6$\sigma$, so the
  background contribution would not need to be large to cause the excess to be
  significant. Therefore we do not include this source as a warm dust candidate.

\item HIP 81870 (\textbf{HD 150697}): Noted as a possible warm dust source from IRAS
  photometry by \citet{2000ApJ...538L.155F}, but with the caution that the star lies on
  the edge of the $\rho$ Oph star-forming region. With a distance of 133pc, it may in
  fact be a young star associated with $\rho$ Oph, but was assigned an age of 3.2Gyr by
  \citet{2009A&A...501..941H}. We find that the IRAS 12$\mu$m excess is 2.98$\sigma$
  significant and that the W3 images appear clean, so the excess appears reliable. Dust
  temperature is 234K.

\item HIP 83006 (HD 153525): Bright star, W1 underestimated, no excess.

\item HIP 83301 (HD 153592): Near edge of region with no $W3$ coverage, image has major
  artefacts so excess very unlikely to be real.

\item HIP 83877 (\textbf{HD 154593}): WISE images appear reasonably clean though the
  source is not very bright. Very bright carbon star to the North (IRAS 17050-4642) does
  not appear to affect the images at the target position. That an excess is seen in both
  W3 and W4 suggests that it is real, with a dust temperature of 285K.

\item HIP 84720 (HD 156274): Very bright, SED fitting shows that $W1$ is underestimated
  and excess unlikely.

\item HIP 85340 (HD 157792): Bright star, W1 underestimated, no excess.

\item HIP 86596 (HD 161157): Bright star, W1 underestimated, no excess.

\item HIP 86853 (\textbf{HD 160959}): Excess found by \citet{2012MNRAS.421L..97R} and
  Upper Centaurus Lupus membership assigned 50\% probability based on this excess. Disk
  temperature is 205K.

\item HIP 87368 (HD 162186): Has a 3.8$\sigma$ $W3$ excess, but is very near the Galactic
  center ($<5^\circ$) so in a high background region and probably not real. Very bright
  IRAS 60 and 100$\mu$m detections, and a ROSAT X-ray detection 33'' to the SE (1RXS
  J175106.2-321836), which is unlikely to be associated given the 10'' position
  uncertainty. The IRAS excess may be associated with the X-ray source.

\item HIP 88683 (HD165269): No excess. SED fitting shows that $W1$ is underestimated and
  WISE catalogue shows significant variability.

\item HIP 88692 (\textbf{HD 165439}): Excess appears reliable. Age of 11Myr derived by
  \citet{2011MNRAS.410..190T}.

\item HIP 89046 (\textbf{HD 166191}): Large far-IR excess from AKARI and IRAS may be
  associated with nearby red source. Alternatively, far-IR emission could be associated
  with HIP 89046 and it harbours a previously unknown transitional protoplanetary
  disk. Preliminary conclusion from \emph{Herschel} PACS observations favours the
  protoplanetary disk interpretation.

\item HIP 91242 (HD 171452): WISE images appear clean and significant excess at W4. W3
  excess only 2$\sigma$ significant so not considered a warm dust source here.

\item HIP 93940 (HD 178332): No excess. $W1$ underestimated and WISE catalogue shows
  significant variability.

\item HIP 100217 (HD 193632): Bright star, W1 underestimated, no excess.

\item HIP 100464 (\textbf{HD 194931}): Excess appears reliable in WISE images. Dust
  temperature 188K.

\item HIP 107556 (HD 207098): Bright star, $W1$ saturated and $W3$ excess not real.

\item HIP 108431 (HD 208450 ): Bright star, W1 underestimated, no excess.

\item HIP 108456 (HD 209942): About 13'' from the RS CVn secondary V376 Cep (HD
  209943). Both stars have very similar $K_s$ magnitudes of 5.6, meaning that if the WISE
  $W3$ photometry covers the pair (as suggested by the catalogue) it would show a flux
  ratio of about 2. The $W3$ flux ratio is 2.7 (and 2 for IRAS 12$\mu$m), but the WISE
  photometry is also variable in $W1-3$, suggesting that the excess is spurious.

\item HIP 110355 (HD 211729): Bright star, W1 underestimated, no excess.

\item HIP 110785 (HD 212754): Bright star, W1 underestimated, no excess.

\item HIP 111278 (HD 213617): Significant 60$\mu$m excess from IRAS and 70$\mu$m excess
  from \emph{Spitzer} \citep{2011ApJS..193....4M}, WISE 12$\mu$m excess not significant.

\item HIP 114944 (EZ Peg): W3 excess not significant.

\end{enumerate}

\section{Sample age distribution}\label{s:ages}

While stars of a given spectral type may reasonably be assumed to have random
main-sequence ages, our sample spans a sufficiently wide range of spectral types that the
main-sequence lifetimes $\tau_{\rm MS}$ vary significantly. Because our stars are
selected based on their Tycho-2 $B_{\rm T}-V_{\rm T}$ colour, to derive the age
distribution we want a relation between this colour and $\tau_{\rm MS}$.

\begin{figure}
  \begin{center}
    \hspace{-0.5cm} \includegraphics[width=0.5\textwidth]{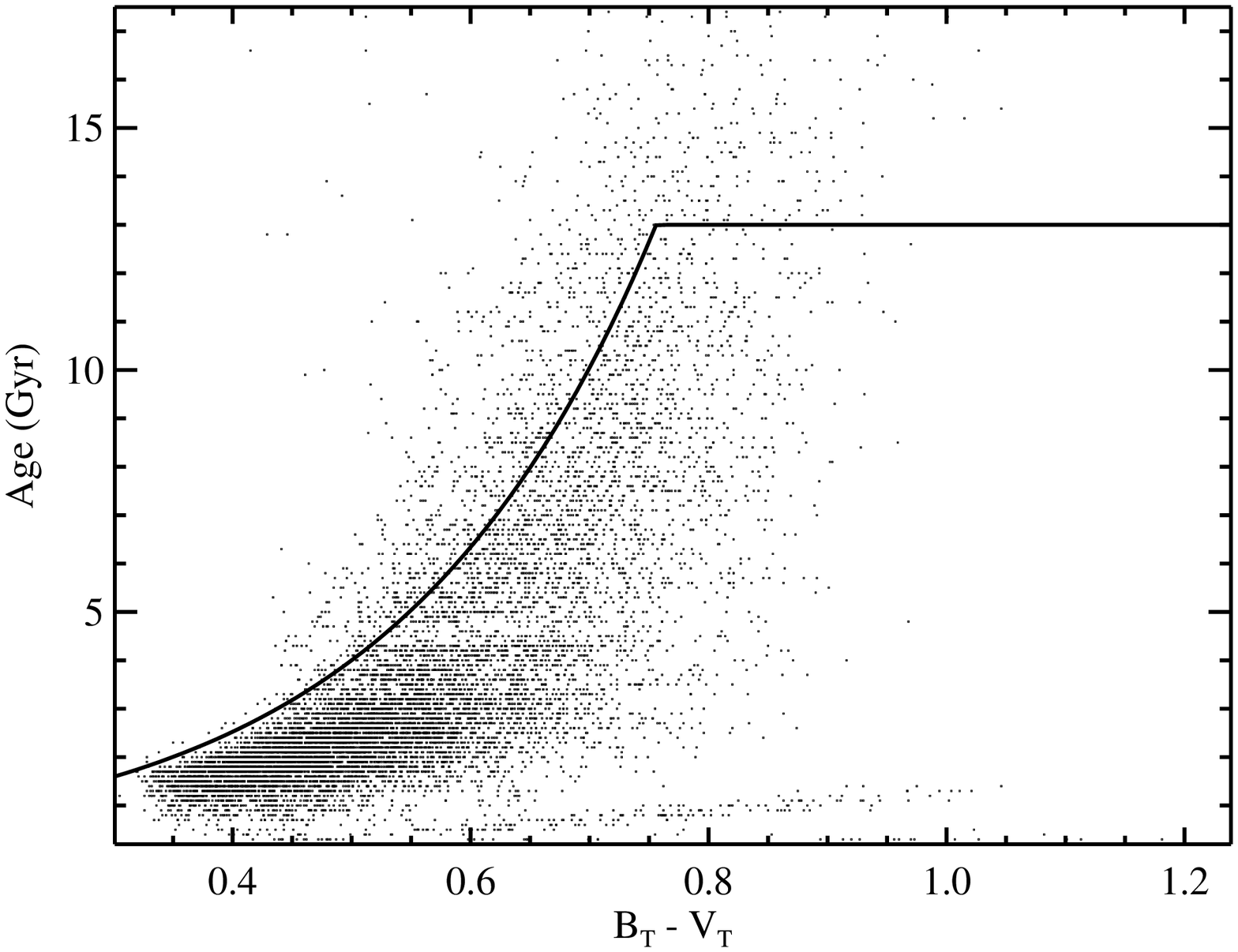}
  \end{center}
  \caption{Age vs. Tycho-2 colour for 10,593 stars in the Geneva-Copenhagen catalogue
    with age estimates. The line shows an empirical estimate for the main-sequence
    lifetime of $\tau_{\rm MS} = 0.4 \times 10^{(B_{\rm T}-V_{\rm T})/0.5}$Gyr, where
    ages are also restricted to be less than 13Gyr.}\label{fig:age}
\end{figure}

About 10,000 of our 24,174 stars actually have ages derived by The Geneva-Copenhagen
survey of the Solar neighbourhood \citep{2004A&A...418..989N,2009A&A...501..941H}. The
ages are derived using isochrones, and therefore have the weakness that stars not
sufficiently evolved from the zero-age main-sequence cannot be assigned ages. Indeed, the
authors state ``that the true age distribution of the full sample cannot be derived
directly from the catalogue; careful simulation of the biases operating on the selection
of the stars and their age determination will be needed to obtain meaningful results in
investigations of this type.'' Despite this limitation however, the catalogue does
provide a distribution of ages as a function of colour, shown in Figure \ref{fig:age},
which can be used to verify an empirical relation between $\tau_{\rm MS}$ and $B_{\rm
  T}-V_{\rm T}$.

\begin{figure}
  \begin{center}
    \hspace{-0.5cm} \includegraphics[width=0.5\textwidth]{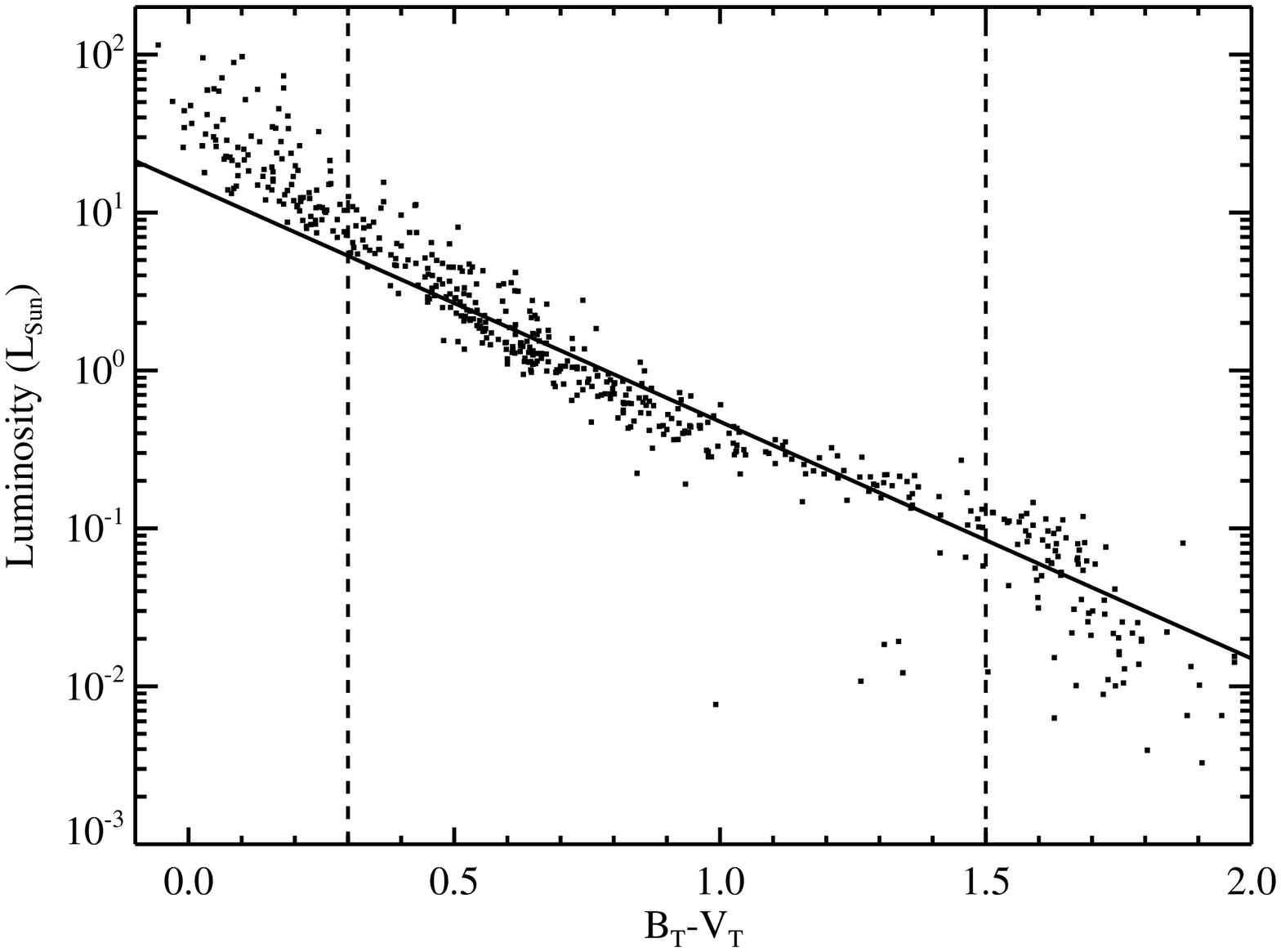}
  \end{center}
  \caption{HR diagram for stars from the UNS sample, where stellar luminosities have been
    derived from SED fitting. The solid line is $L_\star = 15 \times 10^{-1.5 (B_{\rm
        T}-V_{\rm T})}$, and the dashed lines show the $B_T-V_T$ colour range of our
    sample.}\label{fig:lvsbv}
\end{figure}

We first plot absolute luminosities as derived by SED fitting against colour for
main-sequence stars from the Unbiased Nearby Stars (UNS) sample
\citep{2010MNRAS.403.1089P} in Figure \ref{fig:lvsbv}. The correlation in this HR diagram
is reasonably well described as $L_\star \propto 10^{-1.5 (B_{\rm T}-V_{\rm T})}$. This
relation, combined with $\tau_{\rm MS} \propto M_\star/L_\star$ (i.e. assuming all mass
is turned into energy) and $L_\star \propto M_\star^{3.5}$ yields $\tau_{\rm MS} \propto
10^{B_{\rm T}-V_{\rm T}}$. This dependence is only approximate however. Using it as a
starting point, with some experimentation we found that
\begin{equation}
  \tau_{\rm MS} = 0.4 \times 10^{(B_{\rm T}-V_{\rm T})/0.5} {\rm Gyr}
\end{equation}
provides a reasonable match to the maximum ages for early-types in Figure \ref{fig:age},
while giving an age of 9Gyr at $B_{\rm T}-V_{\rm T} = 0.68$, corresponding approximately
to the Sun. We add a further restriction that ages must be less than 13Gyr.

To derive the age distribution of our sample now requires converting the known
distribution of colours. This conversion is done by setting up a series of colour bins,
and selecting ages randomly between zero and $\tau_{\rm MS}$ for the stars in each
bin. This process results in the age distribution for our sample shown in Figure
\ref{fig:agehist}.

To generate this distribution numerically we make use of a fourth order polynomial fit to
the histogram, with coefficients of 0.171124, -0.0528395, 0.00686227, -0.000410586, and
9.29393e-06. We generate a random age in the desired range, and another random number
between 0 and 0.1035. If the second random number is lower than the polynomial function
evaluated for that age, or if the polynomial is greater than 0.1035, we keep the
age. This process is repeated until the desired number of ages is acquired.

Our analysis has been done with both a naive uniform distribution of ages between 0-10Gyr
and the more detailed distribution derived above, and our conclusions are the same for
either distribution.

\section{SEDs for 25 warm dust candidates/systems}\label{s:seds}

\begin{figure*}
  \begin{center}
    \hspace{-0.5cm} \includegraphics[width=0.45\textwidth]{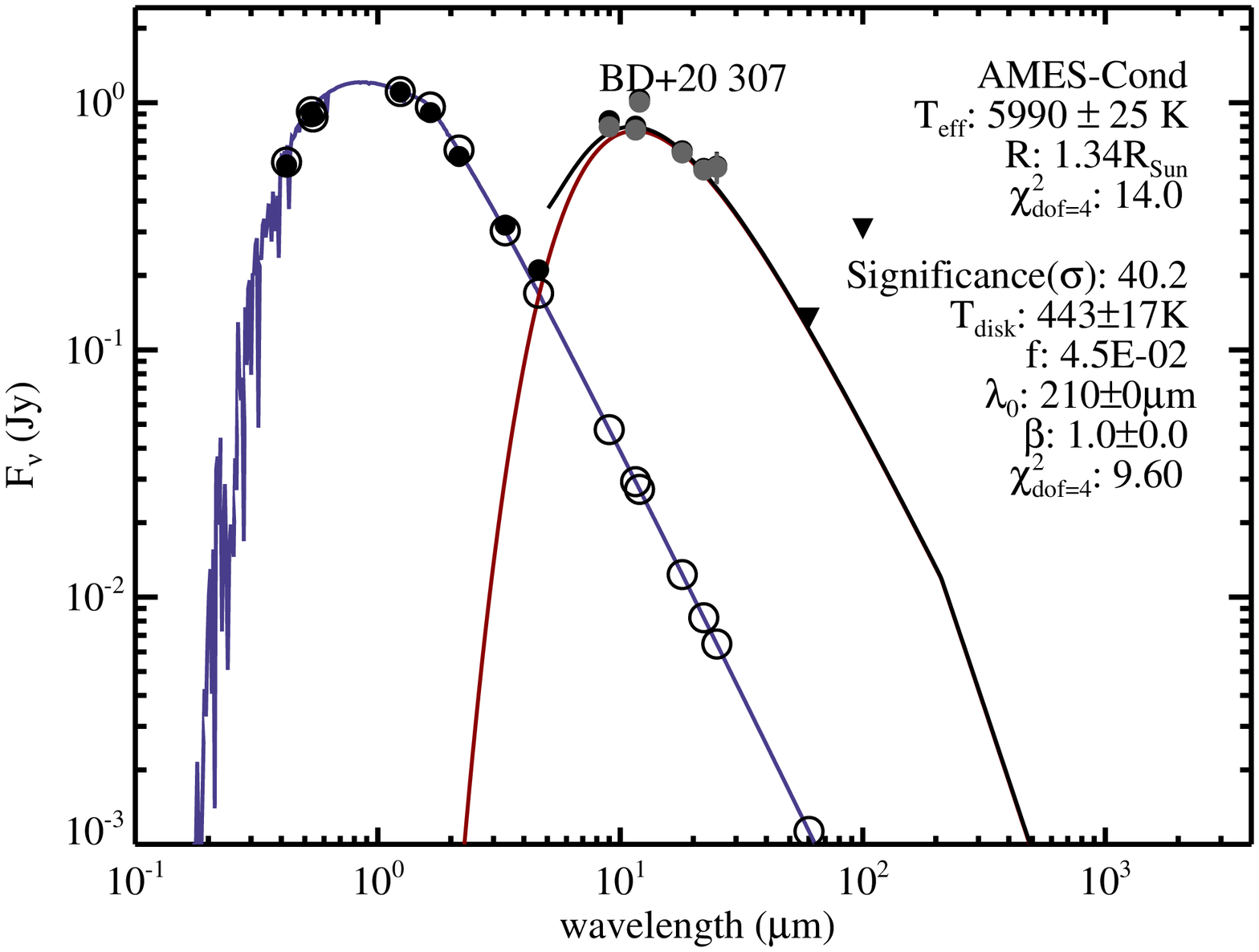}
    \includegraphics[width=0.45\textwidth]{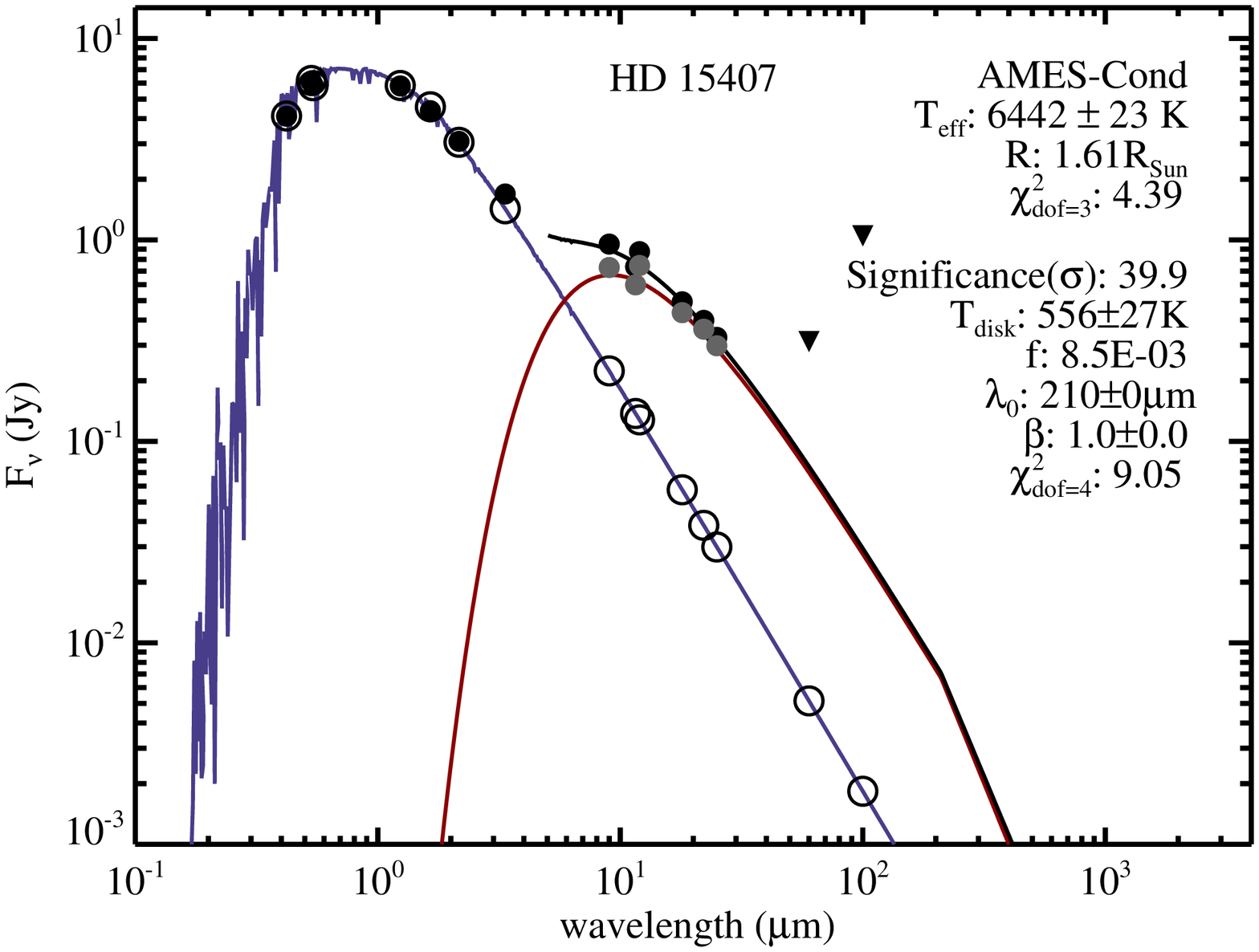}\\
    \hspace{-0.5cm} \includegraphics[width=0.45\textwidth]{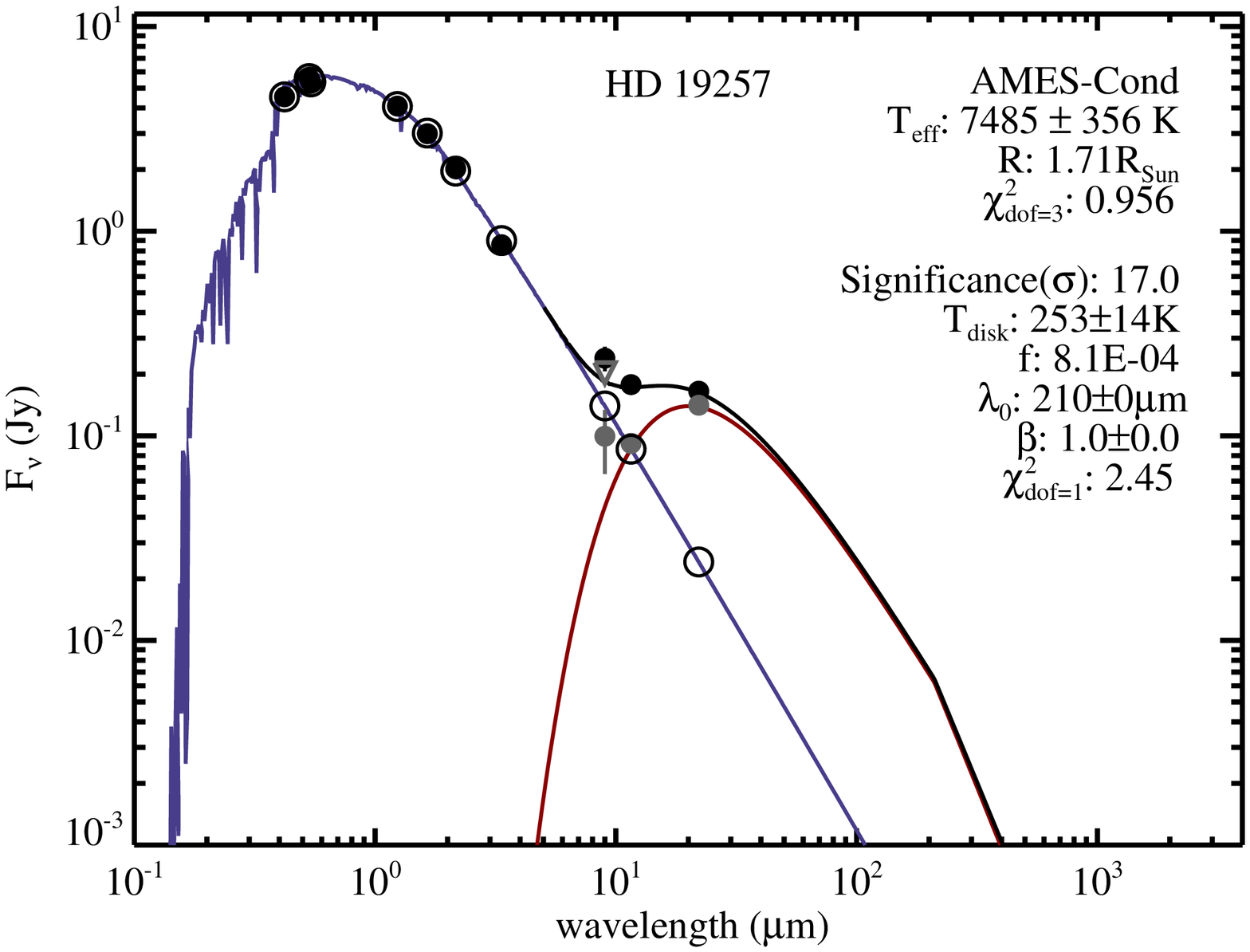}
    \includegraphics[width=0.45\textwidth]{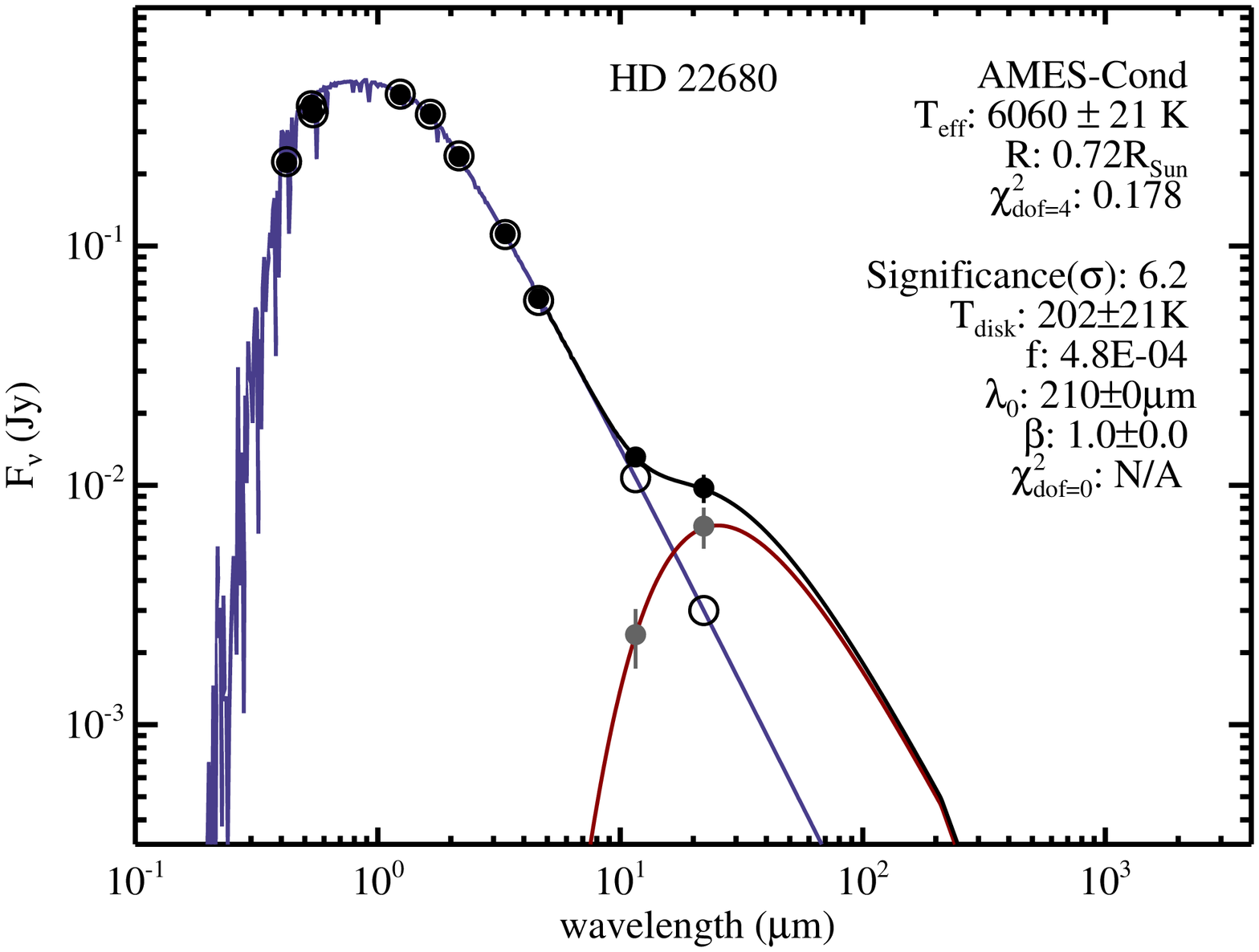}\\
    \hspace{-0.5cm} \includegraphics[width=0.45\textwidth]{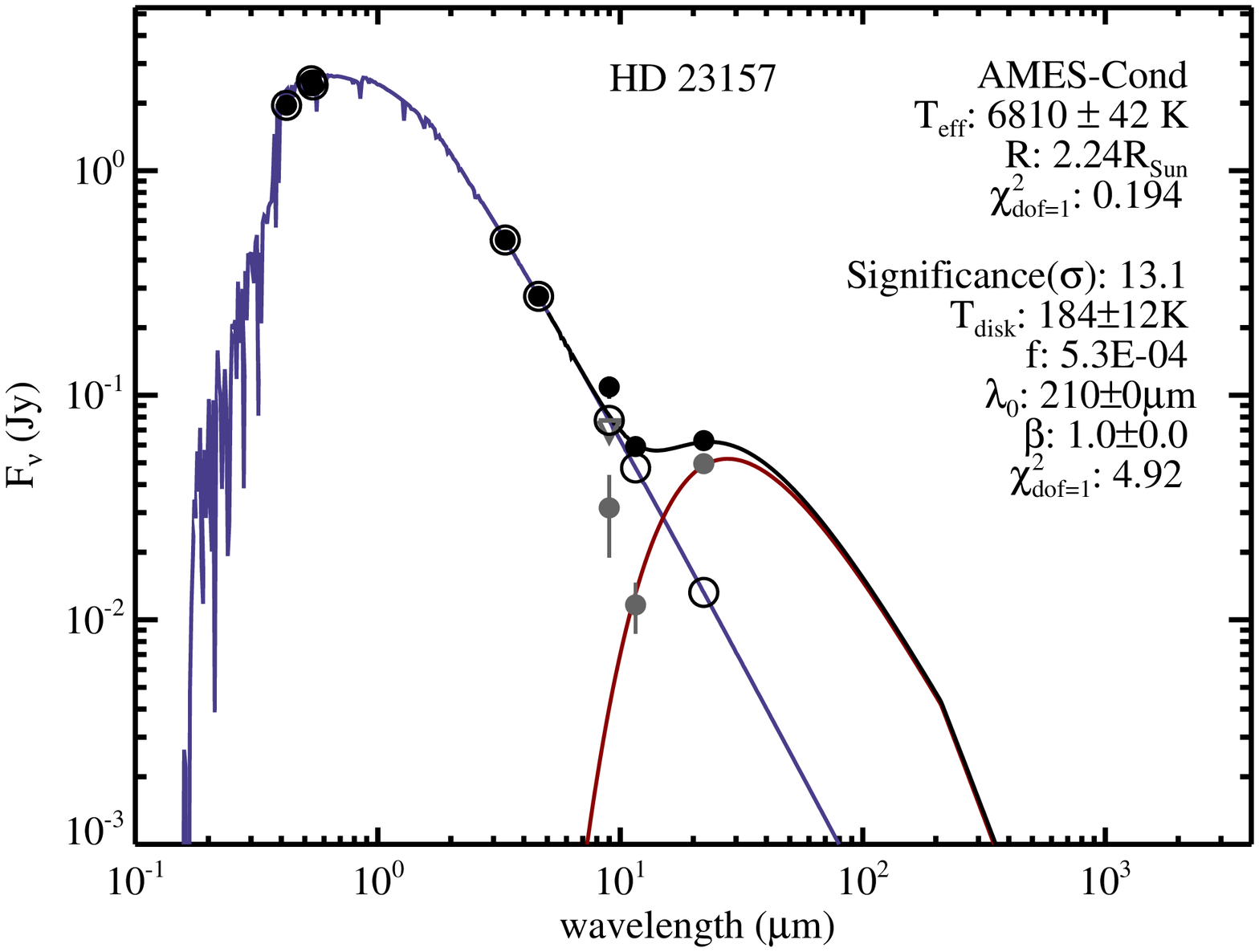}
    \includegraphics[width=0.45\textwidth]{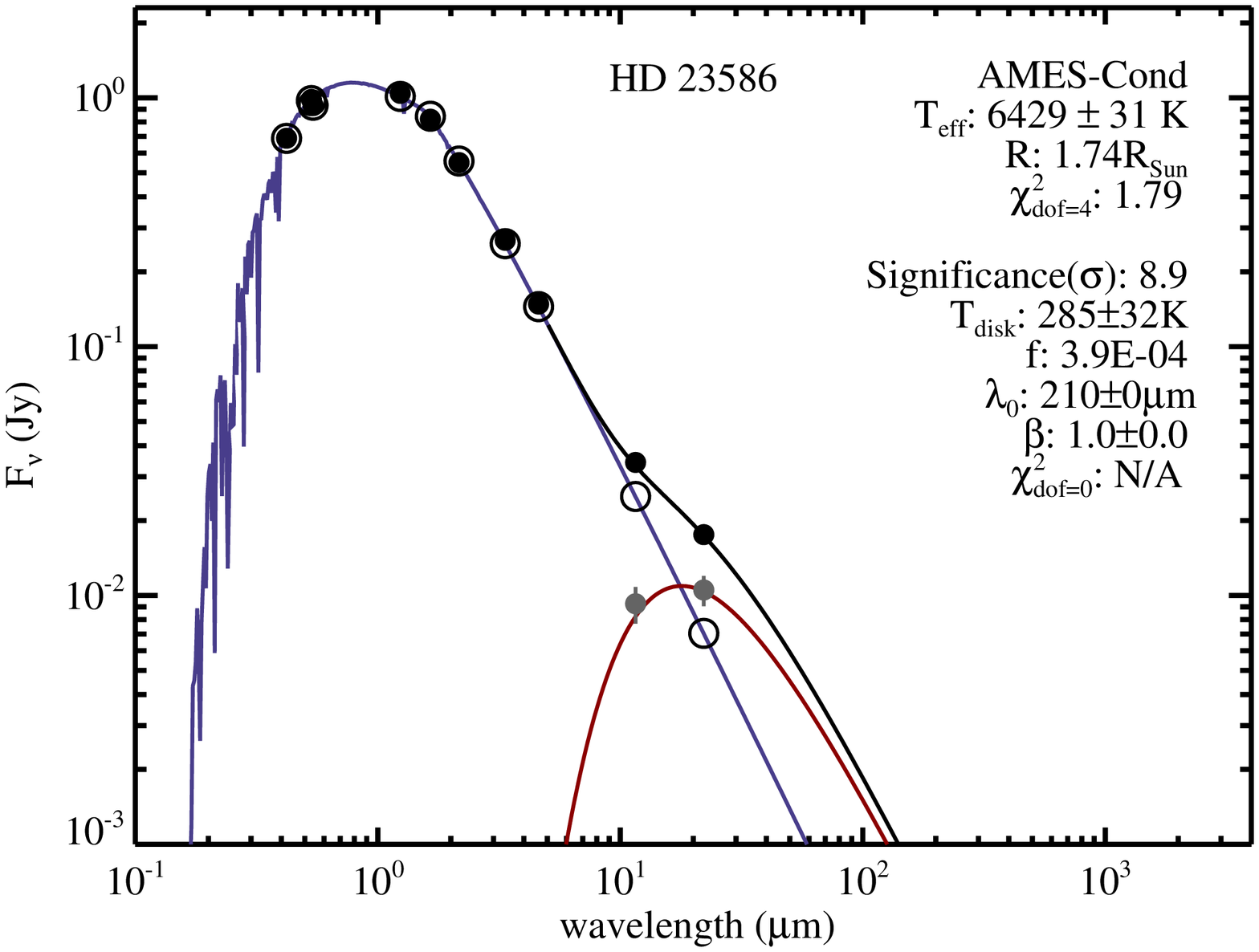}\\
  \end{center}
  \caption{SED modelling for the 25 candidate warm dust systems from Table
    \ref{tab:notes} (i.e. including the three that are excluded from the final list of
    22). Each panel shows PHOENIX stellar (blue), blackbody disk (red), and star+disk
    spectra (black). Photometry is shown as filled dots, and photospheric model fluxes at
    these wavelengths as open circles. Star-subtracted (i.e. disk) fluxes are shown as
    grey filled dots. Upper limits are shown as downward pointing triangles. The legend
    for each panel shows the parameters of the stellar and disk models. Modified
    blackbody parameters of $\lambda_0=210\mu$m and $\beta=1$ have been assumed
    \citep{2008ARA&A..46..339W}.}\label{fig:seds1}
\end{figure*}

\begin{figure*}
  \begin{center}
    \hspace{-0.5cm} \includegraphics[width=0.5\textwidth]{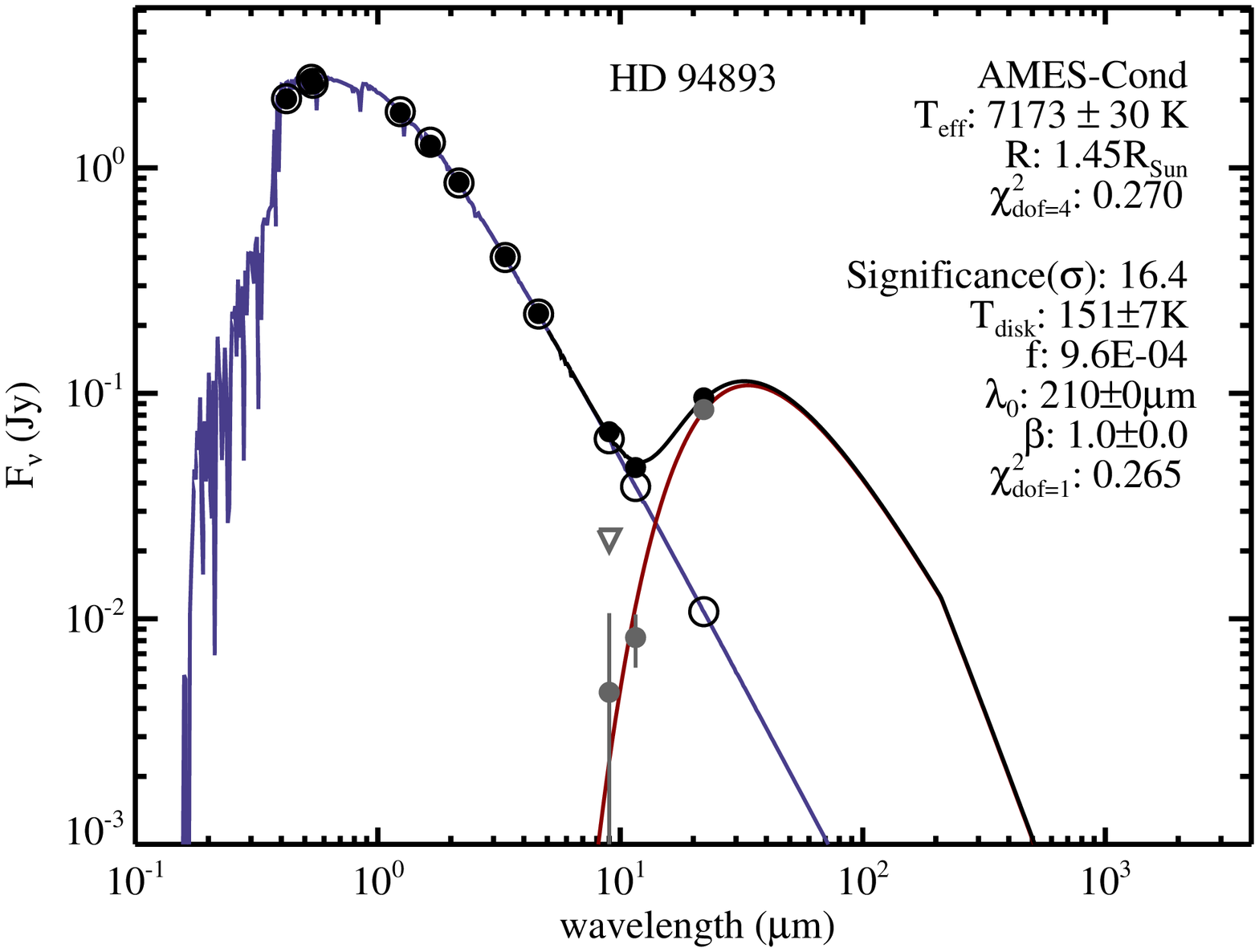}
    \includegraphics[width=0.5\textwidth]{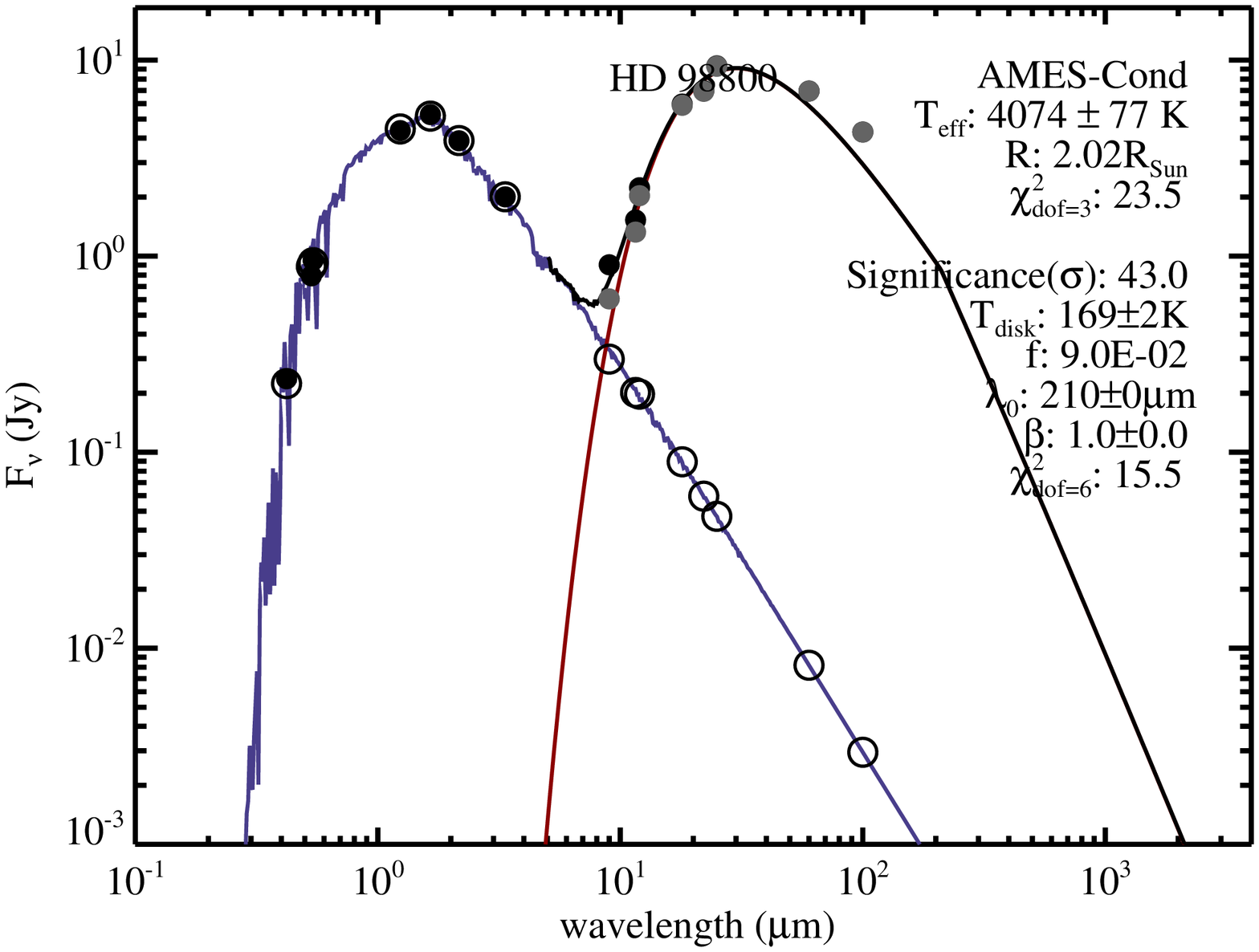}\\
    \hspace{-0.5cm} \includegraphics[width=0.5\textwidth]{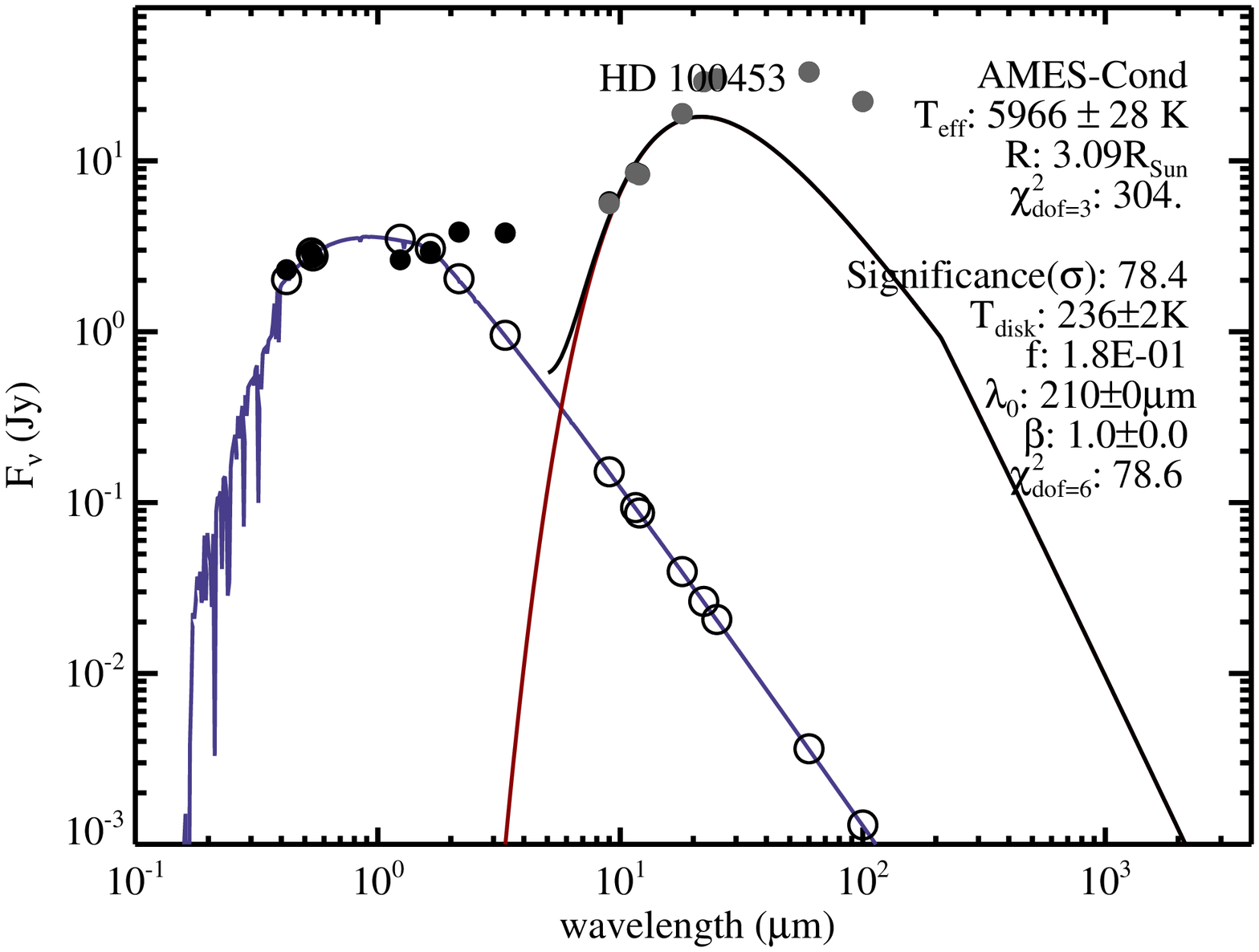}
    \includegraphics[width=0.5\textwidth]{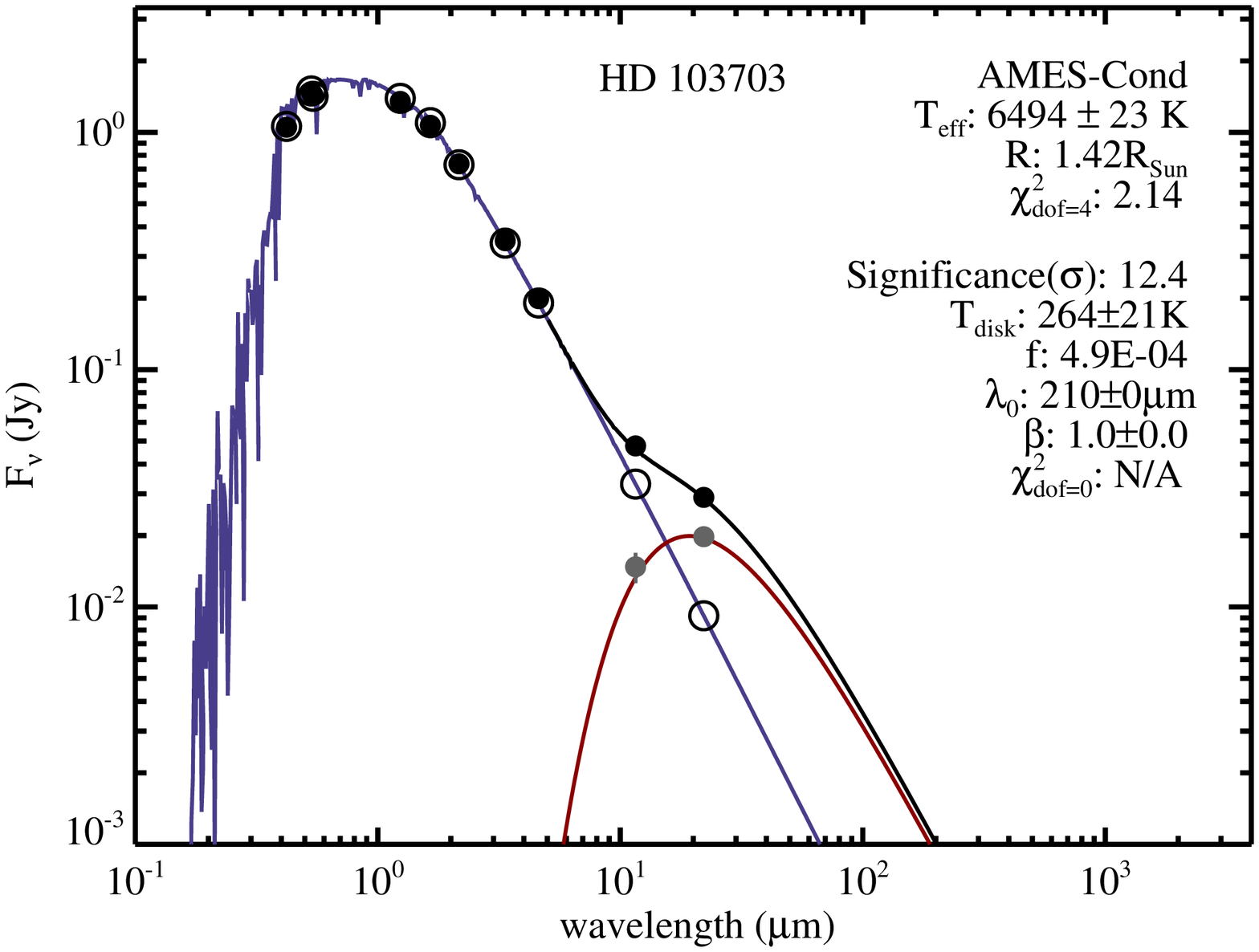}\\
    \hspace{-0.5cm} \includegraphics[width=0.5\textwidth]{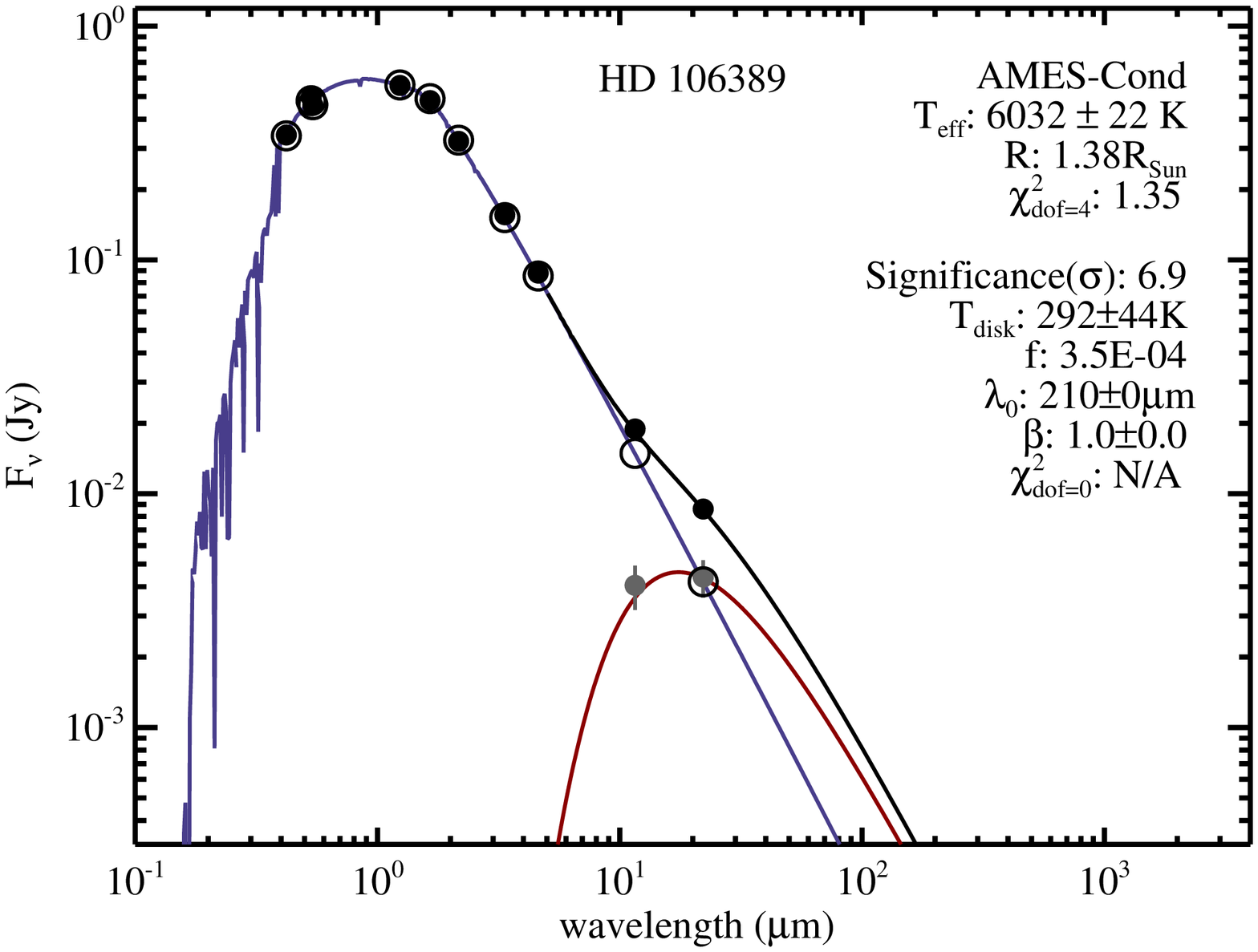}
    \includegraphics[width=0.5\textwidth]{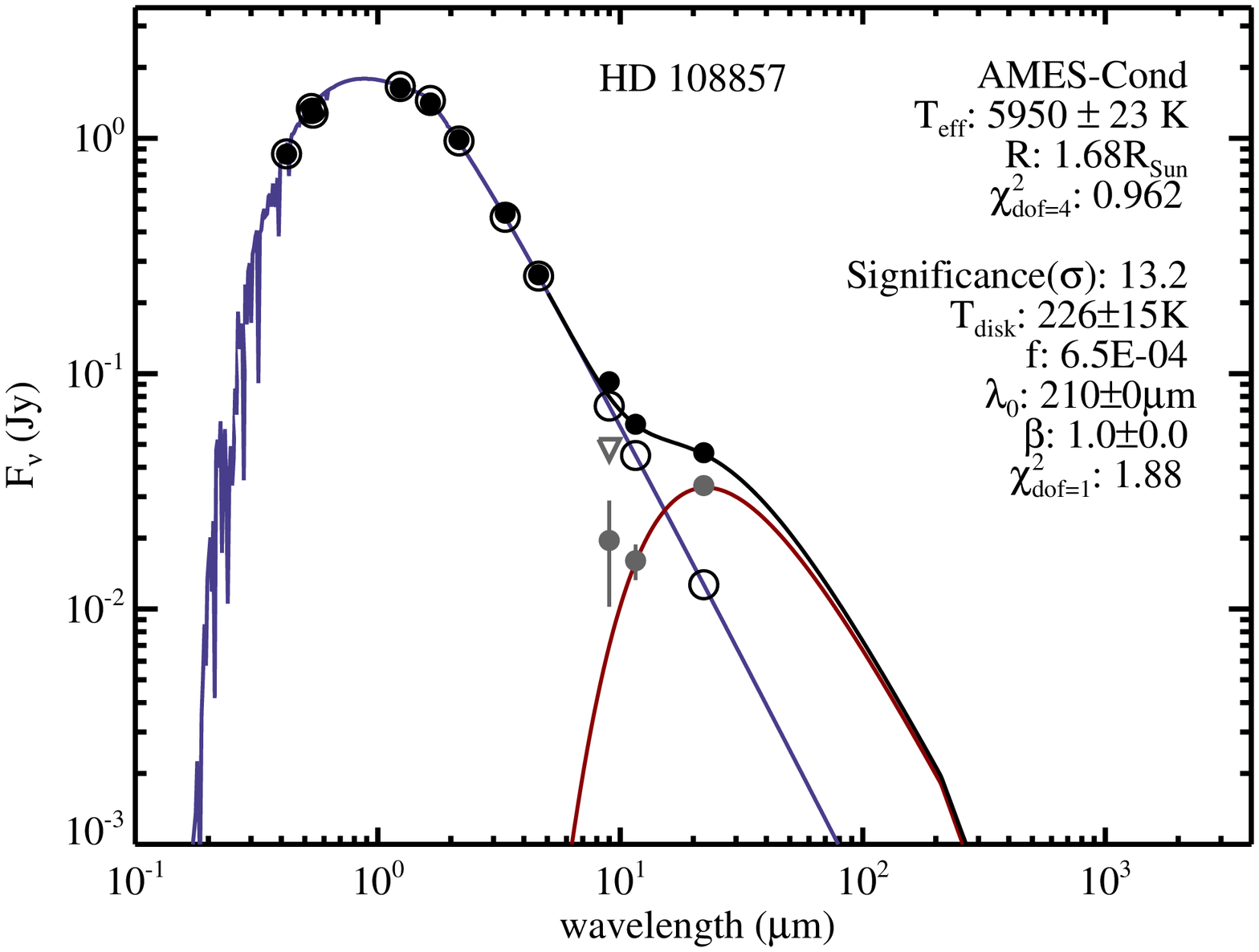}\\
  \end{center}
  \caption{25 warm dust systems cont.}\label{fig:seds2}
\end{figure*}

\begin{figure*}
  \begin{center}
    \hspace{-0.5cm}     \includegraphics[width=0.5\textwidth]{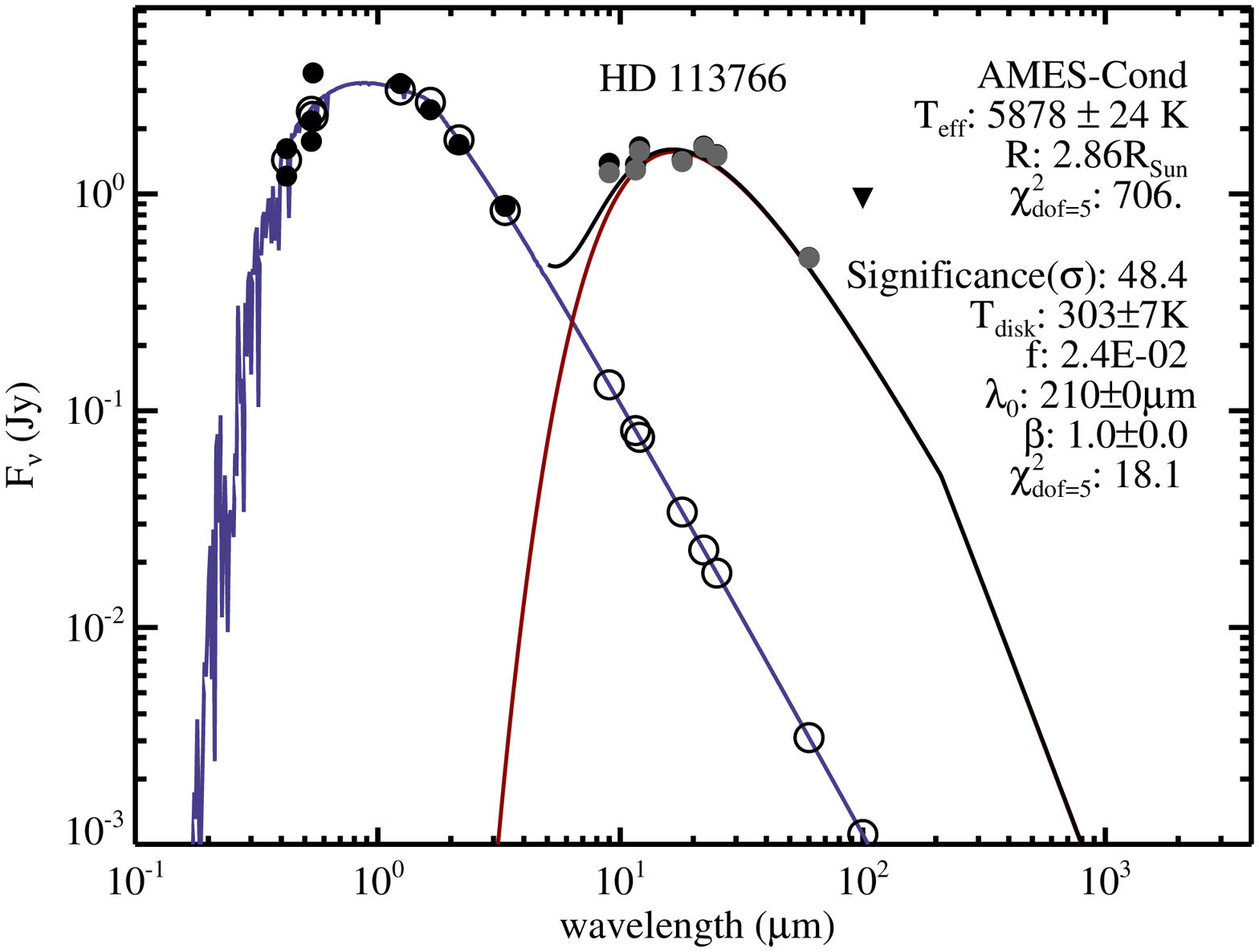}
    \includegraphics[width=0.5\textwidth]{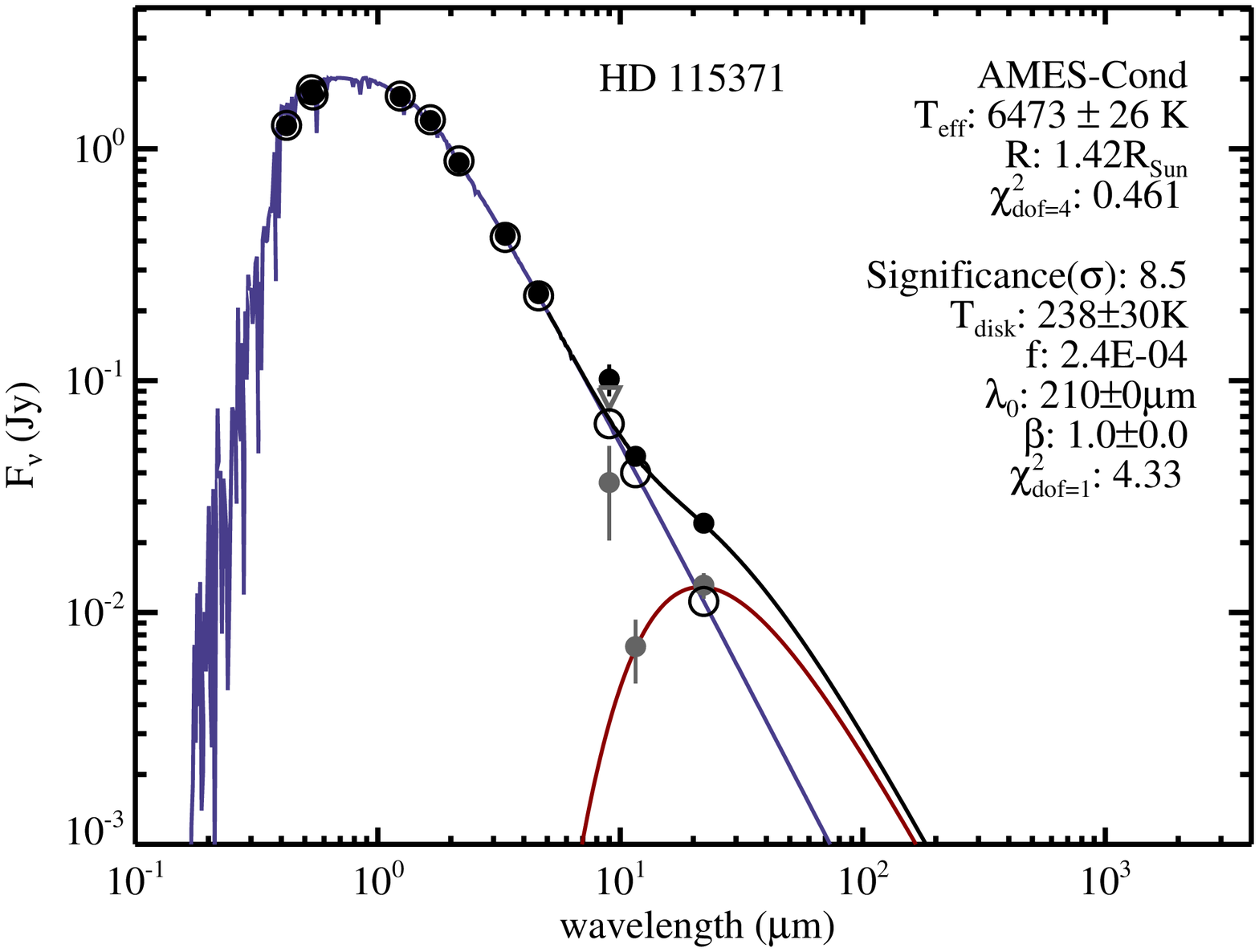}\\
    \hspace{-0.5cm}     \includegraphics[width=0.5\textwidth]{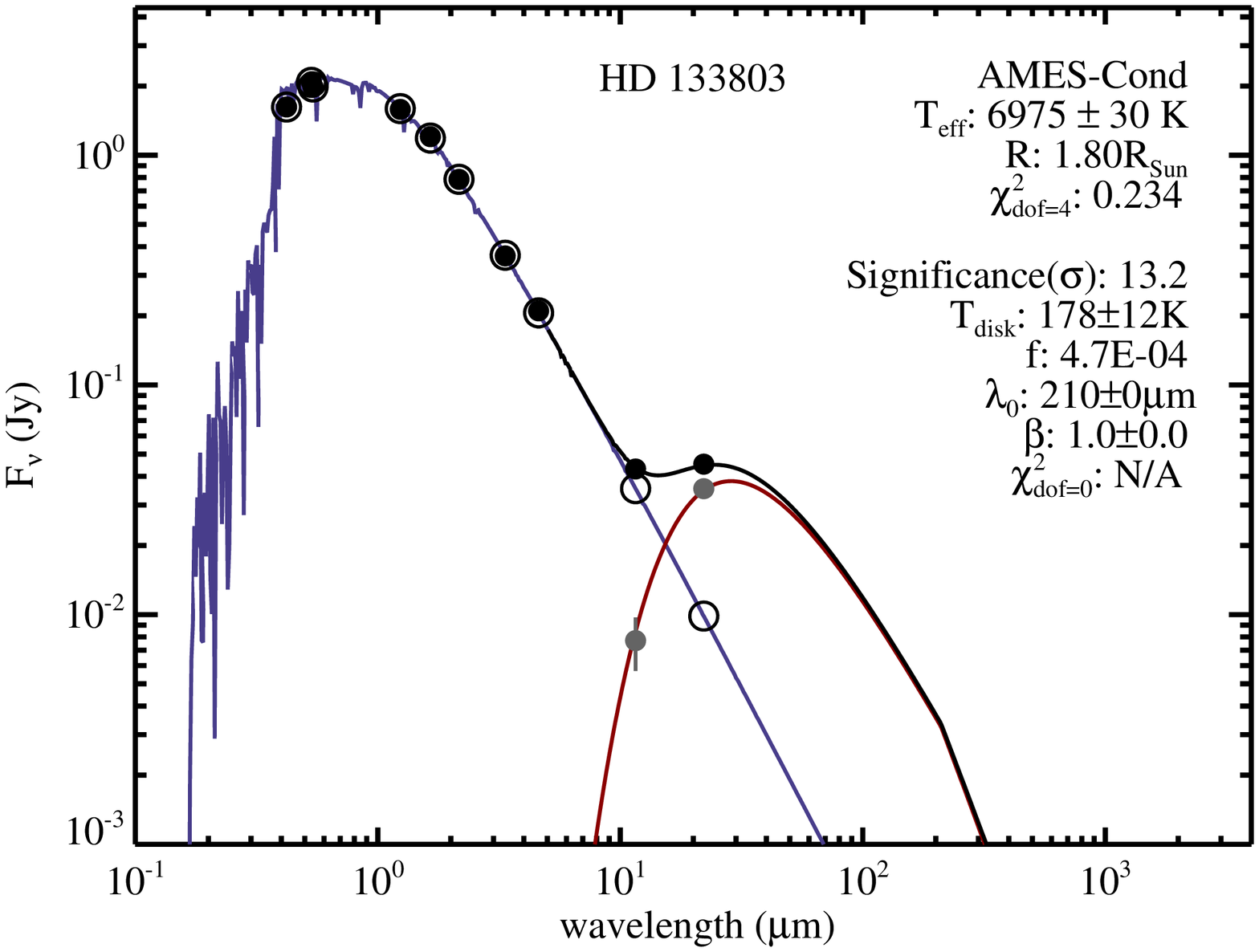}
 \includegraphics[width=0.5\textwidth]{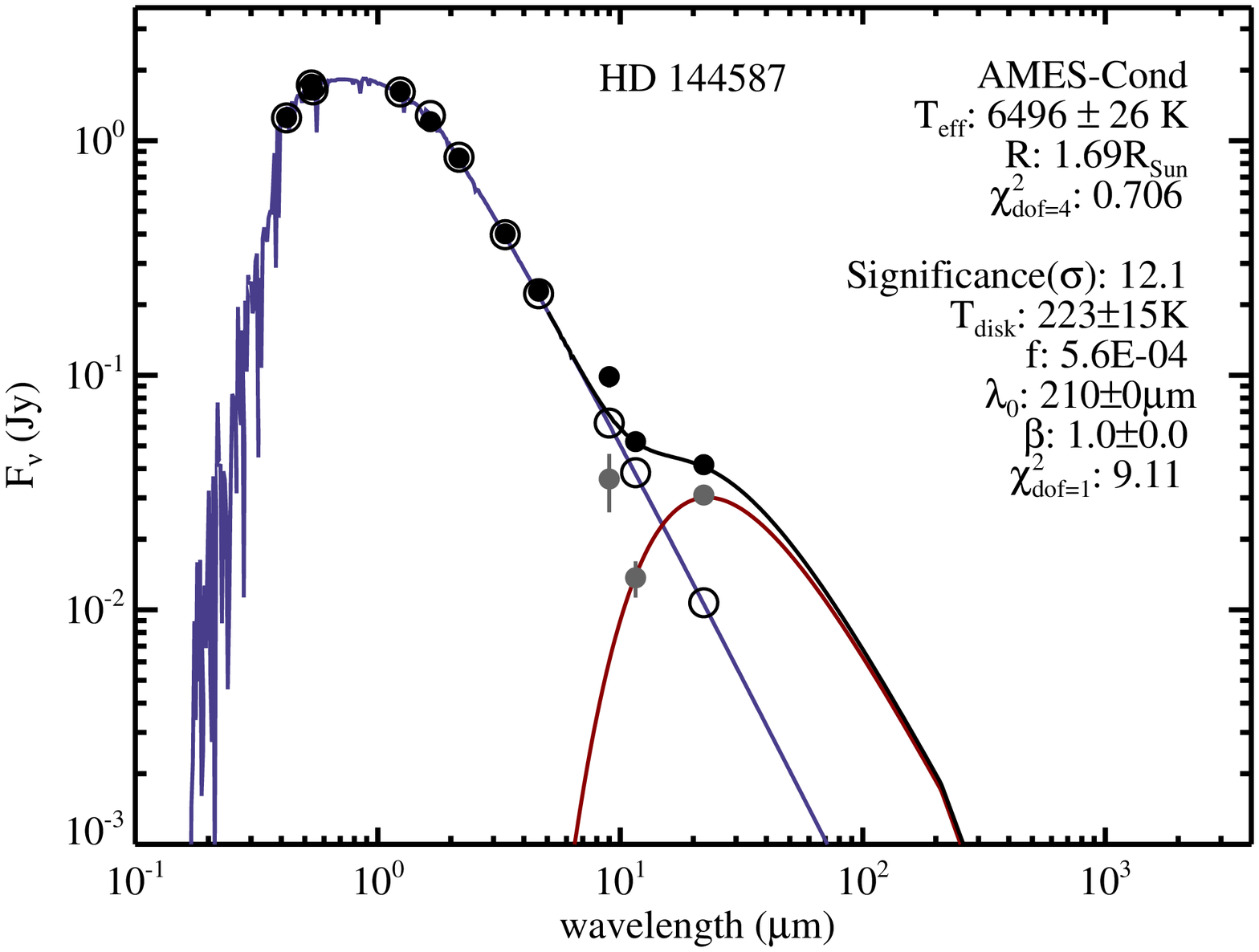}\\
    \hspace{-0.5cm}     \includegraphics[width=0.5\textwidth]{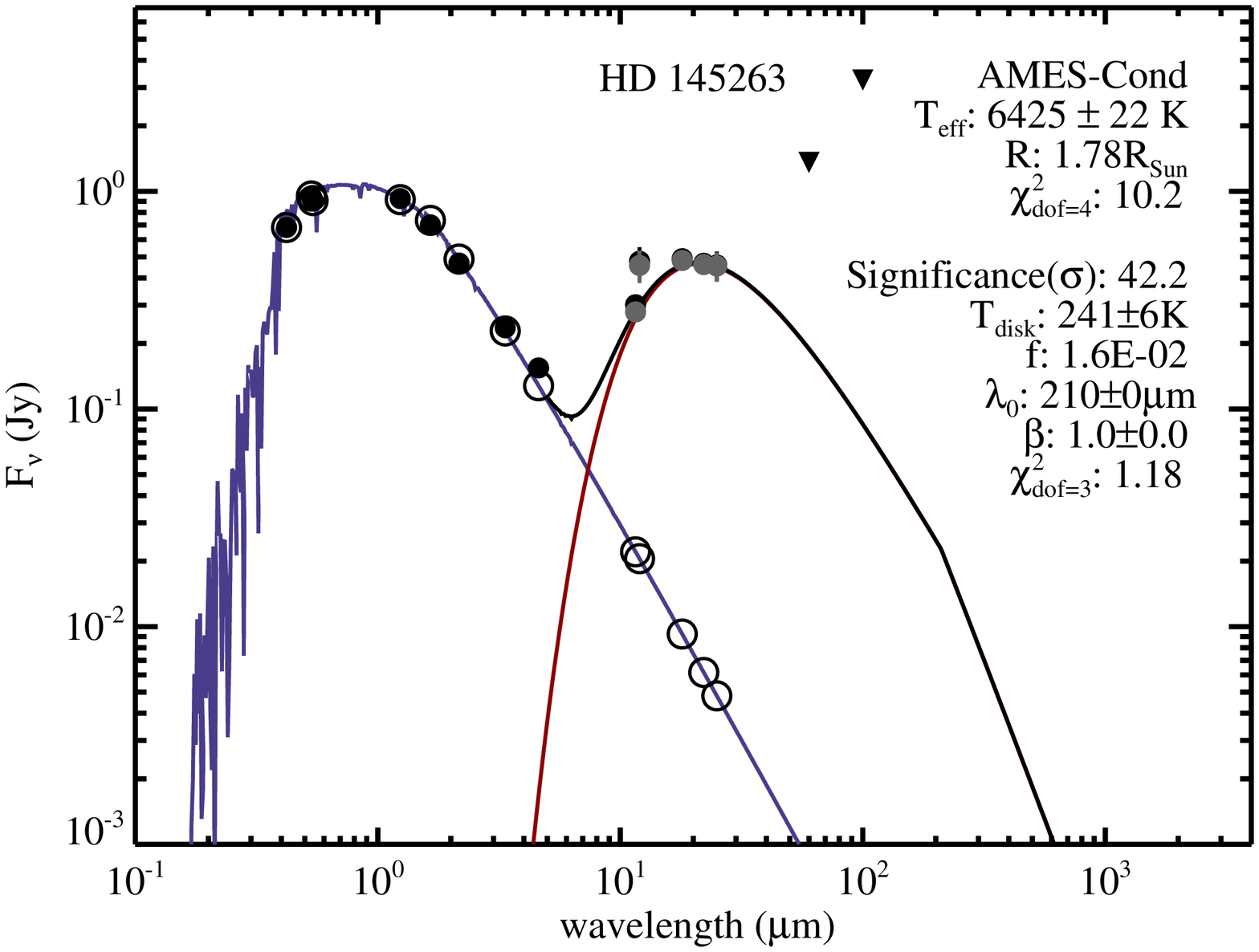}
    \includegraphics[width=0.5\textwidth]{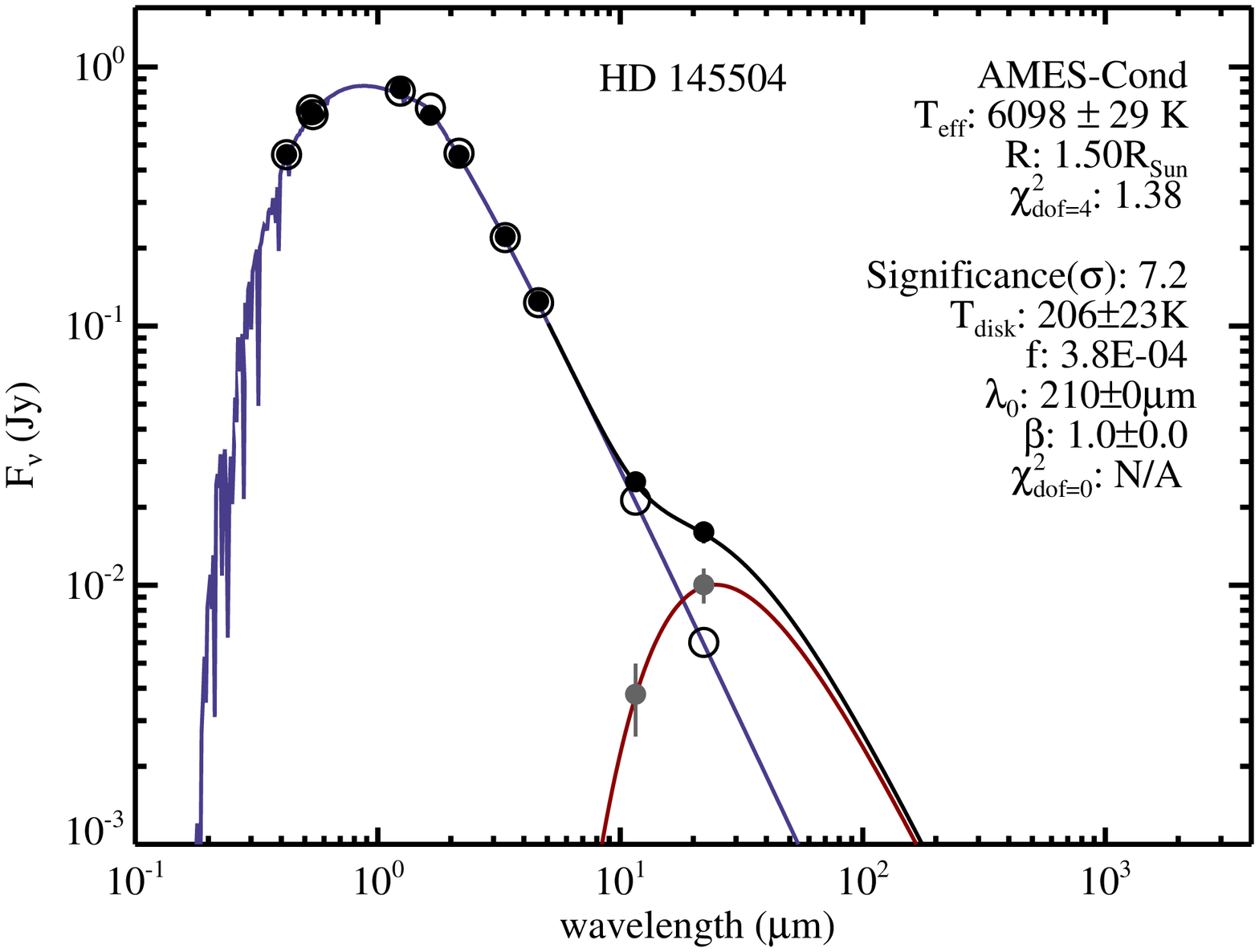}
  \end{center}
  \caption{25 warm dust systems cont.}\label{fig:seds3}
\end{figure*}

\begin{figure*}
  \begin{center}
        \hspace{-0.5cm}  \includegraphics[width=0.5\textwidth]{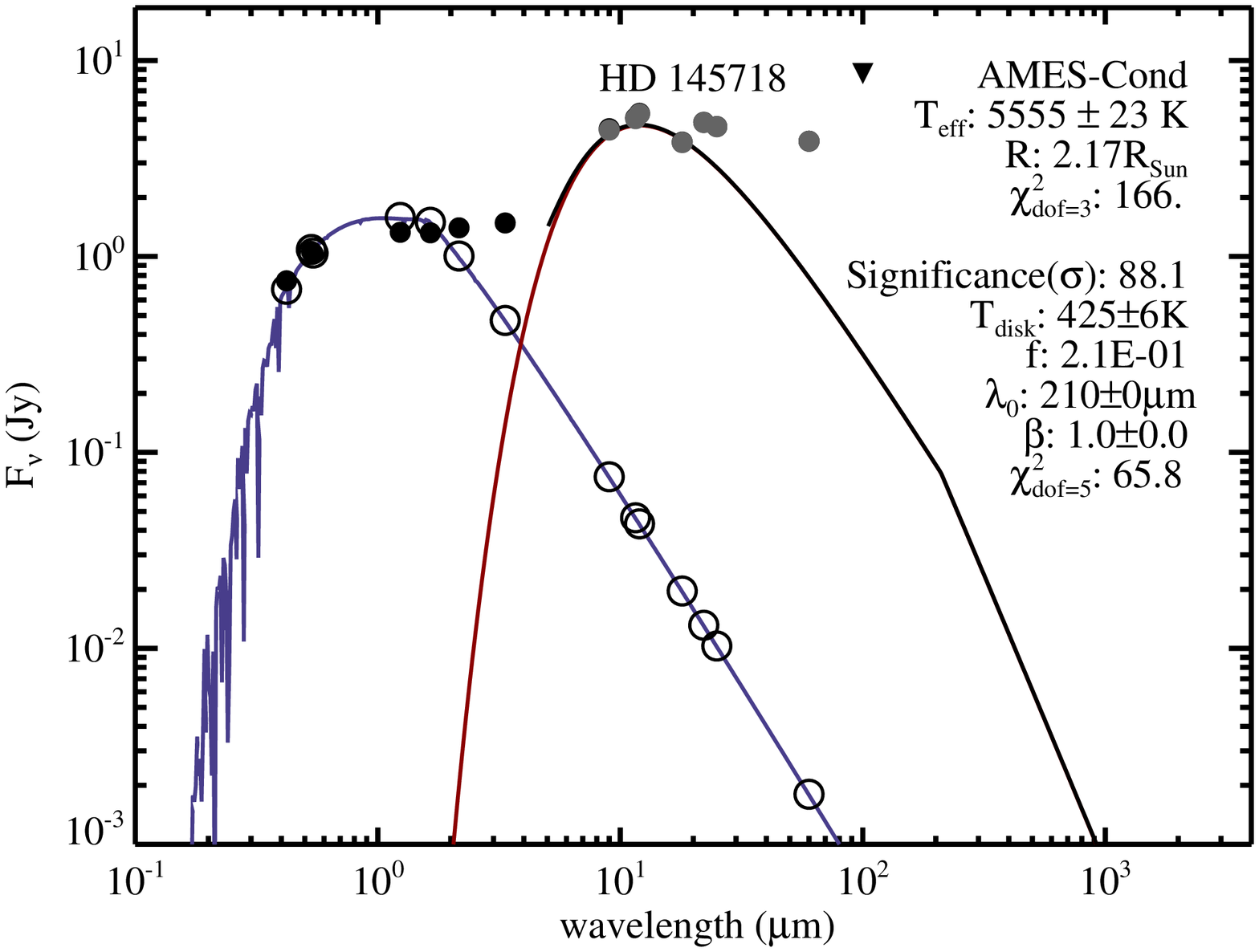}
        \includegraphics[width=0.5\textwidth]{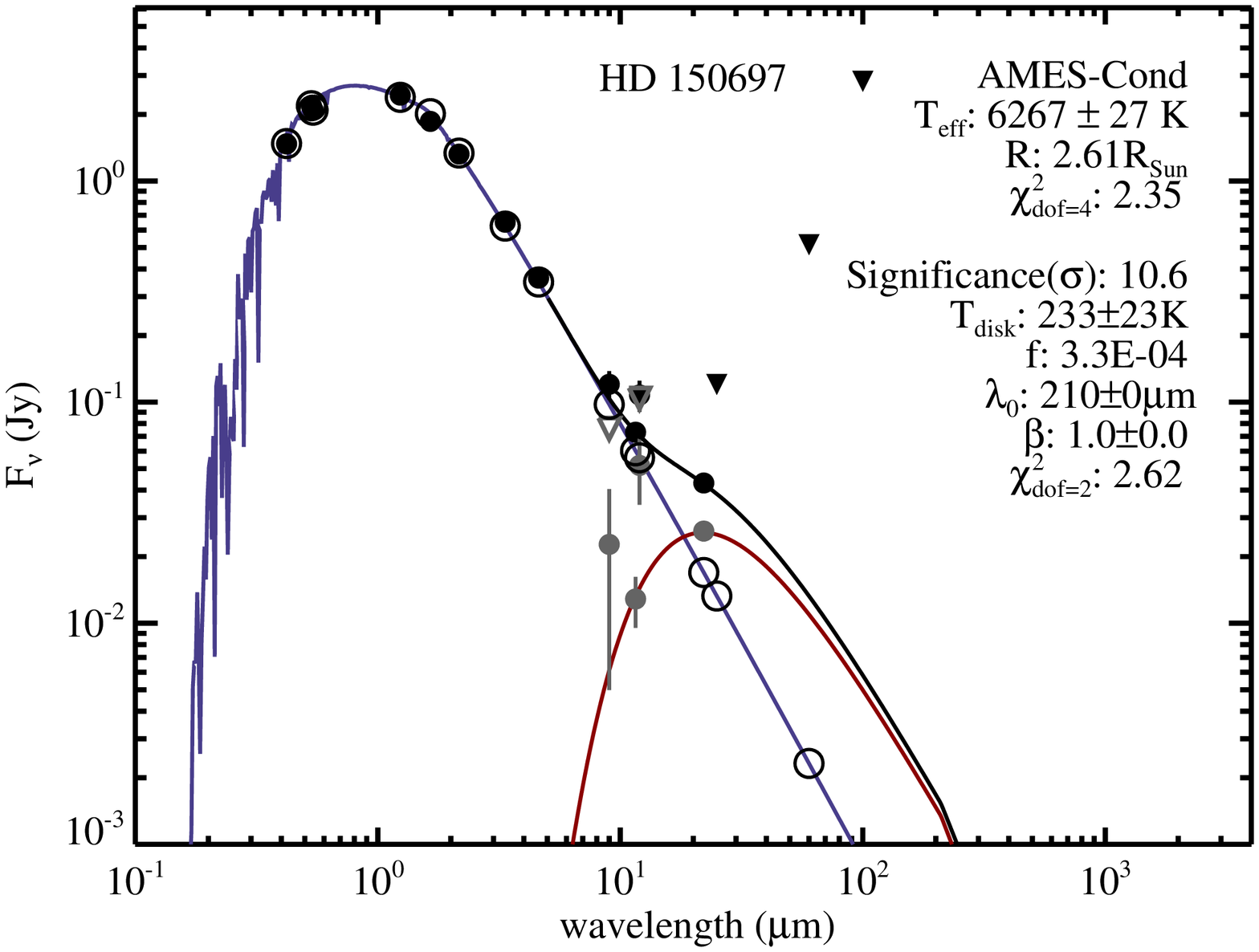}\\
    \hspace{-0.5cm} \includegraphics[width=0.5\textwidth]{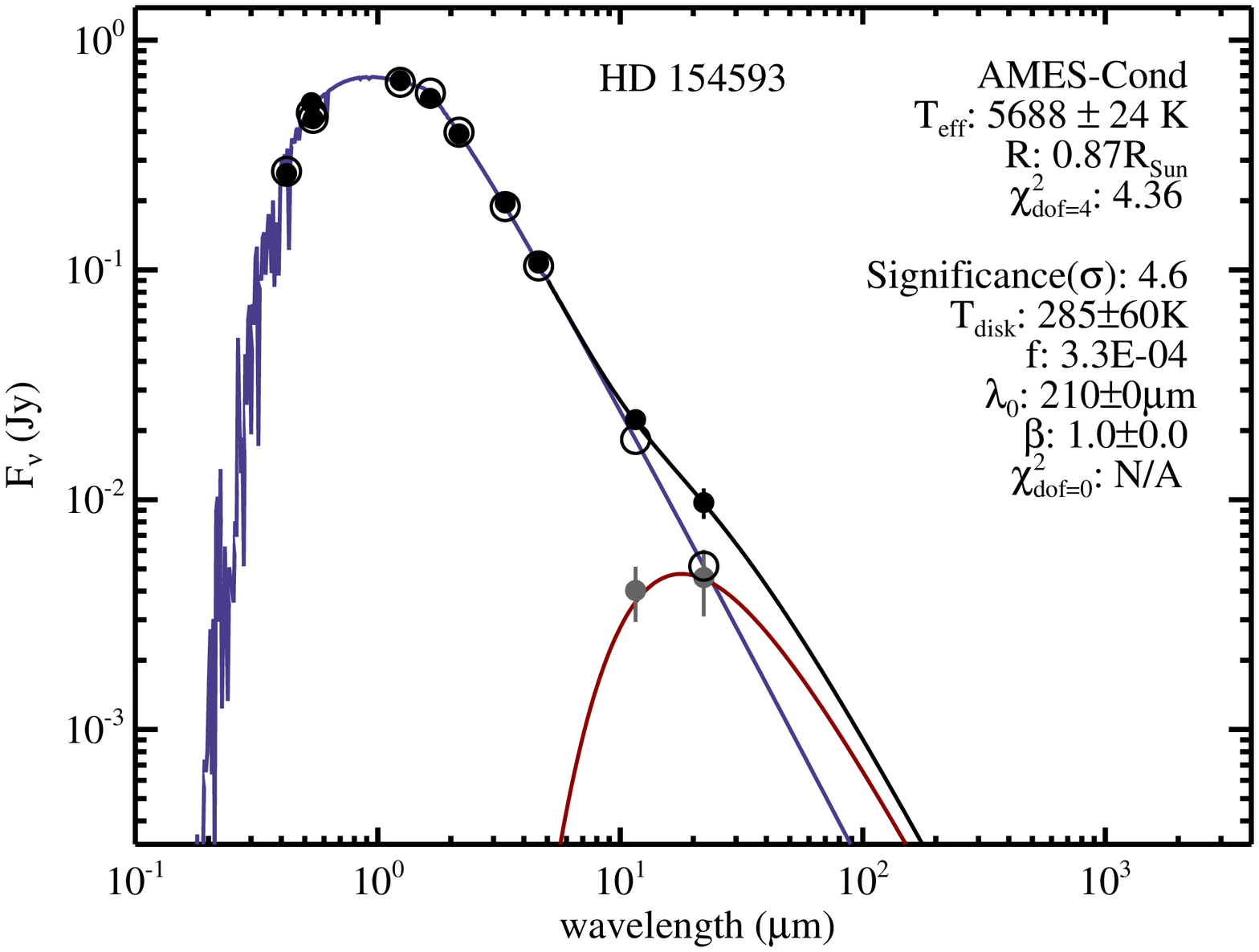}
    \includegraphics[width=0.5\textwidth]{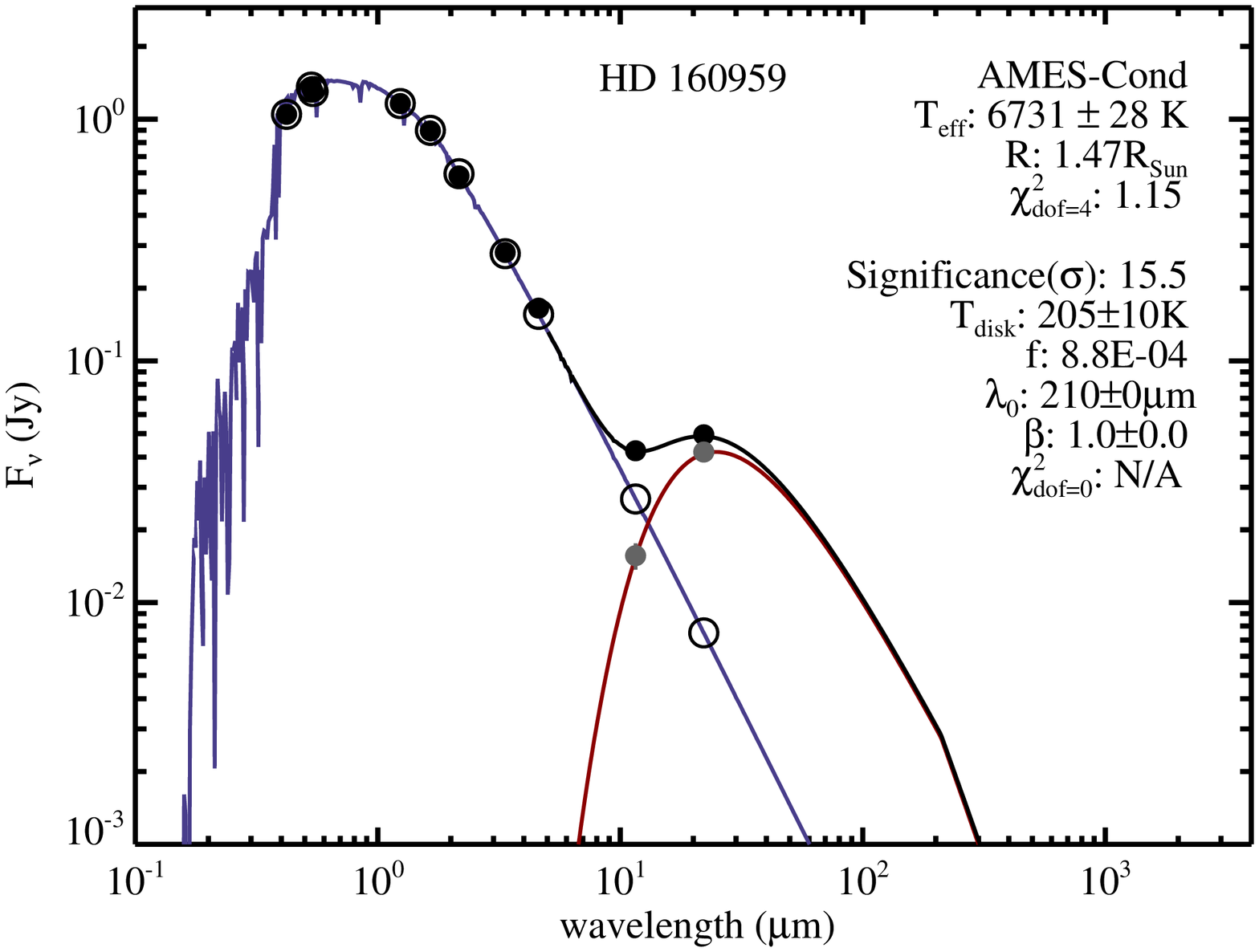}\\
    \hspace{-0.5cm} \includegraphics[width=0.5\textwidth]{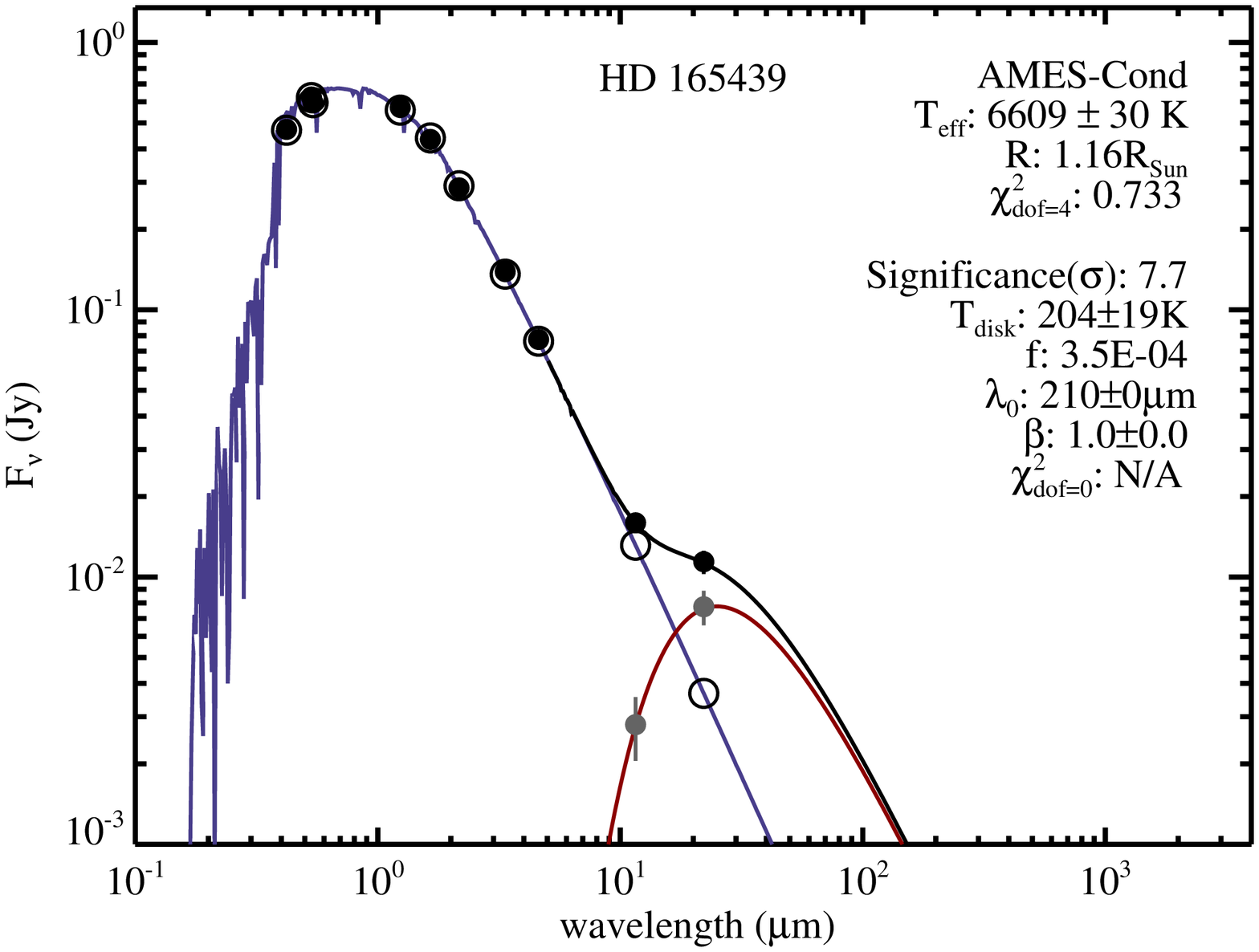}
    \includegraphics[width=0.5\textwidth]{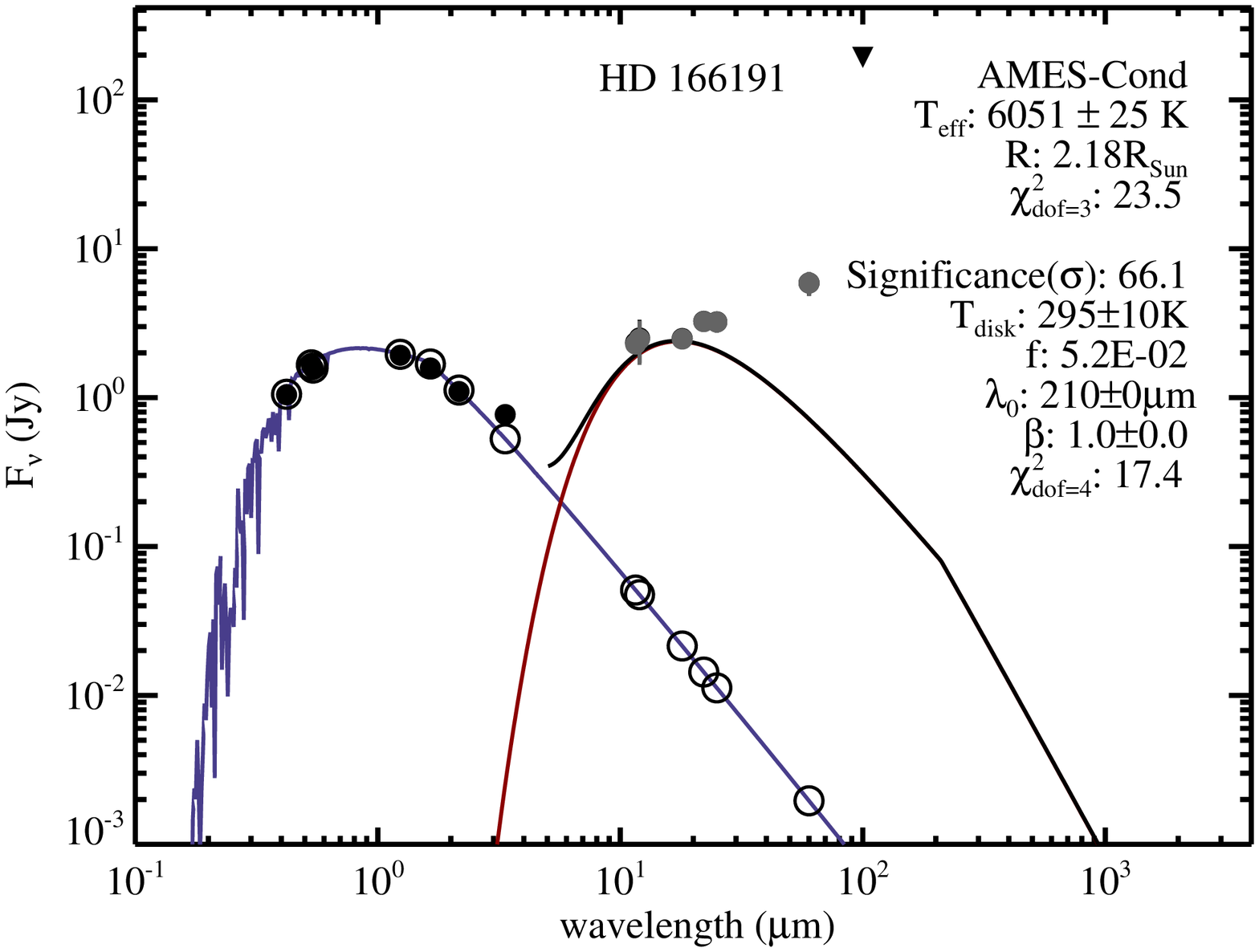}\\
  \end{center}
  \caption{25 warm dust systems cont.}\label{fig:seds4}
\end{figure*}

\begin{figure}
  \begin{center}
    \hspace{-0.5cm} \includegraphics[width=0.5\textwidth]{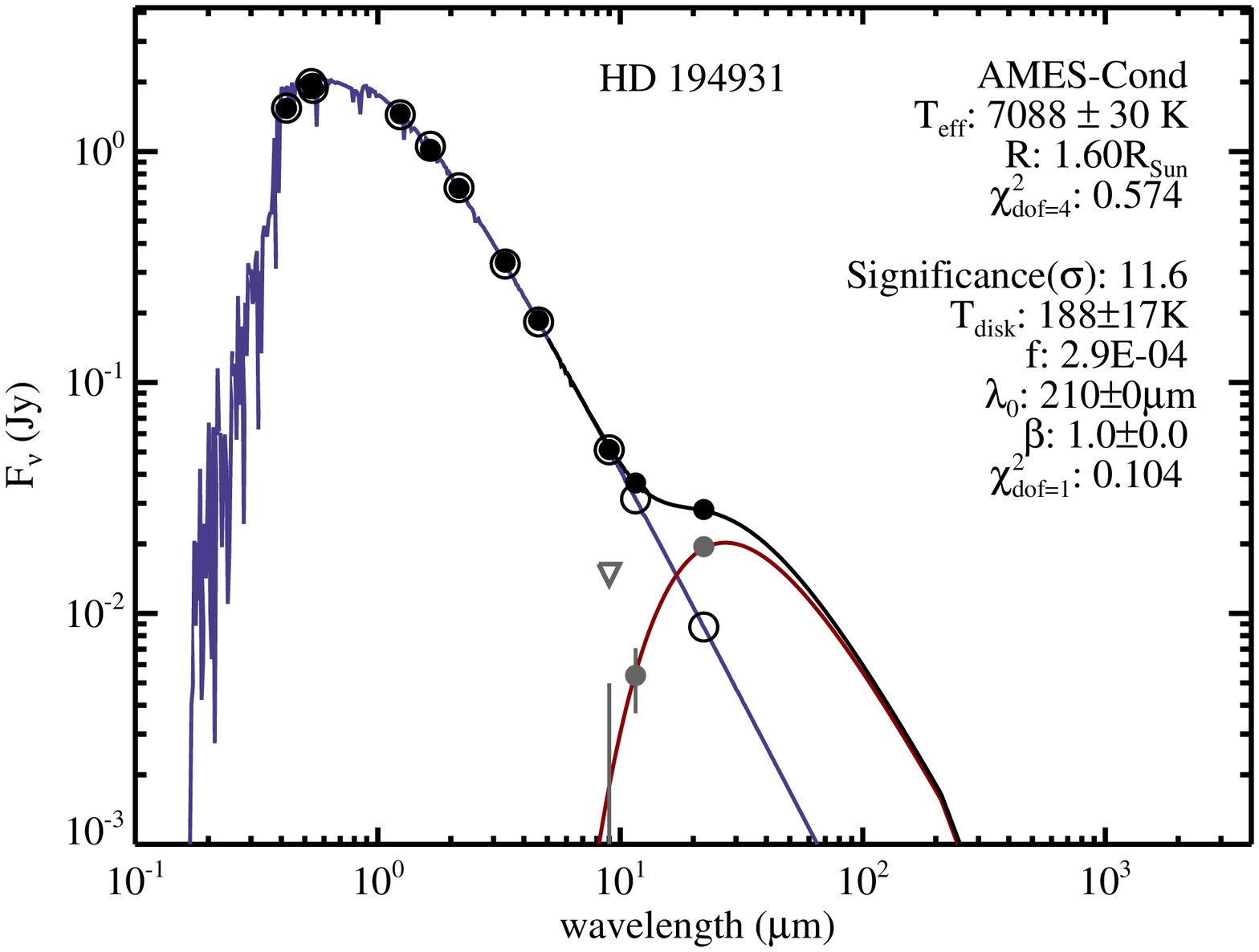}
  \end{center}
  \caption{25 warm dust systems cont.}\label{fig:seds5}
\end{figure}


\end{document}